\documentclass[a4paper,11pt]{article}
\pdfoutput=1
\usepackage{jheppub}
\usepackage[T1]{fontenc}

\usepackage{tensor}
\usepackage{mathtools}
\usepackage{multirow}

\usepackage{graphicx,graphics,multicol}
\usepackage{tikz}
\usetikzlibrary{decorations.pathmorphing}
\usetikzlibrary{decorations.pathreplacing}
\usetikzlibrary{decorations.markings}
\usetikzlibrary{arrows.meta}
\usepgflibrary{shapes.geometric}

\tikzset{
   vector2/.style={decorate, decoration={snake, amplitude=1pt, segment length=6pt}, draw,double},
   vector/.style={decorate, decoration={snake, amplitude=1pt, segment length=6pt}, draw},
	provector/.style={decorate, decoration={snake,amplitude=2.5pt}, draw},
	antivector/.style={decorate, decoration={snake,amplitude=-2.5pt}, draw},
    fermion/.style={draw=black, postaction={decorate},
        decoration={markings,mark=at position .55 with {\arrow[draw=black]{>}}}},
    fermionbar/.style={draw=black, postaction={decorate},
        decoration={markings,mark=at position .55 with {\arrow[draw=black]{<}}}},
    fermionnoarrow/.style={draw=black},
    gluon/.style={decorate, draw=black,
        decoration={coil,amplitude=4pt, segment length=5pt}},
    scalar/.style={dashed,draw=black, postaction={decorate},
        decoration={markings,mark=at position .55 with {\arrow[draw=black]{>}}}},
    scalarbar/.style={dashed,draw=black, postaction={decorate},
        decoration={markings,mark=at position .55 with {\arrow[draw=black]{<}}}},
    scalarnoarrow/.style={dashed,draw=black},
    electron/.style={draw=black, postaction={decorate},
        decoration={markings,mark=at position .55 with {\arrow[draw=black]{>}}}},
	bigvector/.style={decorate, decoration={snake,amplitude=4pt}, draw},
}
\tikzset{cross/.style={cross out, draw, 
         minimum size=2*(#1-\pgflinewidth), 
         inner sep=0pt, outer sep=0pt}}

\usepackage{marvosym}
\usepackage{comment}
\usepackage{ytableau}
\ytableausetup{mathmode,boxsize=.6em}

\usepackage[colorinlistoftodos]{todonotes}
\setuptodonotes{color=cyan}

\title{\boldmath Standard Model EFTs via On-Shell Methods}

\author[a]{Manuel Accettulli Huber}
\author[a]{and Stefano De Angelis}

\affiliation[a]{Centre for Research in String Theory\\ School of Physics and Astronomy\\ Queen Mary University of London,\\Mile End Road, London E1 4NS, United Kingdom}

\emailAdd{m.accettullihuber@qmul.ac.uk}
\emailAdd{s.deangelis@qmul.ac.uk}

\abstract{We present the Standard Model Effective Field Theories (SMEFT) from purely on-shell arguments. Starting from a few basic assumptions such as Poincaré invariance and locality, we classify all the renormalisable and non-renormalisable interactions at the lowest order in the couplings. From these building blocks, we review how locality and unitarity enforce Lie algebra structures to appear in the S-matrix elements together with relations among couplings (and hypercharges). Furthermore, we give a fully on-shell algorithm to compute any higher-point tree-level amplitude (or form factor) in generic EFTs, bypassing BCFW-like recursion relations which are known to be problematic when non-renormalisable interactions are involved. Finally, using known amplitudes techniques we compute the mixing matrix of SMEFT irrelevant interactions up to mass dimension 8, to linear order in the effective interactions.
}

\newcommand{\NN}{\mathcal{N}}
\newcommand{\OO}{\mathcal{O}}
\newcommand{\Aa}{\mathcal{A}}
\newcommand{\Tr}{\mathrm{Tr}}

\newcommand{\agl}[2]{\langle#1\, #2 \rangle}
\newcommand{\sqr}[2]{\lbrack #1\, #2 \rbrack}
\newcommand{\tlambda}{\widetilde{\lambda}}

\begin{document} 
\maketitle
\flushbottom

\section{Introduction}
The Standard Model Effective Field Theories (SMEFT) are a systematic and model independent framework to characterise both experimental deviation from predictions of the Standard Model (SM) and possible extensions beyond it (for a review, see \cite{Brivio:2017vri} and references therein). Indeed, up to now, LHC measurements of cross-sections are compatible with SM theoretical predictions. Nonetheless the SM is expected to be an incomplete description of Nature: many theoretical puzzles are still unsolved, including but not limited to the hierarchy problem, the magnitude of the quartic $\lambda$ coupling of the Higgs, the origin of CP violation in the quark sector, or the unnatural pattern of the Yukawa couplings. More recently, also an experimental deviation from SM predictions has been measured in the $g_\mu-2$ experiment. \cite{Muong-2:2006rrc,Aoyama:2020ynm,Muong-2:2021ojo}.

Then, how should we look for new physics beyond the SM? When considering extensions of the SM, additional heavy modes with mass $\Lambda$ can be integrated out at energie scales $E \ll \Lambda$. This leaves us, in the usual Lagrangian formalism, with some effective interactions which can be organised in terms of their mass dimension as
\begin{equation}
    \mathcal{L}_{\rm SMEFT} = \mathcal{L}_{\rm SM} + \sum_{i,j}\frac{c_{i}^{(j)}}{\Lambda^{j-4}}\,\mathcal{O}^{(j)}_i\ .
\end{equation}
%Then how should we look for new physics beyond the SM? We know that additional heavy modes with mass $M$ at energies $E\ll M$ can be integrated out and leave us with effective interactions in the usual Lagrangian formalism, with mass dimension larger than 4.
The first example are the dimension-5 Weinberg operators \cite{Weinberg:1979sa} which generate light Majorana-like neutrino masses:
\begin{equation}
    \mathcal{L}_{\rm SMEFT} = \mathcal{L}_{\rm SM} + \frac{C_{m n}}{\Lambda_L} \left( \epsilon_{i k} \epsilon_{j l} L_{m}^i L_{n}^j H^k H^l + \epsilon^{i k} \epsilon^{j l}  \bar{L}_{m\, i} \bar{L}_{n\, j} \bar{H}_k \bar{H}_l \right)\ ,
\end{equation}
where $m,n=1,2,3$ are flavour indices and $i,j$ are $SU(2)$ indices, and $\Lambda_L$ is the natural cut-off of the effective theory. More precisely $\Lambda_L$ is the cut-off scale for effective interactions which violate lepton and barion number, as opposed to the scale $\Lambda$ associated to lepton/barion number preserving interactions. Experimental constraints on the neutrino masses put the lower bound on the cut-off scale at $\Lambda_L/C_{m n} \gtrsim 10^{15}\, {\rm GeV}$, which are scales currently impracticable for the observation of new physics.
%Our knowledge of neutrino masses puts a very high lower bound on the cut-off $\Lambda_L/C_{m n} \gtrsim 10^{15}\, {\rm GeV}$. Then in the SMEFT approach, we usually consider two different scales, $\Lambda$ for effective interaction which do not violate lepton and barion numbers and $\Lambda_L$ for those interaction which actually do violate them.
%In fact, the leading contributions to the SMEFT come from dimension-6 operators which have been studied extensively. A first operator basis was found already in \cite{Buchmuller:1985jz}, only later the complete one-loop mixing of the operators was obtained (Manohar, Trott Alonso) and just very recently the structure of the two-loop anomalous dimensions (in an $SU(N)$ model) was studied (Bern). Despite their relevance, dimension-6 operators are clearly not the end of the story, and recently  
In fact, the leading contributions to the SMEFT come from dimension-6 operators \cite{Henning:2014wua,Brivio:2017btx,Brivio:2019myy,Dawson:2020oco,David:2020pzt,Ellis:2020unq,Trott:2021vqa}, but there are interesting processes for which the dominant contribution comes from even higher-dimensional operators. Some examples include the light-by-light scattering \cite{Ellis:2017edi}, the light production via gluon fusion \cite{Ellis:2018cos} and the neutral bosons production \cite{Ellis:2020ljj} and even, in some scenarios, the $g_\mu -2$ \cite{Arkani-Hamed:2021xlp} and Higgs production in association with a $W$ boson \cite{Hays:2018zze}, which receive the first contribution from dimension-8 operators. Dimension-8 operators can play a relevant role even when appearing as subleading contributions \cite{Corbett:2021cil}, and recently studies of their impact on SMEFT have been performed \cite{Alioli:2020kez,Boughezal:2021tih,Corbett:2021eux,Corbett:2021jox,Martin:2021vwf}.

In general, the classification of irrelevant operators in the Standard Model is a complicated task. The counting of non-redundant operators can be performed via the Hilbert series method, as shown in \cite{Lehman:2015via,Henning:2015daa,Henning:2015alf}, however an explicit construction of the SMEFT operators is rather involved. Traditional techniques require to take care separately of many source of redundancy, {\it e.g.} Bianchi identities and IBP identities of operators with derivative insertions, field redefinitions and Fierz identities. More recently, a more direct way of constructing this basis has been proposed, which relies on the classification of the independent effective interactions directly from their S-matrix elements \cite{Shadmi:2018xan,Ma:2019gtx,Aoude:2019tzn,Durieux:2019eor,Falkowski:2019zdo,Durieux:2019siw,Falkowski:2020fsu,Durieux:2020gip}, and has been used to classify all the SMEFT operators up to mass dimension 9 \cite{Li:2020gnx,Li:2020xlh}.
%In general, the classification of irrelevant operators in the Standard Model is a complicated task. The counting of non-redundant operators is usually performed via Hilbert series method, as shown in \cite{Lehman:2015via,Henning:2015daa,Henning:2015alf}. An explicit construction of SMEFT operators is more involved, and traditional techniques require to take care separately of many source of redundancy ({\it e.g.} Bianchi identities and IBP identities of operators with derivative insertions, field redefinitions, Fierz identities). More recently, a more direct way to construct this basis explicitly has been proposed, which goes through the classification of the independent effective interaction directly from their S-matrix elements \cite{Shadmi:2018xan,Ma:2019gtx,Durieux:2019eor,Falkowski:2019zdo,Falkowski:2020fsu,Durieux:2020gip}. This allowed to classify SMEFT operators up to mass dimension 9 \cite{Li:2020gnx,Li:2020xlh}.

Following this line of reasoning, in this paper we will present a fully on-shell construction not only of the effective interactions but of the SM itself, avoiding any mention of the Lagrangian formalism other than for comparison purposes. In the recent years, on-shell methods have proven to be the most powerful techniques in a variety of settings, such as collider physics \cite{Ellis:2009zyy,Berger:2010zx}, the study of the ultraviolet (UV) behaviour of $\mathcal{N}=8$ supergravity \cite{Bern:2017ucb,Bern:2018jmv}, the study of the inspiral phase of binary systems of celestial objects \cite{Bjerrum-Bohr:2013bxa,Cachazo:2017jef,Guevara:2017csg,Kosower:2018adc,Chung:2018kqs,Maybee:2019jus,Bern:2019nnu,Bern:2021dqo}, the perturbative exploration of supersymmetric gauge theories \cite{Arkani-Hamed:2013jha,Caron-Huot:2019vjl,Caron-Huot:2020bkp} and also the perturbative study of off-shell quantities such as form factors \cite{Brandhuber:2010ad,Bork:2010wf,Brandhuber:2011tv,Bork:2011cj,Bork:2012tt,Boels:2012ew,Penante:2014sza,Bianchi:2018peu,Bianchi:2018rrj,Dixon:2020bbt}. Besides the classification of effective field theory (EFT) interactions themselves, S-matrix properties, such as unitarity, causality and analyticity, have been used to constrain Wilson coefficients associated to EFTs \cite{Adams:2006sv,Arkani-Hamed:2021xlp,Bern:2021ppb,Chiang:2021ziz}, including the SMEFT \cite{Remmen:2019cyz,Remmen:2020vts,Trott:2020ebl}. Moreover, on-shell techniques also provide powerful strategies to study the UV mixing in (non-supersymmetric) EFTs, as first pointed out in \cite{Caron-Huot:2016cwu} using techniques developed for the study of the anomalous dimension of operators in $\mathcal{N}=4$ super-Yang-Mills \cite{Minahan:2002ve,Beisert:2003jj,Beisert:2003yb,Ferretti:2004ba} (for a review, see \cite{Beisert:2010jr} and references therein) from scattering amplitudes and form factors \cite{Zwiebel:2011bx,Wilhelm:2014qua,Nandan:2014oga,Brandhuber:2014pta,Brandhuber:2015boa,Loebbert:2015ova,Frassek:2015rka,Brandhuber:2015dta,Brandhuber:2016fni,Loebbert:2016xkw,Brandhuber:2017bkg,Brandhuber:2018kqb,Brandhuber:2018xzk} and recently applied to the SMEFT \cite{EliasMiro:2020tdv,Baratella:2020lzz,Jiang:2020mhe,Bern:2020ikv}. Furthermore, on-shell techniques also provided a good understanding of the mysterious pattern of zeros in the one-loop anomalous dimension matrix of the SMEFT \cite{Cheung:2015aba,Bern:2019wie,Jiang:2020rwz}.

The first systematic and complete computation of the one-loop anomalous dimension matrix for dimension-six operators in the SMEFT has been carried out in \cite{Jenkins:2013zja,Jenkins:2013wua,Alonso:2013hga}. So far, the study of the anomalous dimension of SMEFT interactions has been completed only partially in the literature for operators up to dimension 8 \cite{Antusch:2001ck,Alonso:2014zka,Liao:2016hru,Davidson:2018zuo,Liao:2019tep,Chala:2021juk,Chala:2021pll}. In this paper, we present the general on-shell set-up which will allow us to fully compute the one-loop anomalous dimension matrix for all the operators in the SMEFT up to mass dimension 8. As a proof of concept, we reproduce know results for the mixing matrix of operators of dimension $5,6$ and $7$ and we present for the first time the mixing matrix of dimension 8 operators for the SMEFT considering a single flavour family $N_f = 1$. In the present work, we compute the anomalous dimension matrix to linear order in the Wilson coefficients, {\it i.e.} we ignore the mixing between dimension-6 and dimension-8 operators, which are however partially known in the literature \cite{Chala:2021pll}.

The present paper is organised as follows. In Section \ref{sec::4points} we describe the complete construction of the SM from on-shell principles, beginning with the classification of all the possible three-point amplitudes. % compatible with Poincaré invariance and locality in a renormalisable theory allowing for spin 0, 1/2 and 1 particles. 
From there we review how locality and unitarity constraints on the four-point tree-level amplitudes enforce the Lie algebra structure of the SM along with non-trivial relations among the couplings and charge conservation, whereas the same conditions at one-loop impose relations among the hypercharges usually found from anomaly cancellation requirements. 
In Section \ref{sec:classificationSMEFTops} we discuss the classification of the possible independent SMEFT interactions for fixed mass dimension. In particular we present an original take on the problem of finding the independent kinematic structures, which are then combined with colour singlets in order to obtain all the possible effective interactions, which are in one-to-one correspondence with the irrelevant operators.
Next, in Section \ref{sec::UVmixing}, we review the computation of the one-loop anomalous dimension matrix from on-shell data through unitarity. We make use of the presented techniques to reproduce known results for dimension 5, 6, and 7 operators as well as to compute for the first time the mixing of dimension 8 operators at linear order in the Wilson coefficients and leading (quadratic) order in the renormalisable couplings. These general results are made available in separate ancillary files: the file \texttt{AllMinimalAmplitudes.wl} contains the basis of operators of dimension 5,6,7 and 8 build using the algorithm presented in Section \ref{sec:classificationSMEFTops} with an arbitrary number of flavours, while the files \texttt{dimension6.wl}, \texttt{dimension7.wl} and \texttt{dimension8.wl} contains the mixing matrices for all the operators at dimension 6,7 and 8 respectively in the SMEFT with $N_f = 1$. On the other hand here we present the explicit mixing coefficients for dimension 6 and 8 operators relevant for Higgs production with a $W$ boson.
Finally, in Section \ref{sec::bootstraphigherpoints}, we present a completely on-shell algorithm which allows to compute tree-level amplitudes (and form factors) in a generic EFT, which will prove crucial in extending our results beyond leading order. This algorithm is based on factorisation properties of the tree-level amplitudes, and allows to bypass the use of recursion relations which can be problematic when non-renormalisable interactions are involved. Furthermore, the computed amplitudes are manifestly local, which is particularly well suited for example when computing loop-level results through generalised unitarity.

\section{The Standard Model from on-shell techniques}
\label{sec::4points}

In this section we are going to present a perturbative on-shell construction of the Standard Model, through the consistency of its S-matrix elements, under the following assumptions:
\begin{itemize}
    \item The scattering amplitudes are invariant under Poincaré transformations but transform under some representation of the Little Group specified by their particle content. In four dimensions, under Little Group transformations each massless state transforms with a phase $e^{i\, h_i \phi}$ where $h_i$ is the helicity of the $i^{\rm th}$-state. These assumptions make the Spinor Helicity variables, briefly reviewed in Appendix \ref{sec::spinorhelicity}, the most suited for the description of scattering amplitudes.
    \item In natural units, the {\it mass dimension} of an $n$-point scattering amplitude, at any loop order $L$\footnote{In the following, when the number of loops is not specified as superscript, we mean tree-level.}, is
    \begin{equation}\label{eq:massdimamp}
        \left[ \Aa_{\, n}^{(L)} \right] = 4 - n \> .
    \end{equation}
    \item {\it Locality}: the non-analytic terms of the scattering amplitudes correspond to intermediate particles going on-shell. In particular, simple poles correspond to single-particle exchanges with the intermediate particle going on-shell.
    \item {\it Unitarity}: the discontinuities of the amplitudes are given by a proper sum of products of lower-point (and lower-loop) amplitudes. In particular, the residues on the simple poles are given by
    \begin{equation}\label{eq:factorisation}
        -i\, \underset{s_{1 \dots m}}{\rm Res}\, \mathcal{A}_n (p_1^{h_1} \dots p_n^{h_n}) = f \sum_{s_{\rm I},h_{\rm I}} \mathcal{A}_{m+1} (p_1^{h_1} \dots p_m^{h_m}, p_I^{h_I}) \mathcal{A}_{n-m+1} (p_{I}^{h_I} \to p_{m+1}^{h_{m+1}} \dots p_n^{h_n})\ ,
    \end{equation}
    where $f=(-1)^{\Delta s}$ with $\Delta s$ the respective signature of the fermion ordering between the LHS and the RHS, $s_I$ and $h_I$ are the type and the helicity of the intermediate state propagating\footnote{We adopt the following convention: we indicate with $\mathcal{A}_n (p_1^{h_1} \dots p_n^{h_n})$ an $n$-point scattering amplitude with all the momenta outgoing and with $\mathcal{A}_{n} (p_{1}^{h_1} \dots p_{m}^{h_m} \to p_{m+1}^{h_{m+1}} \dots p_n^{h_n})$ an $n$-point amplitude with $m$ incoming and $n-m$ outgoing states.}.
\end{itemize}

Gauge invariance is not assumed a priori. Indeed, it has been proven that the Lie algebra structures are required by consistent factorisation of the four-point tree-level amplitude \cite{Benincasa:2007xk}. We will briefly review and extend this considerations to the Standard Model in section \ref{sec::4pointamplitudes}. Moreover, since we work purely on-shell (in four dimensions) with spinor helicity variable, there is no need for polarisation tensors or Ward identities.

All the three-point scattering amplitudes in the Standard Model can be fixed by symmetry, helicity weight and mass dimension considerations (up to a constant) \cite{Benincasa:2007xk}. In particular, the kinematic part of any massless three-point amplitude can be be written as
\begin{align}\label{eq:threeptansatz}
    \Aa (1^{h_1},2^{h_2},3^{h_3}) = 
    \begin{cases}
        i\,g\, \agl{1}{2}^{h_1+h_2-h_3} \agl{2}{3}^{h_2+h_3-h_1} \agl{3}{1}^{h_3+h_1-h_2} & \sum_i h_i = -1 \\[.4em]
        i\,g\, \sqr{1}{2}^{h_1+h_2-h_3} \sqr{2}{3}^{h_2+h_3-h_1} \sqr{3}{1}^{h_3+h_1-h_2} & \sum_i h_i = 1
    \end{cases}\ ,
\end{align}
where the mass dimension of the coupling constant is zero $\left[g\right]=0$, because we are assuming only renormalisable interactions for the moment. A list of all the tree-level three-point amplitudes in the Standard Model, with the proper colour and flavour structures, are presented in appendix \ref{sec::3points}. We will show that this is enough to fully describe, at the perturbative level, the Standard Model in terms of its S-matrix elements.

\subsection{Four-point amplitudes from factorisation}\label{sec:4pt}
\label{sec::4pointamplitudes}
All the 4-point amplitudes in the Standard Model, but $\Aa(\bar{H}^i,\bar{H}^j,H^k,H^l)$, can be completely fixed by factorisation. This will be proven in \ref{sec:ansatz} but we assume it for the moment. Consistency between different factorisation channels at tree-level for four-point amplitudes then constrains many of the structures in the three-point amplitude. These constraints fix the (gauge-invariant) structures appearing and impose relations between couplings.

The constraints imposed by factorisation are completely equivalent to those found when we construct a consistent gauge-invariant Lagrangian describing a unitary QFT of self-interacting vector bosons \cite{Yang:1954ek} and their minimal coupling to fermions and scalars, {\it i.e.} the Lie algebra structures and the universality of Yang-Mills coupling (see, for example, \cite{Peskin:1995ev}). Moreover, we generalise this argument and find that factorisation also imposes relations between the hypercharges associate to the minimal coupling of matter with (non-self-interacting) $U(1)$-vectors, which are equivalent from a Lagrangian perspective to the requirement that the Yukawa interactions are $U(1)_{\rm Y}$ invariant, {\it i.e.} scattering amplitudes are non zero only for hypercharge-conserving processes.

\subsection{Lie algebras from tree-level unitarity %\texorpdfstring{$\Aa (g_{-}^{A},g_{-}^{B},g_{+}^{C},g_{+}^{D})$}{Four-gluon} and \texorpdfstring{$\Aa (g_{-}^{A},g_{+}^{B},\bar{u}^{a},u^{b})$}{Two-gluon two-Q}
}

%\subsubsection{\texorpdfstring{$\Aa (G_{-}^{A},G_{-}^{B},G_{+}^{C},G_{+}^{D})$}{4-gluon amplitude}}\label{sec:4gluon}
\subsubsection{Jacobi identities from factorisation}\label{sec:4gluon}

In this subsection we review the observations in \cite{Benincasa:2007xk}.
 We consider the three-gluon amplitudes\footnote{The relative minus sign between the so called MHV and $\overline{\rm MHV}$ amplitudes is fixed by requiring parity invariance of the theory (at the perturbative level).}
\begin{equation}
    \Aa (G_{-}^{A},G_{-}^{B},G_{+}^{C}) = g_{3}\, f^{A B C} \frac{\agl{1}{2}^3}{\agl{2}{3}\agl{3}{1}}\ , \qquad \Aa (G_{-}^{A},G_{+}^{B},G_{+}^{C}) =- g_{3}\, f^{B C A} \frac{\sqr{2}{3}^3}{\sqr{1}{2}\sqr{3}{1}}\ ,
\end{equation}
where $f^{A B C} = f^{[A B C]}$\footnote{In principle, this assumption could be lifted and would follow from factorisation as well, but for simplicity we keep it.} to satisfy Bose-Einstein symmetry of the three-point amplitude and we try to bootstrap the four-gluon amplitude from factorisation.
The most generic (slightly redundant) ansatz for the four-point amplitude which is compatible with locality and unitarity is
\begin{equation}\label{eq:gluon4ptans}
    \begin{split}
        \frac{\Aa (G_{-}^{A},G_{-}^{B},G_{+}^{C},G_{+}^{D})}{\agl{1}{2}^2 \sqr{3}{4}^2} &= \frac{f^{A B E} f^{C D E}}{s_{12}} \left(\frac{c_{1}}{s_{13}}+\frac{c_{2}}{s_{14}}\right) + \frac{f^{A C E} f^{B D E}}{s_{13}} \left(\frac{c_{3}}{s_{12}}+\frac{c_{4}}{s_{14}}\right)\\[.2em]
        &+ \frac{f^{A D E} f^{B C E}}{s_{14}} \left(\frac{c_{5}}{s_{12}}+\frac{c_{6}}{s_{13}}\right)\ .
    \end{split}
\end{equation}

The coefficients $c_i$ can be fixed from factorisation using \eqref{eq:factorisation} which in the 4-point case reduces to%\footnote{We are omitting on the RHS possible minus signs due to the reordering of fermions in the amplitudes and eventually an $i$ due from crossing when a fermion is exchanged. This subtlety will be relevant in the computations of the following sections.}:
\footnote{We remind the reader that when fermions are present in the amplitudes, the RHS of \eqref{eq::4pointfact} might get a minus sign contribution from fermion reordering and a further factor of $-i$ when crossing a fermion from initial to final state. This subtlety will be relevant in the computations of the following sections.}
\begin{equation}
\label{eq::4pointfact}
    -i \underset{s_{i j} = 0}{\rm Res}\, \Aa_4 = \Aa_3 \cdot \Aa_3\ .
\end{equation}
Imposing this constraint for all the three distinct channels, we find
\begin{equation}\label{eq:4gluonconsttraints}
    \begin{cases}
        f^{A B E} f^{C D E} (c_1 -c_2) + f^{A C E} f^{B D E} c_3 - f^{A D E} f^{B C E} c_5 = -g_{3}^2\,f^{A B E} f^{C D E}\\
        f^{A B E} f^{C D E} c_1 + f^{A C E} f^{B D E} (c_3 - c_4) - f^{A D E} f^{B C E} c_6 = -g_{3}^2\,f^{A C E} f^{B D E}\\
        f^{A B E} f^{C D E} c_2 - f^{A C E} f^{B D E} c_4 + f^{A D E} f^{B C E} (c_5 - c_6) = -g_{3}^2\,f^{A D E} f^{B C E}
    \end{cases}\ .
\end{equation}
This linear system in general has no solutions, unless we impose the following quadratic relations among the constants $f^{A B C}$:
\begin{equation}
    f^{A B E} f^{C D E} + f^{B C E} f^{A D E} + f^{C A E} f^{B D E} = 0\ ,
\end{equation}
which can be recognised as the Jacobi identities for the structure constants of a Lie algebra.

%\subsubsection{\texorpdfstring{$\Aa (G_{-}^{A},G_{+}^{B},\bar{u}^{a},u^{b})$}{4-gluon amplitude}}
\subsubsection{Lie algebras from factorisation}
We can apply the same reasoning to scalars and fermions coupled to the non-abelian spin-1 particles and find that also their minimal coupling is tightly constrained by locality and unitarity \cite{Arkani-Hamed:2017jhn}. We consider as an example the four-point amplitude $\Aa (G_{-}^{A},G_{+}^{B},\bar{u}^{a},u^{b})$. The three-point minimal coupling is fixed by little group and in principle can take the general form
\begin{equation}
    \Aa (G^A_{-},\bar{u}^a,u^b) = i\,g_{3,m}\,\tau\indices{^{A\, a}_{b}} \frac{\agl{1}{2}^2}{\agl{2}{3}}\ , \qquad \Aa (G^A_{+},\bar{u}^a,u^b) = i\,g_{3,m}\,\tau\indices{^{A\, a}_{b}} \frac{\sqr{1}{3}^2}{\sqr{2}{3}}\ ,
\end{equation}
where, for the moment, $\tau\indices{^{A\, a}_{b}}$ is some generic matrix encoding the interaction properties of the fermions $u_a$ ($\bar{u}^a$) and the vector bosons, and we factored out an overall numerical coefficient. The most general ansatz for the four-point is then
\begin{equation}
    \begin{split}
        \frac{\Aa (G_{-}^{A},G_{+}^{B},\bar{u}^{a},u^{b})}{\agl{1}{3}^2 \sqr{2}{3} \sqr{2}{4}} &= \frac{f^{A B C} \tau\indices{^{C\, a}_{b}}}{s_{12}} \left(\frac{c_{1}}{s_{13}}+\frac{c_{2}}{s_{14}}\right) + \frac{\tau\indices{^{A B\, a}_{b}}}{s_{13}} \left(\frac{c_{3}}{s_{12}}+\frac{c_{4}}{s_{14}}\right)\\[.2em]
        &+ \frac{\tau\indices{^{B A\, a}_{b}}}{s_{14}} \left(\frac{c_{5}}{s_{12}}+\frac{c_{6}}{s_{13}}\right)\ 
    \end{split}
\end{equation}
where $\tau\indices{^{A B\, a}_{b}} = \tau\indices{^{A\, a}_{c}} \tau\indices{^{B\, c}_{b}}$. Again taking the residues and matching with the factorisation channels as in equation \eqref{eq::4pointfact}, we find:
\begin{equation}
    \begin{cases}
        f^{A B C} \tau\indices{^{C\, a}_{b}} (c_1 -c_2) + \tau\indices{^{A B\, a}_{b}} c_3 - \tau\indices{^{B A\, a}_{b}} c_5 = i\,g_{3}\, g_{3,m}\, f^{A B C} \tau\indices{^{C\, a}_{b}}\\[.2em]
        f^{A B C} \tau\indices{^{C\, a}_{b}} c_1 + \tau\indices{^{A B\, a}_{b}} (c_3 - c_4) - \tau\indices{^{B A\, a}_{b}} c_6 = g_{3,m}^{\,2} \tau\indices{^{A B\, a}_{b}}\\[.2em]
        f^{A B C} \tau\indices{^{C\, a}_{b}} c_2 - \tau\indices{^{A B\, a}_{b}} c_4 + \tau\indices{^{B A\, a}_{b}} (c_5 - c_6) = g_{3,m}^{\,2} \tau\indices{^{B A\, a}_{b}}
    \end{cases}\ ,
\end{equation}
This linear system has solutions if and only if
\begin{align}
    g_{3,m}&=g_3\ ,\\[.2em]
    \tau\indices{^{A B\, a}_{b}} - \tau\indices{^{B A\, a}_{b}}&=i\,f^{A B C} \tau\indices{^{C\, a}_{b}}\ ,
\end{align}
{\it i.e.} iff the coupling constant of the interaction is universal and the matrices $\tau\indices{^{A\, a}_{b}}$ are representations of the elements of a Lie algebra, with $f^{A B C}$ the structure constants.
%\subsubsection{\texorpdfstring{$\Aa (g_{-}^{A},g_{+}^{B},\bar{u}^{a},u^{b})$}{Two-gluon two-Q} and the gauge minimal coupling}

\subsubsection{Charge conservation and Yukawa coupling}
Last we generalise the procedure of the previous sections to the minimal coupling of the abelian vectors with scalars and fermions interacting via Yukawa coupling. Unitarity and locality will then imply that the hypercharge associated to the minimal coupling of the matter states to the abelian vector is conserved. The relevant three-point amplitudes are
\begin{align}
    \Aa (B_{-}, \bar{e}, e) &= i\, g_1 Y_e \frac{\agl{1}{2}^2}{\agl{2}{3}}\ ,\\
    \Aa (B_{-}, \bar{L}^i, L^j) &= i\, g_1 Y_L \delta^{j}_{i} \frac{\agl{1}{2}^2}{\agl{2}{3}}\ ,\\
    \Aa (B_{-}, \bar{H}^i, H^j) &= i\, g_1 Y_H \delta^{j}_{i} \frac{\agl{1}{2}\agl{3}{1}}{\agl{2}{3}}\ ,\\
    \Aa (L^i,e,\bar{H}^j) &= i\,\bar{\mathcal{Y}}^{(3)} \delta^{i}_{j} \sqr{1}{2}\ ,
\end{align}
where $Y_i$ is the hypercharge associated to the $i$-th state, and $\mathcal{Y}^{(3)}$ is the Yukawa coupling matrix for the electron family, with $\bar{\mathcal{Y}}^{(3)} = \left(\mathcal{Y}^{(3)}\right)^\dagger$. The most generic ansatz consistent with locality and unitarity is
\begin{equation}
    \frac{\Aa (B_{-},L^i,e,\bar{H}^j)}{\agl{1}{2}\agl{1}{3}\sqr{2}{3}^2} = \delta^{i}_{j} \left(\frac{c_1}{s_{12} s_{13}} + \frac{c_2}{s_{12} s_{14}} + \frac{c_3}{s_{13} s_{14}}\right)\ ,
\end{equation}
and probing the three different factorisation channels we find the system:
\begin{equation}
    \begin{cases}
        c_1 - c_2 = - g_1 \bar{\mathcal{Y}}^{(3)} Y_{L}\\
        c_1 - c_3 = + g_1 \bar{\mathcal{Y}}^{(3)} Y_{e}\ ,\\
        c_2 - c_3 = + g_1 \bar{\mathcal{Y}}^{(3)} Y_{H}
    \end{cases}
\end{equation}
which has solutions if and only if we impose the hypercharge conserving condition:
\begin{equation}
    Y_L = Y_H - Y_e\ .
\end{equation}
Analogously, one can also find the charge conservation conditions for the processes involving quarks, instead of leptons:
\begin{align}
    Y_Q &= Y_H - Y_d\ ,\\
    Y_Q &= - Y_H - Y_u\ .
\end{align}

\subsection{Hypercharge constraints from gauge anomalies}

On top of the relations we found so far, it would be nice to be able to further relate $Y_e$ and $Y_u$ as is done by the anomaly cancellation condition $Y_L+3Y_Q=0$.
%In the usual presentation of the Standard Model, a relation between $Y_e$ and $Y_u$ is required by anomaly cancellation conditions.\todo{Come mai hai aggiunto questo commento? per introdurre la sezione, la prima cosa che dobbiamo spiegare è perché facciamo le cose.}
Indeed, it has long been known that in gauge theories with chiral fermions anomalies arise from fermion loops \cite{Adler:1969,Bell:1969ts}. These gauge anomalies impose consistency conditions on the theory, which in the case of the SM translate into relations among the hypercharges of the fermions. %More recently such anomaly constraints have been obtained in a completely on-shell way for a chiral QCD theory in four spacetime dimensions as well as QCD, QED and gravity in six dimensions \cite{Huang:2013vha,Chen:2014eva}.
Interestingly, as first noticed in \cite{Huang:2013vha,Chen:2014eva}, the same cancellation conditions are required from a purely on-shell point of view by a clash of unitarity and locality in some one-loop amplitudes. In this section we apply this method to recover the SM anomaly cancellation conditions.
%From a purely on-shell point of view, anomalies cancellation conditions are required by a clash of unitarity and locality in one-loop amplitudes, as first noticed in \cite{Huang:2013vha,Chen:2014eva}.

The core of the idea is that one-loop amplitudes can be computed and entirely fixed using generalised unitarity methods \cite{Bern:1994zx,Bern:1994cg,Bern:1995db,Bern:1997sc,Britto:2004nc,Brandhuber:2005jw,Anastasiou:2006jv,Britto:2006sj,Anastasiou:2006gt,Mastrolia:2006ki,Mastrolia:2008jb,Mastrolia:2009dr,Mastrolia:2012an,Forde:2007mi}, up to rational terms which have no branch points. Such amplitudes by construction are unitary, however locality is not guaranteed (spurious poles can appear in the final result) and needs to be restored by appropriately fixing the rational terms to which the unitarity methods are blind. These rational terms might in turn introduce new corrections to the factorisation of the four-point amplitude, which is inconsistent with the fact that the three-point amplitudes are tree-level exact and fixed by helicity and mass dimension. When this happens additional properties of the theory need to be required for these terms to vanish. In particular, in this section we will show that for the Standard model this leads to well known anomaly constraints on the fermion hypercharges. %(They have always been in the paper) In this section, we restore $Y_L$ and $Y_Q$ as couplings in the amplitudes in order to recover the cancellation conditions in their usual form.

\begin{figure}
    \centering
    \begin{tikzpicture}[scale=15]
        \def\x{0};
        \def\y{0};

        \node at (-1.5pt+\x,1.5pt+\y) (A){};
        \node at (1.5pt+\x,1.5pt+\y) (B){};
        \node at (1.5pt+\x,-1.5pt+\y) (C){};
        \node at (-1.5pt+\x,-1.5pt+\y) (D){};
        \node at (-2.5pt+\x,2.5pt+\y) (A2){$1^+$};
        \node at (2.5pt+\x,2.5pt+\y) (B2){$2^-$};
        \node at (2.5pt+\x,-2.5pt+\y) (C2){$3^+$};
        \node at (-2.5pt+\x,-2.5pt+\y) (D2){$4^-$};

        \node at (5.5pt+\x,\y) {$c_4^f=-\dfrac{s_{12} ^4 s_{14}^2}{2 \, s_{13}^4}$};
        \node at (-2.2pt+\x,0.55pt+\y){\small $+\frac{1}{2}$};
        \node at (-2.2pt+\x,-0.55pt+\y){\small $-\frac{1}{2}$};

        \draw [->,>=Latex] (A.center)--(B.center);
        \draw [->,>=Latex] (B.center)--(C.center);
        \draw [->,>=Latex] (C.center)--(D.center);
        \draw [->,>=Latex] (D.center)--(A.center);

        \draw[] (A.center) -- (A2);
        \draw[] (B.center) -- (B2);
        \draw[] (C.center) -- (C2);
        \draw[] (D.center) -- (D2);

        \draw [dashed] (0+\x,0.5pt+\y) -- (0+\x,2.5pt+\y);
        \draw [dashed] (0.5pt+\x,\y) -- (2.5pt+\x,\y);
        \draw [dashed] (0+\x,-0.5pt+\y) -- (0+\x,-2.5pt+\y);
        \draw [dashed] (-0.5pt+\x,\y) -- (-2.5pt+\x,\y);

        \def\x{13.5pt};
        \def\y{0};

        \node at (-1.5pt+\x,1.5pt+\y) (A){};
        \node at (1.5pt+\x,1.5pt+\y) (B){};
        \node at (1.5pt+\x,-1.5pt+\y) (C){};
        \node at (-1.5pt+\x,-1.5pt+\y) (D){};
        \node at (-2.5pt+\x,2.5pt+\y) (A2){$1^+$};
        \node at (2.5pt+\x,2.5pt+\y) (B2){$2^-$};
        \node at (2.5pt+\x,-2.5pt+\y) (C2){$3^+$};
        \node at (-2.5pt+\x,-2.5pt+\y) (D2){$4^-$};

        \node at (5.5pt+\x,\y) {$c_4^{\overline{f}}=-\dfrac{s_{12} ^2 s_{14}^4}{2\, s_{13}^4}$};
        \node at (-2.2pt+\x,0.55pt+\y){\small $-\frac{1}{2}$};
        \node at (-2.2pt+\x,-0.55pt+\y){\small $+\frac{1}{2}$};

        \draw [->,>=Latex] (A.center)--(B.center);
        \draw [->,>=Latex] (B.center)--(C.center);
        \draw [->,>=Latex] (C.center)--(D.center);
        \draw [->,>=Latex] (D.center)--(A.center);

        \draw[] (A.center) -- (A2);
        \draw[] (B.center) -- (B2);
        \draw[] (C.center) -- (C2);
        \draw[] (D.center) -- (D2);

        \draw [dashed] (0+\x,0.5pt+\y) -- (0+\x,2.5pt+\y);
        \draw [dashed] (0.5pt+\x,\y) -- (2.5pt+\x,\y);
        \draw [dashed] (0+\x,-0.5pt+\y) -- (0+\x,-2.5pt+\y);
        \draw [dashed] (-0.5pt+\x,\y) -- (-2.5pt+\x,\y);

        \def\x{0};
        \def\y{-7pt};

        \node at (-1.5pt+\x,1.5pt+\y) (A){};
        \node at (1.5pt+\x,\y) (B){};
        \node at (-1.5pt+\x,-1.5pt+\y) (D){};
        \node at (-2.5pt+\x,2.5pt+\y) (A2){$2^-$};
        \node at (2.5pt+\x,1.5pt+\y) (B2){$3^+$};
        \node at (2.5pt+\x,-1.5pt+\y) (C2){$4^-$};
        \node at (-2.5pt+\x,-2.5pt+\y) (D2){$1^+$};

        \node at (5.5pt+\x,\y) {$c_{3,(12)}^f=\dfrac{s_{12}^4  s_{14}}{2\, s_{13}^4}$};
        \node at (-2.2pt+\x,0.55pt+\y){\small $+\frac{1}{2}$};
        \node at (-2.2pt+\x,-0.55pt+\y){\small $-\frac{1}{2}$};

        \draw [->,>=Latex] (A.center)--(B.center);
        \draw [->,>=Latex] (B.center)--(D.center);
        \draw [->,>=Latex] (D.center)--(A.center);

        \draw[] (A.center) -- (A2);
        \draw[] (B.center) -- (B2);
        \draw[] (B.center) -- (C2);
        \draw[] (D.center) -- (D2);

        \draw [dashed] (-0.3pt+\x,0.2pt+\y) -- (0.3pt+\x,2pt+\y);
        \draw [dashed] (-0.3pt+\x,-0.2pt+\y) -- (0.3pt+\x,-2pt+\y);
        \draw [dashed] (-0.8pt+\x,\y) -- (-2.5pt+\x,\y);

        \def\x{13.5pt};
        \def\y{-7pt};

        \node at (-1.5pt+\x,1.5pt+\y) (A){};
        \node at (1.5pt+\x,\y) (B){};
        \node at (-1.5pt+\x,-1.5pt+\y) (D){};
        \node at (-2.5pt+\x,2.5pt+\y) (A2){$2^-$};
        \node at (2.5pt+\x,1.5pt+\y) (B2){$3^+$};
        \node at (2.5pt+\x,-1.5pt+\y) (C2){$4^-$};
        \node at (-2.5pt+\x,-2.5pt+\y) (D2){$1^+$};

        \node at (5.5pt+\x,\y) {$c_{3,(12)}^{\overline{f}}=\dfrac{s_{12}^2  s_{14}^3}{2 \, s_{13}^4}$};
        \node at (-2.2pt+\x,0.55pt+\y){\small $-\frac{1}{2}$};
        \node at (-2.2pt+\x,-0.55pt+\y){\small $+\frac{1}{2}$};

        \draw [->,>=Latex] (A.center)--(B.center);
        \draw [->,>=Latex] (B.center)--(D.center);
        \draw [->,>=Latex] (D.center)--(A.center);

        \draw[] (A.center) -- (A2);
        \draw[] (B.center) -- (B2);
        \draw[] (B.center) -- (C2);
        \draw[] (D.center) -- (D2);

        \draw [dashed] (-0.3pt+\x,0.2pt+\y) -- (0.3pt+\x,2pt+\y);
        \draw [dashed] (-0.3pt+\x,-0.2pt+\y) -- (0.3pt+\x,-2pt+\y);
        \draw [dashed] (-0.8pt+\x,\y) -- (-2.5pt+\x,\y);

        \def\x{4pt};
        \def\y{-12.5pt};

        \node at (-1.5pt+\x,\y) (A){};
        \node at (1.5pt+\x,\y) (B){};
        \node at (-2.5pt+\x,1.5pt+\y) (A2){$2^-$};
        \node at (2.5pt+\x,1.5pt+\y) (B2){$3^+$};
        \node at (2.5pt+\x,-1.5pt+\y) (C2){$4^-$};
        \node at (-2.5pt+\x,-1.5pt+\y) (D2){$1^+$};

        \node at (-0.5pt+\x,-1.3pt+\y) {\small $+\frac{1}{2}$};
        \node at (0.5pt+\x,-1.3pt+\y) {\small $-\frac{1}{2}$};

        \node at (10pt+\x,\y) {$c_{2,(12)}^{f}=\dfrac{s_{14}(2 s_{14}^2+11s_{12}^2+7 s_{12}s_{14})}{6 s_{13}^3}$};

        \draw [->,>=Stealth] (A.center) to[out=60,in=120] (B.center);
        \draw [->,>=Stealth] (B.center) to[out=-120,in=-60] (A.center);

        \draw[] (A.center) -- (A2);
        \draw[] (B.center) -- (B2);
        \draw[] (B.center) -- (C2);
        \draw[] (A.center) -- (D2);

        \draw [dashed] (\x,1.5pt+\y) -- (\x,-1.5pt+\y);

        \def\x{4pt};
        \def\y{-18pt};

        \node at (-1.5pt+\x,\y) (A){};
        \node at (1.5pt+\x,\y) (B){};
        \node at (-2.5pt+\x,1.5pt+\y) (A2){$2^-$};
        \node at (2.5pt+\x,1.5pt+\y) (B2){$3^+$};
        \node at (2.5pt+\x,-1.5pt+\y) (C2){$4^-$};
        \node at (-2.5pt+\x,-1.5pt+\y) (D2){$1^+$};

        \node at (-0.5pt+\x,-1.3pt+\y) {\small $-\frac{1}{2}$};
        \node at (0.5pt+\x,-1.3pt+\y) {\small $+\frac{1}{2}$};

        \node at (10pt+\x,\y) {$c_{2,(12)}^{\overline{f}}=\dfrac{s_{14}(2 s_{14}^2-s_{12}^2-5 s_{12}s_{14})}{6 s_{13}^3}$};

        \draw [->,>=Stealth] (A.center) to[out=60,in=120] (B.center);
        \draw [->,>=Stealth] (B.center) to[out=-120,in=-60] (A.center);

        \draw[] (A.center) -- (A2);
        \draw[] (B.center) -- (B2);
        \draw[] (B.center) -- (C2);
        \draw[] (A.center) -- (D2);

        \draw [dashed] (\x,1.5pt+\y) -- (\x,-1.5pt+\y);

    \end{tikzpicture}
    \caption{Kinematic coefficients from generalised unitarity \cite{Forde:2007mi}, here a kinematic contribution of the type $\frac{\agl{2}{4}^2 \sqr{1}{3}^2}{s_{12}s_{14}}$ has been factored out.}
    \label{fig:cut}
\end{figure}
We will specifically consider a fermion loop coupled to four external gauge bosons in the MHV configuration. The full one-loop amplitudes in the Standard Model can be schematically written as
\begin{equation}
    \Aa^{\rm 1-loop}= \Aa^{\rm 1-loop}_{\rm vec} + \Aa^{\rm 1-loop}_{\rm ferm} + \Aa^{\rm 1-loop}_{\rm scal} \>,
\end{equation}
where the three contributions correspond respectively to vector bosons, fermions or scalars running in the internal loop, the specific type of these particles depending on the external states. We want to focus here on the fermion loop contributions, which are infrared finite and are the only part contributing to the chiral anomaly.%, since the latter is due to the coupling of vector bosons (which will be chosen as external states) to chiral fermions.
%The one loop amplitudes we consider have an internal fermion loop coupled to four external gauge bosons in alternating helicity configuration.
The kinematic information of these amplitudes is entirely captured by the coefficients of Figure \ref{fig:cut} with cyclic rotations providing the other orderings. For later convenience we define the following kinematic combinations, which turn out to be ubiquitous in the one-loop amplitudes
\begin{equation}
    K_{\rm even} \coloneqq \frac{\agl{2}{4}^2 \sqr{1}{3}^2}{s_{12}s_{14}} \,\sum_{i,j} (c_{i,j}^f+c_{i,j}^{\overline{f}})I_i(j) \>, \hspace{0.5cm}    K_{\rm odd} \coloneqq \frac{\agl{2}{4}^2 \sqr{1}{3}^2}{s_{12}s_{14}} \, \sum_{i,j} (c_{i,j}^f-c_{i,j}^{\overline{f}})I_i(j) \>,
\end{equation}
with $i=2,3,4$ and $j=s_{12},s_{14}$, and $I_2$, $I_3$ and $I_4$ being the bubble, triangle and box integrals given in Appendix \ref{sec:integrals}. Notice that in the chosen helicity configuration in the one-loop amplitude there are no discontinuities in the $s_{13}$ channel, because all the tree-amplitudes entering the fermion loop contribution in the generalised unitarity calculation vanish in this channel.
%Notice that in this section we do not allow for further constraints on the colour structure of the theory: here we consider the gauge-group structure as a priori given and to be those of the SM, thus we focus on the kinematics associated to individual colour structures and omit the latter for the rest of this section. In particular the expressions shown are those associated to the ordered external particles, which means that the only physical singularities are simple poles in $s_{12}$ and $s_{14}$.

%Consider for example the case of QCD, where four gluons couple to an internal loop of $Q$, $u$ and $d$,
%the corresponding part of the one loop amplitude is given by
%\begin{equation}
%    \mathcal{A}^{1-loop}_{ferm}(g_+^A,g_-^B,g_+^C,g_-^D) \Big|_{\tau^{ABCD}}=
%\end{equation}
%\todo[inline, color=purple]{Stefano: the color-ordering has to go! Done}
Then we consider as a first example the one-loop amplitude with two $W$s and two $B$s as external states, and consequently $Q/\bar{Q}$ and $L/\bar{L}$ as the only possible fermions running through the loop. We find
\begin{equation}\label{eq:firstexample}
    \mathcal{A}^{\rm 1-loop}_{\rm ferm}(W_+^I,B_-,W_+^J,B_-) \Big|_{\rm cut}= \, g_1^2 g_2^2 (Y_L^2+3Y_Q^2) \, \delta^{IJ}  \, K_{\rm even} \>.
\end{equation}
The presence of only $K_{\rm even}$ was to be expected due to the interplay of the colour part with the kinematics. The $SU(3)$ colour part is trivial being absent in the case of the $L/\overline{L}$ circulating in the loop and contributing a numeric factor $\delta^a_a=3$ for the $Q/\overline{Q}$ loop. The $SU(2)$ part on the other hand contributes with a factor of $\Tr \, \sigma^{I}\sigma^{J}=\frac{1}{2}\delta^{IJ}$ in both the $s_{12}$ and $s_{14}$ channels, which then leads to an additive combination of the kinematic parts into $K_{\rm even}$.
Studying the behaviour of $K_{\rm even}$ in the small-$s_{13}$ limit one finds that
\begin{equation}\label{eq:spurious1}
    K_{\rm even} \xrightarrow{s_{13}\to 0} \frac{\agl{2}{4}^2 \sqr{1}{3}^2}{s_{12}s_{14}} \, \left(-\dfrac{s_{12}^2}{s_{13}^2}-\dfrac{s_{12}}{s_{13}} +\mathcal{O}(s_{13}^0)\right) \>,
\end{equation}
thus, in order to restore locality, this amplitude requires a rational term whose kinematic part is of the form
\begin{equation}
    R_{\rm even}=-\frac{\agl{2}{4}^2\sqr{1}{3}^2}{s_{13}^2} \>,
\end{equation}
which cancels both the spurious poles of \eqref{eq:spurious1} and does not produces any modification to the
residues in the $s_{12}$ and $s_{14}$ channels. Adding together the cut-constructible and rational piece one gets the
complete fermion loop contribution
\begin{equation}
    \mathcal{A}^{\rm 1-loop}_{\rm ferm}(W_+^I,B_-,W_+^J,B_-)= \, g_1^2 g_2^2 (Y_L^2+3Y_Q^2) \, \delta^{IJ} \, \left(K_{\rm even} + R_{\rm even}\right)  \>.
\end{equation}

On the other hand, considering three external $W$ and a single $B$, one ends up with 
\begin{equation}\label{eq:secondexample}
    \mathcal{A}^{\rm 1-loop}_{\rm ferm}(W_+^I,W_-^J,W_+^K,B_-) \Big|_{\rm cut}= \frac{i}{2} \, g_1 g_2^3 (Y_L+3Y_Q) \, \epsilon^{IJK} \, K_{\rm odd} \>,
\end{equation}
where once again the $SU(2)$ colour structure, which is $\Tr\, \sigma^{I} \sigma^{J} \sigma^{K} = \frac{i}{4} \epsilon^{IJK}$ in the $s_{12}$ channel and $\Tr\, \sigma^{I} \sigma^{K} \sigma^{J} = - \frac{i}{4} \epsilon^{IJK}$ in the $s_{14}$ channel, is responsible for the relative sign among the kinematic structures and the combination into $K_{\rm odd}$.

Now $K_{\rm odd}$ in the small-$s_{13}$ limit goes as
\begin{equation}\label{eq:spurious2}
    K_{\rm odd} \xrightarrow{s_{13}\to 0} \frac{\agl{2}{4}^2 \sqr{1}{3}^2}{s_{12}s_{14}} \,  \left(-\dfrac{s_{12}}{s_{13}} +\mathcal{O}(s_{13}^0)\right) \>,
\end{equation}
requiring a compensating rational term of the form
\begin{equation}
    R_{\rm odd}=\agl{2}{4}^2\sqr{1}{3}^2 \, \frac{s_{12}-s_{14}}{2s_{12}s_{13}s_{14}} \>,
\end{equation}
which would lead to a complete fermion loop contribution of
\begin{equation}
    \mathcal{A}^{\rm 1-loop}_{\rm ferm}(W_+^I,W_-^J,W_+^K,B_-) = \frac{i}{2} \, g_1 g_2^3 (Y_L+3Y_Q) \, \epsilon^{IJK} \, \left(K_{\rm odd}+R_{\rm odd}\right)  \>.
\end{equation}
However, $R_{\rm odd}$ introduces (unphysical) corrections to the residues in the $s_{12}$ and $s_{14}$ channels, because the one-loop four-point amplitude cannot have any factorisation channel and thus it cannot appear in the one loop amplitude\footnote{Three-point amplitudes are exact at tree-level and fixed by helicity and mass dimensions consideration. This make the poles of four-point amplitudes tree-level exact, \textit{i.e.} there are no loop corrections to the residues of these poles.}. In order to get an answer which satisfies both unitarity and locality we must then enforce the coefficient of the amplitude to vanish, which means imposing
\begin{equation}
    Y_L+3Y_Q=0 \>.
\end{equation}
In a similar fashion, when looking at the one-loop interaction of three gluons with a single $B$ we get the condition
\begin{equation}\label{eq:condition2}
    2Y_Q=Y_u+Y_d \>,
\end{equation}
which is necessary for the fermion-loop contribution to recombine in the physically meaningful form
\begin{equation}\label{eq:oneloopamp2}
    \mathcal{A}^{\rm 1-loop}_{\rm ferm}(G_+^A,G_-^B,G_+^C,B_-) \Big|_{\tau^{ABC}}=-2g_1 g_3^3 (Y_u+Y_d) \, \left(K_{even} +R_{even}\right) \>.
\end{equation}
%This case is somewhat peculiar compared to the previous ones, in the fact that instead of requiring the condition on the hypercharges to cancel a $K_{\rm odd}$ contribution, it is necessary for a correct recombination into $K_{\rm even}$. The $Q$ ($\overline{Q}$) part, which gets a factor of 2 from a trace over an $SU(2)$ delta, comes together with the $\overline{u}$ ($u$) and $\overline{d}$ ($d$) contributions, and upon requiring \eqref{eq:condition2} leads to \eqref{eq:oneloopamp2}. Notice the crucial interplay between the fermions $Q$ and the anti-fermions $\overline{u}$, $\overline{d}$ (and vice versa) which transform alike under $SU(3)$: if the $SU(3)$ interaction was chiral this could not have happened. Not only that, one would find additional cancellation conditions arising for example from the four-gluon one-loop interaction, where this interplay is equally relevant.

Finally, in order to obtain the additional textbook constraint on the hypercharges
\begin{equation}\label{eq:gravityconstraints}
    \left(2Y_L^3-Y_e^3\right)+3\left(2Y_Q^3-Y_u^3-Y_d^3\right)=0 \>,
\end{equation}
we need to look at four-point amplitudes involving a fermion loop with three external $B$ and a boson which can couple universally to all the fermions, in other words a graviton $g$. Similarly, considering the fermionic contribution to the one-loop interaction of three gravitons with a single $B$ will lead to the anomaly cancellation condition
\begin{equation}
    \left(2Y_L-Y_e\right)+3\left(2Y_Q-Y_u-Y_d\right)=0 \>.
\end{equation}

\section{The on-shell classification of SMEFT operators}\label{sec:classificationSMEFTops}

In this section we are going to extend the on-shell methods to the classification of effective interactions \cite{Shadmi:2018xan,Falkowski:2019zdo,Durieux:2020gip} in the SMEFT \cite{Ma:2019gtx,Aoude:2019tzn,Durieux:2019eor,Durieux:2019siw}, corresponding in the Lagrangian formalism to insertions of irrelevant operators \cite{Grzadkowski:2010es,Lehman:2014jma,Murphy:2020rsh,Liao:2020jmn}. First we are going to classify all the independent kinematic structures in a generic theory in four dimensions introducing a new algorithm in terms of graphs and then we will consider the specific case of the Standard Model, combining these with the colour structures\footnote{The approach presented in this section has been formulated by one of the authors and coded in \texttt{Mathematica} \cite{Mathematica}. The code and an example notebook are available at the link \url{https://github.com/StefanoDeAngelis/SMEFT-operators}.}.

\subsection{Kinematic structures from spinor helicity variables}\label{sec:kinematics}

Each effective interaction will be identified by its {\it minimal} amplitude, {\it i.e.} the amplitude at leading order which does not vanish in free theory (if we switch off all the other interactions). This has to be a contact term, {\it i.e.} there are no intermediate modes propagating.

As a first step in the classification procedure, we fix the mass-dimension $\left[\mathcal{O}\right]$ of the irrelevant operators for which we want to find a complete basis. From the minimal amplitudes we strip off the coupling of the effective interaction, which is related to the dimension of the corresponding irrelevant operator by
\begin{equation}
    \left[g_{\mathcal{O}}\right] = 4-\left[\mathcal{O}\right]\ .
\end{equation}
What we are looking for are the kinematic structures which have mass dimension
\begin{equation}
\label{eq::dimcontact}
    \left[\mathcal{O}\right]-n\geq 0\ ,
\end{equation}
where $n$ is the number of external legs in the corresponding minimal amplitude. Equation \eqref{eq::dimcontact} provides a constraint on $n$ which can be further refined by taking into account which types of particles are found in the amplitudes. In fact, in order to get helicity weights right, each vector in the minimal amplitude will contribute at least with two spinor variables and each fermion at least with one. This leads to the stronger constraint\footnote{This condition is not only necessary but also sufficient for having local interactions.}
\begin{equation}
\label{eq::particleconstraint}
 \left[\mathcal{O}\right] - n \geq 2 \times \frac{1}{2} \times n_g + \frac{1}{2}\times n_f \, \hspace{0.5cm} \implies \hspace{0.5cm}
    2 n_g + \frac{3}{2} n_f + n_s \leq \left[\mathcal{O}\right]\ ,
\end{equation}
where $n_g$, $n_f$ and $n_s$ are respectively the number of vectors, fermions and scalars and clearly $n=n_g+n_f+n_s$.
Next, we need to take into account the constraints coming from the condition that our kinematic structures must be ${\rm SL}(2,\mathbb{C})$ invariant. This requires to further distinguish between helicities of the different particles, and to find all the $(n_{g^-},n_{g^+},n_{f^-},n_{f^+},n_s)$\footnote{The superscript of the subscript specify the helicity of the particles: $n_g = n_{g^-}+n_{g^+}$ and  $n_f = n_{f^-}+n_{f^+}$.} compatible with the constraint \eqref{eq::particleconstraint}.
Once $n_g$, $n_f$ and $n_s$ are fixed, we take into account that every state can contribute to the kinematic structures with powers of its momentum, which correspond to derivates in the operator language. The total number of momenta $n_\partial$ is fixed by saturating the mass dimension constraint to
\begin{equation}
    n_\partial = \left[ \mathcal{O}\right] - 2 n_g - \frac{3}{2}n_f - n_s \> .
\end{equation}
A simple way of finding all the possible structures is to identify them with an oriented multigraph, where each vertex is associated to a particle, and the edges correspond to angle (red) or square (blue) ${\rm SL}(2,\mathbb{C})$ invariants. The orientation of the edges then keeps track of the ordering of particles in the brackets and thus provides potential minus signs.

The valence of each vertex is given by two natural numbers $v^{i}=(v^i_a,v^{i}_s)$ such that $v^i_s-v^i_a =2 h_i$ is the helicity of the $i^{\rm th}$ particle (see, for example, Figure \ref{fig:2graphexample}). Finally, for reasons which will become clear in the next section, we consider a circular embedding for our graphs, in other words we take all the nodes to be  ordered points on a circle.
This method has proven to be a computationally efficient way of finding a basis of independent structures up to Schouten and momentum conservation identities. %In particular, the former act separately on angle and square invariants, while the latter mixes the two structures. We are going to show how to deal with this identities in terms of above mentioned multigraphs.
Notice that the former act separately on angle and square invariants, while the latter mixes the two structures. In the following sections we are going to show how to deal with these identities in terms of above mentioned multigraphs.

\begin{figure}
\label{fig::multigraphEx}
	\begin{multicols}{2}
	    \begin{center}
		    \begin{tikzpicture}[scale=2.5,>=Stealth]
		    
			    \node at (-0.05,-0.05) {$4$};
			    \node at (-0.05,1.05) {$1$};
			    \node at (1.05,1.05) {$2$};
			    \node at (1.05,-0.05) {$3$};
			    
			    \draw [fill] (0,0) circle [radius = 0.02];
			    \draw [fill] (1,0) circle [radius = 0.02];
			    \draw [fill] (0,1) circle [radius = 0.02];
			    \draw [fill] (1,1) circle [radius = 0.02];
			    
			    \draw [<-,thick,red](0,0) -- (0,1);
			    
			    \draw [->,thick,red](0,1) to [out=-10,in=-170] (1,1);
			    \draw [->,thick,red](0,1) to [out=10,in=-190] (1,1);
			    
			    \draw [<-,thick,red](1,1) to [out=-100,in=100] (1,0);
			    \draw [->,thick,blue](1,1) to [out=-80,in=80] (1,0);
			    
			    \draw [->,thick,blue](0,1) -- (1,0);
			    \draw [<-,thick,blue](0,1) to [out=-55,in=145] (1,0);
			    \draw [->,thick,blue](0,1) to [out=-35,in=125] (1,0);
	
		    \end{tikzpicture}
	    \end{center}
	    \begin{center}
	        \begin{tikzpicture}[scale=1.5,>=Stealth]
	            
	            \node at (-0.08,-0.08) {$5$};
			    \node at (-0.08,1.08) {$1$};
			    \node at (0.95,1.43) {$2$};
			    \node at (1.58,0.5) {$3$};
			    \node at (0.95,-0.43) {$4$};
			    
			    \draw [fill] (0,0) circle [radius = 0.02];
			    \draw [fill] (0,1) circle [radius = 0.02];
			    \draw [fill] (0.95,1.3) circle [radius = 0.02];
			    \draw [fill] (1.45,0.5) circle [radius = 0.02];
			    \draw [fill] (0.95,-0.3) circle [radius = 0.02];
			    
			    %\draw [->,thick,red](0,1) -- (1.45,0.5);
			    %\draw [->,thick,red](0,1) -- (0.95,-0.3);
			    \draw [->,thick,red](0.95,1.3) -- (0,0);
			    
			    \draw [->,thick,red](0.41,0.45) -- (0.95,-0.3);
			    \draw [thick,red](0,1) -- (0.315,0.56);
			    \draw [thick,red] (0.41,0.45) arc (-60:150:0.08);
			    %\draw [fill] (0.315,0.56) circle [radius = 0.02];
			    %\draw [fill] (0.41,0.45) circle [radius = 0.02];
			    %\draw [fill] (0.66,0.77) circle [radius = 0.02];
			    %\draw [fill] (0.51,0.82) circle [radius = 0.02];
			    \draw [->,thick,red](0.66,0.77) -- (1.45,0.5);
			    \draw [thick,red](0,1) -- (0.51,0.82);
			    \draw [thick,red] (0.66,0.77) arc (-30:165:0.08);
			    
			    \draw [->,thick,blue](0.95,1.3) -- (1.45,0.5);
			    \draw [->,thick,blue](1.45,0.5) -- (0.95,-0.3);
	            
	        \end{tikzpicture}
	    \end{center}
	\end{multicols}
    \caption{The graph associate to the kinematic structures $\textcolor{red}{\agl{1}{2}^2 \agl{1}{4} \agl{3}{2}} \textcolor{blue}{\sqr{2}{3}\sqr{1}{3}^2 \sqr{3}{1}}$ and $\textcolor{red}{\agl{1}{3} \agl{1}{4} \agl{2}{5}} \textcolor{blue}{\sqr{2}{3}\sqr{3}{4}}$ respectively.}
	\label{fig:2graphexample}
\end{figure}

\subsubsection{Schouten identities}

Schouten identities for angle and square brackets read
\begin{equation}\label{eq:Schouten}
\begin{aligned}
    \agl{1}{2}\agl{3}{4} + \agl{2}{3}\agl{1}{4} + \agl{3}{1}\agl{2}{4} &= 0\ ,\\
    \sqr{1}{2}\sqr{3}{4} + \sqr{2}{3}\sqr{1}{4} + \sqr{3}{1}\sqr{2}{4} &= 0\ .
\end{aligned}
\end{equation}

Thinking of the kinematic structures in terms of graphs, specifically using the already mentioned circular embedding, one way of implementing \eqref{eq:Schouten} is by untying crossing edges as shown in Figure \ref{fig::Schouten}. In a generic graph, this can be applied recursively until, after a finite number of steps, we end up with graphs which do not have any crossing. It is then clear that a basis of kinematic structures which are independent under Schouten identities can be obtained by building a basis of planar graphs only.

%\textcolor{blue}{If we represent these relations in terms of graph, this is equivalent to loosening a crossing graph, as shown in Figure \ref{fig::Schouten}. In a generic graph, this can be applied recursively until we get a sum of graphs which do not have any crossing. Then a basis of independent structures up to momentum conservation identities is given by digraphs for which each of the graphs is planar.\todo{Attenzione al digrafph e al multigraph}}
%\textcolor{green}{If we consider a circular embedding for our graphs, in other words we take all the ordered nodes to be equidistant points on a circle, one way of implementing \eqref{eq:Schouten} is by disentangling crossing edges as shown in Figure \ref{fig::Schouten}. Clearly, applying this procedure a finite number of times to any given graph will lead to a linear combination of planar graphs, where none of the edges cross each other\footnote{The concept of planar graphs and crossing edges must always be interpreted in light of the above mentioned circular embedding, and only makes sense in such a context.}. This means that if we are able to find a basis of kinematic structures corresponding to planar graphs, this will be independent under Schouten identities.}

\begin{figure}
    \centering
		    \begin{tikzpicture}[scale=2.5,>=Stealth]
		    
			    \node at (-0.08,-0.08) {$4$};
			    \node at (-0.08,1.08) {$1$};
			    \node at (1.08,1.08) {$2$};
			    \node at (1.08,-0.08) {$3$};
			    \node at (1.25,0.5) {$=$};
			    
			    \draw [fill] (0,0) circle [radius = 0.02];
			    \draw [fill] (1,0) circle [radius = 0.02];
			    \draw [fill] (0,1) circle [radius = 0.02];
			    \draw [fill] (1,1) circle [radius = 0.02];
			    
			    \draw (1,0) [<-,thick,red] -- (0.55,0.45);
			    \draw [thick, red] (0.45,0.55) arc (135:-45:0.07);
			    \draw [thick, red] (0.45,0.55) -- (0,1);
			    \draw [->,thick,red](1,1) -- (0,0);
	
		    \end{tikzpicture}
		    \begin{tikzpicture}[scale=2.5,>=Stealth]
		    
			    \node at (-0.08,-0.08) {$4$};
			    \node at (-0.08,1.08) {$1$};
			    \node at (1.08,1.08) {$2$};
			    \node at (1.08,-0.08) {$3$};
			    \node at (1.25,0.5) {$+$};
			    
			    \draw [fill] (0,0) circle [radius = 0.02];
			    \draw [fill] (1,0) circle [radius = 0.02];
			    \draw [fill] (0,1) circle [radius = 0.02];
			    \draw [fill] (1,1) circle [radius = 0.02];
			    
			    \draw [<-,thick,red](1,0) -- (1,1);
			    \draw [->,thick,red](0,1) -- (0,0);
	
		    \end{tikzpicture}
		    \begin{tikzpicture}[scale=2.5,>=Stealth]
		    
			    \node at (-0.08,-0.08) {$4$};
			    \node at (-0.08,1.08) {$1$};
			    \node at (1.08,1.08) {$2$};
			    \node at (1.08,-0.08) {$3$};
			    
			    \draw [fill] (0,0) circle [radius = 0.02];
			    \draw [fill] (1,0) circle [radius = 0.02];
			    \draw [fill] (0,1) circle [radius = 0.02];
			    \draw [fill] (1,1) circle [radius = 0.02];
			    
			    \draw [->,thick,red](0,1) -- (1,1);
			    \draw [->,thick,red](1,0) -- (0,0);
	
		    \end{tikzpicture}
	\caption{Schouten identities are equivalent to untying crossings for both the two graphs in the multigraph: $\textcolor{red}{\agl{1}{3}\agl{2}{4}} = \textcolor{red}{\agl{1}{4}\agl{2}{3}} + \textcolor{red}{\agl{1}{2}\agl{3}{4}}$.}
    \label{fig::Schouten}
\end{figure}

\subsubsection{Momentum conservation}

In general, we consider an $n$-point amplitude with $n_\partial > 0$. Each momentum in the amplitude can be assigned to any of the $n$ particles, which increases the valence of the corresponding vertex by $(1,1)$. The number of momenta associated to each vertex is then $\min\{v^i_a,v^i_s\}$.

We can take into account most of the relations coming from momentum conservation just by excluding the momentum of the $n^{\rm th}$-particle from the previous assignment. Then the $n^{\rm th}$-vertex will have valence $(\frac{|h_n|+h_n}{2},\frac{|h_n|-h_n}{2})$.

There are however $n$\footnote{One of which can be written as a linear combination of the other $n-1$.} additional relations coming from momentum conservation which do not explicitly involve the momentum of the $n^{\rm th}$-particle:
\begin{align}
\label{eq::momcons2}
    0 = 
    \begin{cases}
        \sum\limits_{j=1}^{n-1} \agl{i}{j} \sqr{j}{n} &  h_n>0\\[.5em]
        \sum\limits_{j=1}^{n-1} \agl{n}{j} \sqr{j}{i} &  h_n<0\\
    \end{cases}
\end{align}
which are a consequence of the Dirac equation $p_{n\, \alpha \dot{\alpha}}\, \tlambda_{n}^{\dot{\alpha}}=\, 0 \, = \lambda^{\alpha}_n\, p_{n\, \alpha \dot{\alpha}}$, and
\begin{equation}
\label{eq::momcons1}
    \sum_{i=1}^{n-2} \sum_{j=i+1}^{n-1} s_{i j} = 0\ ,\\
\end{equation}
Some observations are in order:
\begin{itemize}
    \item The Schouten identities do not change the valences of vertices in the multigraph, so they do not change the number of momenta associated to each vertex.
    %\item Since We want a basis of planar graphs, we solve most of the \eqref{eq::momcons2} for one of the momenta which maximises the number of planar multigraphs, the natural choice being either $p_1$ or $p_{n-1}$ (a different choice would give an over-counting of the independent structures) . Most of the relations in \eqref{eq::momcons2} are taken into account discarding all the structures with $\agl{i}{n-1} \sqr{n-1}{n}$ or $\agl{n}{n-1} \sqr{n-1}{i}$ according to the helicity of the $n^{\rm th}$-particle (or equivalently $\agl{i}{1} \sqr{1}{n} \big/ \agl{n}{1} \sqr{1}{i}$).
    %\item Among \eqref{eq::momcons2}, there is one relation which do not involve neither $p_n$ nor $p_{n-1}$. This is taken into account discarding those structures where $\agl{n-1}{1} \sqr{1}{n} \big/ \agl{n}{1} \sqr{1}{n-1}$ appears (or $\agl{1}{n-1} \sqr{n-1}{n} \big/ \agl{n}{n-1} \sqr{n-1}{1}$).
    \item Since we want a basis of planar graphs, we solve all but one of the \eqref{eq::momcons2} for one of the momenta which maximises the number of planar multigraphs, the natural choice being either $p_1$ or $p_{n-1}$ (a different choice would give an over-counting of the independent structures). The considered identities are then taken into account by simply discarding all the structures involving $\agl{i}{n-1} \sqr{n-1}{n}$ or $\agl{n}{n-1} \sqr{n-1}{i}$ according to the helicity of the $n^{\rm th}$-particle (or equivalently $\agl{i}{1} \sqr{1}{n} \big/ \agl{n}{1} \sqr{1}{i}$).
    \item Among \eqref{eq::momcons2}, there is one relation which does not involve neither $p_n$ nor $p_{n-1}$. This is taken into account by discarding those structures where $\agl{n-1}{1} \sqr{1}{n} \big/ \agl{n}{1} \sqr{1}{n-1}$ appears (or $\agl{1}{n-1} \sqr{n-1}{n} \big/ \agl{n}{n-1} \sqr{n-1}{1}$).
    \item Finally, the constraint \eqref{eq::momcons1} forces us to discard the terms proportional to $s_{1\, n-1}$.
\end{itemize}
This algorithm classifies efficiently all the ${\rm SL}(2,\mathbb{C})$-invariant structures which are polynomial in the spinor variables with fixed mass dimension and helicity configuration, associated to each $(n_{g^-},n_{g^+},n_{f^-},n_{f^+},n_s)$. It also provides a very simple way of writing the dependent structures as linear combinations of the independent ones. We also notice that this algorithm an be applied also beyond gauge theories. Furthermore, the generalisation of this algorithm to massive spinors is possible and it will be discussed in future works.

\subsection{The classification of SMEFT interactions}
The classification of the helicity structures is completely theory-independent and is indeed not limited to gauge theories, but can be applied to effective field theories of gravity, with (massive and spinning) matter as well. Information about the Standard Model enters only in the $SU(3)\times SU(2) \times U(1)$ (invariant) structures associated to the chosen set of particles.

\subsubsection{The gauge group structures}
\label{sec::gaugestructures}
The classification of the invariant structures of the gauge groups can be worked out using standard group theory techniques. In particular
\begin{itemize}
    \item $U(1)$: to each $(n_{g^-},n_{g^+},n_{f^-},n_{f^+},n_s)$ structure we associate all the possible combinations of Standard Model states for which the total hypercharge is zero.
    \item $SU(2)$: we notice that the algorithm presented in the previous section can be generalised to the the case of $SU(2)$ invariants with a single graph associated to the invariants. Each oriented edge from the $n^{\rm th}$ to the $m^{\rm th}$ vertices correspond to an $\epsilon^{i_n i_m}$ tensors and the valence of each vertex $v_i$ is fixed by the representation of the $i^{\rm th}$-particle, labelled by its dimension $\mathbf{v_i + 1}$. The indices associate to the same vertex must be taken as completely symmetric. In the case of the $SU(2)$ group there is no analogous of momentum conservation, so the independent structures can be taken to be in one-to-one correspondence with planar graphs.
    \item $SU(3)$: the $SU(N)$ invariants have been studied a lot both in the mathematics and in the physics literature (see, for example, \cite{Dittner:1971fy,Dittner:1972hm,deAzcarraga:1997ya}), so we will not go into further details here. In our algorithm we adopt the standard Littlewood-Richardson rule \cite{littlewood1934,Robinson1938} as suggested in \cite{Li:2020gnx,Li:2020xlh}.
\end{itemize}

Once the kinematic structures associated to $(n_{g^-},n_{g^+},n_{f^-},n_{f^+},n_s)$ have been generated and a compatible set of gauge singlets was found, we combine all the invariants in order to find a basis of independent structures enclosing information about both the kinematics and the colour. If no identical fields are present, these structures coincide with the minimal amplitudes, else one needs to impose Bose-Einstein and Dirac-Fermi statistics as explained in the next section.

\subsubsection{Repeated fields and Young projectors}
There are cases for which the minimal amplitude involves identical states, for example for $\left[g_{\OO}\right]=-2$ we could have minimal amplitudes with $(G^+,G^+,G^+)$ or $(Q,Q,u,d)$. The treatment of this subtlety has been systematically taken into account in \cite{Fonseca:2019yya,Li:2020gnx}. Starting from their classification, we take a slightly different approach, since we deal with minimal amplitudes and not with operators. We distinguish between identical bosons and fermions at the level of the minimal amplitude and impose Bose-Einstein statistic to the former and Dirac-Fermi statistic for the latter. In practise, we consider all the previously classified independent structures and we act on them with a proper Young projector over the labels of the identical states:
\begin{itemize}
    \item in the case of $n$ identical bosons we act on the structures with the symmetriser projector
    \begin{equation}
        \mathcal{Y}_{\begin{ytableau} {\scriptstyle 1} \\  \none[{\scriptscriptstyle \vdots}] \\ {\scriptstyle n} \end{ytableau}} = \frac{1}{n!} \sum_{i=1}^{n!} p_i\ ,
    \end{equation}
    where $p_i$ are all the permutations of the $n$ labels associated to the identical bosons.
    \item in the case of $n$ identical fermions we act on the structures with the total anti-symmetriser projector
    \begin{equation}
        \mathcal{Y}_{\begin{ytableau} {\scriptstyle 1} &  \none[{\scriptscriptstyle \cdots}] & {\scriptstyle n} \end{ytableau}} = \frac{1}{n!} \sum_{i=1}^{n!} s_{i}\ p_i\ ,
    \end{equation}
    where $s_i$ is the signature of the permutations $p_i$.
\end{itemize}
Once, we apply the Young projectors to the independent minimal amplitudes, we will end up with a sum over terms which will not necessarily belong to the basis of independent structures chosen. In order to find the minimal amplitudes, we need to re-write these symmetrised amplitudes in terms of elements of our structure basis and check if they are linearly independent from each other (which in general will not be the case, some structures will even be automatically zero after projection).

A further subtlety arise in the case of the Standard Model, due to the {\it flavour} of fermions: to each particle we can associate a further $SU(N_f)$ index, where $N_f$ is the number of flavours. The independent minimal amplitudes can then be classified in terms of inequivalent irreducible representations of $SU(N_f)$, which are in one-to-one correspondence with the irreducible representations of the symmetric group $S_n$, where $n$ is the number of identical fermions in the same family. For example, for dimension 6 operators we can consider the barion number violating effective interactions with $(Q,Q,Q,L)$ ($n=3$). Then we have a basis of four independent structures: 
\begin{align}
    \epsilon^{a_1 a_2 a_3} \epsilon^{i_1 i_4} \epsilon^{i_2 i_3}\, \agl{1}{2} \agl{3}{4}\ ,\\
    \epsilon ^{a_1a_2a_3} \epsilon ^{i_1i_2} \epsilon ^{i_3i_4} \agl{1}{2} \agl{3}{4}\ ,\\
    \epsilon ^{a_1a_2a_3} \epsilon^{i_1i_4} \epsilon^{i_2i_3} \agl{1}{4} \agl{2}{3}\ ,\\
   \epsilon ^{a_1a_2a_3} \epsilon ^{i_1i_2} \epsilon ^{i_3i_4} \agl{1}{4} \agl{2}{3}\ .
\end{align}
There are three inequivalent representations of $S_3$, corresponding to the Young diagrams $\ydiagram{3}$, $\ydiagram{2,1}$ and $\ydiagram{1,1,1}$. Then we can act on the independent structure with the projectors associated to the standard Young tableaux $\begin{ytableau} {\scriptstyle 1} &  {\scriptstyle 2} & {\scriptstyle 3} \end{ytableau}$, $\begin{ytableau} {\scriptstyle 1} &  {\scriptstyle 2} \\ {\scriptstyle 3} \end{ytableau}$, $\begin{ytableau} {\scriptstyle 1} \\  {\scriptstyle 2} \\ {\scriptstyle 3} \end{ytableau}$ \footnote{The fourth standard tableau $\begin{ytableau} {\scriptstyle 1} &  {\scriptstyle 3} \\ {\scriptstyle 2} \end{ytableau}$ would not give an independent minimal amplitude, because it could be obtained from the second one by relabelling: $\mathcal{Y}_{\, \begin{ytableau} {\scriptstyle 1} &  {\scriptstyle 3} \\ {\scriptstyle 2} \end{ytableau}} = (2\medspace 3)\circ \mathcal{Y}_{\, \begin{ytableau} {\scriptstyle 1} &  {\scriptstyle 2} \\ {\scriptstyle 3} \end{ytableau}} \circ (2\medspace 3)$, where $(2\medspace 3)$ is the permutation of the labels $2$ and $3$.}. There is a unique linearly independent structure associated to each irreducible representation:
\begin{align}
    &C^{\{3\},\{1\}}_{m_1 m_2 m_3,m_4} \ \mathcal{Y}_{\begin{ytableau} {\scriptstyle 1} \\  {\scriptstyle 2} \\ {\scriptstyle 3} \end{ytableau}} \circ \epsilon^{a_1 a_2 a_3} \epsilon^{i_1 i_4} \epsilon^{i_2 i_3}\, \agl{1}{2} \agl{3}{4}\ ,\\
    &C^{\{2,1\},\{1\}}_{m_1 m_2 m_3,m_4} \ \mathcal{Y}_{\begin{ytableau} {\scriptstyle 1} &  {\scriptstyle 2} \\ {\scriptstyle 3} \end{ytableau}} \circ \epsilon^{a_1 a_2 a_3} \epsilon^{i_1 i_4} \epsilon^{i_2 i_3}\, \agl{1}{2} \agl{3}{4}\ ,\\
    &C^{\{1,1,1\},\{ 1\}}_{m_1 m_2 m_3,m_4}\ \mathcal{Y}_{\begin{ytableau} {\scriptstyle 1} &  {\scriptstyle 2} & {\scriptstyle 3} \end{ytableau}} \circ \epsilon^{a_1 a_2 a_3} \epsilon^{i_1 i_4} \epsilon^{i_2 i_3}\, \agl{1}{2} \agl{3}{4}\ ,
\end{align}
where $C^{\pi,\{ 1\}}_{m_1 m_2 m_3,m_4}$ is a Wilson coefficient tensor associated to each effective minimal amplitude, with $\pi$ being the integer partition corresponding to the Young diagram for the $Q$ fields. Notice that Dirac-Fermi statistics forces the Wilson coefficient tensor to have the ``opposite'' symmetry properties with respect to the Young tableau associated to the projector: {\it e.g.} $C^{\{3\},\{1\} }_{m_1 m_2 m_3,m_4}= C^{\{3\},\{1\}}_{(m_1 m_2 m_3),m_4}$, $C^{\{2,1\},\{1\}}_{m_1 m_2 m_3,m_4}=C^{\{2,1\},\{1\}}_{[m_1 m_2] m_3,m_4}$, $C^{\{2,1\},\{1\}}_{[m_1 m_2 m_3],m_4} = 0$ and $C^{\{1,1,1\},\{1\}}_{m_1 m_2 m_3,m_4} = C^{\{1,1,1\},\{1\}}_{([m_1 m_2 m_3],m_4}$. %Then in this case we will find\footnote{The counting can be performed using the Hook Content Formula.} $\frac{N_f \, (N_f -1)\, (N_f -2)}{6}$, $\frac{(N_f +1)\, N_f \, (N_f -1)}{3}$ and $\frac{(N_f +2)\, (N_f +1)\, N_f}{6}$ independent operators respectively.
The number of independent operators for this specific case is\footnote{The counting can be performed using the Hook Content Formula.} $\frac{(N_f +2)\, (N_f +1)\, N_f}{6}$, $\frac{(N_f +1)\, N_f \, (N_f -1)}{3}$ and $\frac{N_f \, (N_f -1)\, (N_f -2)}{6}$ for each tensor respectively.

\section{The UV anomalous mass dimension matrix at leading order}
\label{sec::UVmixing}

In the previous two sections we argued that any four-point amplitude in the Standard Model can be fully determined from its factorisation channels (more details on this will be given in the following section), and we gave a general algorithm to find all the SMEFT interactions. These are all the ingredients needed to compute the UV mixing matrix for the SMEFT interactions $\gamma^{\rm UV}_{d\to d}$\footnote{In principle, we could consider $\gamma^{\rm UV}_{d_1\to d_2}$, with $d_1\neq d_2$, which involves amplitudes which are non-linear in the effective couplings. In this work we are not considering such contributions, but it is worth stressing that conceptually their treatment is very much the same.}, where $d$ is the mass dimension of the corresponding effective operators for $d=5,6,7,8$. In this section we restrict to the case of $N_f=1$ and we leave the treatment of full flavour dependence for future work.
The results for $\gamma^{\rm UV}_{d\to d}$ are provided in separate ancillary files, whereas at the end of this section we provide as an example the mixing coefficients of the dimension 6 and 8 operators relevant for Higgs plus $W$ production.

\subsection{Review of the method}
\label{sec::mixing}
Sticking to the notation introduced in Section \ref{sec::4points}, we write the effective amplitudes as $\mathcal{F}_{n,d,i} (p_1^{a_1, h_1}, \ldots , p_n^{a_n, h_n})$, where $d$ is the dimension of the operator and $i$ labels the minimal interactions (for example, in the case of $N_f=1$ and $d=6$, $i=1,\dots , 84$), in order to distinguish it from renormalisable amplitudes $\Aa$. The central formula for our computations has been presented in \cite{Caron-Huot:2016cwu} and gives the action of the {\it dilatation operator} $D=\frac{\partial}{\partial \log \mu}$ on the amplitude in terms of its discontinuity\footnote{This formula has been first presented in \cite{Caron-Huot:2016cwu} for $\mathcal{F}$ being a form factor, but it trivially holds for (effective) amplitudes as well, by setting $q^\mu = 0$ in the form factor.}:
\begin{equation}
\label{eq::dilatationDisc}
    e^{-i \pi D}\, \mathcal{F}^* = S \otimes  \mathcal{F}^*\ ,
\end{equation}
where $S$ is the full S-matrix and on the RHS the product has to be interpreted as a matrix product weighted over a proper Lorentz phase space integral which, via the Optical Theorem, correspond to the discontinuity of the effective amplitude.

The dilatation operator is linked to the UV mixing matrix $\gamma^{\rm UV}_{i\to j}$ by the Callan-Symanzik equation \cite{Callan:1969sn,Symanzik:1970rt,Symanzik:1971vw}:
\begin{equation}
\label{eq::CallanSymanzik}
    D \mathcal{F}_{i} = \left(\gamma^{\rm UV}_{j\to i}-\gamma^{\rm IR}_i \,\delta_{i j} + +\beta(g_k^2) \frac{\partial}{\partial g_k^2}\, \delta_{i j}\right) \mathcal{F}_{j}\ ,
\end{equation}
where $\beta(g_k^2)$ is the beta-function for the coupling $g_k$ and $\gamma^{\rm IR}_i$ is the IR contribution to the anomalous dimension of the amplitude $\mathcal{F}_i$ which depends only on its external states.

Combining \eqref{eq::dilatationDisc} and \eqref{eq::CallanSymanzik}, expanding to leading order in the coupling and at linear order in the effective interactions, we find\footnote{In order not to clutter up the notation with factors of 2 and $\pi$, we will provide the result in the ancillary files in terms of the matrix $\gamma_{i j}^{\rm UV} \equiv 16 \pi^2 \gamma^{\rm UV}_{j\to i}$, where we factored out the usual loop factor $\frac{1}{16 \pi^2}$.}
\begin{equation}
    \begin{split}
        \gamma^{\rm UV}_{j\to i} \ \mathcal{F}_{j}(p_{1}^{h_1} \dots p_{n}^{h_n}) &= -\frac{1}{\pi} \sum_{\substack{l=1\\l<m}}^n \sum_{\{l_1,l_2\}} \int \frac{{\rm d} \Omega_2}{32 \pi^2} \left[\Aa_4 (p_{l_1}^{h_{l_1}} p_{l_2}^{h_{l_2}}\to p_{l}^{h_l} p_{m}^{h_m})-\sum_{k=1}^{3}\frac{ g_k^2\ T_{k,l_1}\cdot T_{k,l_2}}{\cos^2\theta\, \sin^2\theta}\right]\\
        &\hspace{7cm} \cdot \mathcal{F}_{i}(\dots p_{l_1}^{h_{l_1}}\dots p_{l_2}^{h_{l_2}}\dots)\\
        &+ \mathcal{F}_{i}(p_{1}^{h_1} \dots p_{n}^{h_n}) \cdot \sum_{l=1}^{n}\frac{\gamma_{\rm coll}^{(l)}}{16\pi^2} \ ,
    \end{split}
\end{equation}
where
\begin{equation}
    \int \frac{{\rm d} \Omega_2}{4 \pi}= \int_{0}^{2\pi} \frac{{\rm d}\phi}{2\pi} \int_{0}^{\frac{\pi}{2}} {\rm d} \theta\, 2 \cos\theta \sin\theta
\end{equation}
is the Lorentz phase space integral, the sum over $\{l_1, l_2\}$ is over the species and the helicity configurations of the internal particles, $\gamma_{\rm coll}^{(l)}$ is the IR collinear anomalous dimension associated to the $l^{\rm th}$-particle and the term with $T_{k,l_1}\cdot T_{k,l_2}$ takes care of the subtraction of the (divergent) IR cusp anomalous dimension (the label $k$ runs over the three factors of the gauge group $U(1)\times SU(2)\times SU(3)$). In particular, the latter is non-zero if the in- and out-states of the four-point amplitude are the same and, if this is the case, it is a proper contraction of the Lie algebra generators (or the product of the hypercharges in the case of $U(1)$) associated to the outgoing (or equivalently incoming) particles. For example, if the four-point amplitude is $\Aa_4 (\bar{Q}_{l_1} \,Q_{l_2} \to \bar{Q}_{l} \,Q_{m})$, then
\begin{equation}
    \sum_{k=1}^3 g_{k}^2\, T_{k,l}\cdot T_{k,m} = \left( -\frac{1}{6}\right)\cdot \frac{1}{6}\, g_1^2 + g_2^2\, \sigma^{I}\,^{i_m}_{j}\, \sigma^{I}\,^{j}_{i_l} +g_3^2\, \tau^{A}\,^{a_m}_{b}\, \tau^{A}\,^{b}_{a_l}\ .
\end{equation}
The helicity variables associated to the internal momenta, on the cut configuration, can be written in terms of the Lorentz phase space angles $\theta$ and $\phi$ and the external momenta $p_l$ and $p_m$, as first shown in \cite{Zwiebel:2011bx}:
\begin{equation}
    \begin{pmatrix}
        \lambda_{l_1} \\
        \lambda_{l_2}
    \end{pmatrix}
    =
    \begin{pmatrix}
        \cos \theta & -\sin\theta \, e^{i \phi} \\
        \sin\theta \, e^{-i \phi} & \cos \theta \\
    \end{pmatrix}
    \begin{pmatrix}
        \lambda_{l} \\
        \lambda_{m} \\
    \end{pmatrix} \> ,
\end{equation}
together with the complex conjugate rotation for the spinors $\widetilde{\lambda}_{l_1}$ and $\widetilde{\lambda}_{l_2}$.
The collinear anomalous dimensions for the particles in the Standard Model can be obtained by studying the anomalous dimension of UV protected operators, such as the stress-tensor as emphasised in \cite{Caron-Huot:2016cwu}:
\begin{equation}
\label{eq::collinear}
    \begin{split}
        \langle p_{1}^{h_1} p_{2}^{h_2} | T^{\mu \nu} | 0 \rangle \cdot \sum_{l=1}^{2}\frac{\gamma_{\rm coll}^{(l)}}{16\pi^2}  = \frac{1}{\pi} \sum_{\{l_1,l_2\}} \int& \frac{{\rm d} \Omega_2}{32 \pi^2} \Bigg[\Aa_4 (p_{l_1}^{h_{l_1}} p_{l_2}^{h_{l_2}}\to p_{1}^{h_1} p_{2}^{h_2})\\
        &-\sum_{k=1}^{3}\frac{ g_k^2\ T_{k,l_1}\cdot T_{k,l_2}}{\cos^2\theta\, \sin^2\theta}\Bigg] \cdot \langle p_{l_1}^{h_{l_1}} p_{l_2}^{h_{l_2}} | T^{\mu \nu} | 0 \rangle\ ,
    \end{split}
\end{equation}
A list of the collinear anomalous dimensions computed from the stress-tensor form factor can be found in Appendix \ref{sec::collinear}.

\subsection{The Higgs production in association with a \texorpdfstring{$W$}{W} boson}

As an illustrative application of the techniques discussed so far, we consider a subset of dimension-six and dimension-eight operators relevant for the Higgs production in association with a $W$ boson via proton scattering, {\it i.e.} the operators contributing to the scattering $p\, p \rightarrow h\, W$ as considered in \cite{Hays:2018zze}, with a technical difference due to the fact that in the mixing problem considered in this work we look at $N_f = 1$. In this section, we will compute the mixing among dimension-six and dimension-eight effective interactions separately. First, we present the relevant minimal amplitudes found using the algorithm presented in Section \ref{sec:classificationSMEFTops}, which are in one-to-one correspondence with the independent operators considered in \cite{Hays:2018zze}. Then, using the techniques just reviewed we compute the two UV mixing matrices, comparing the mixing matrix for dimension-six operators with known results in the literature \cite{Jenkins:2013zja,Jenkins:2013wua,Alonso:2013hga,Elias-Miro:2013mua,EliasMiro:2020tdv,Baratella:2020lzz}. The full mixing matrix for all the operators in the SMEFT up to dimension 8 can be found in the ancillary files.

There are thirteen dimension-six operators (five of which are self-hermitian) contributing to the scattering $p\, p \rightarrow h\, W$ and such counting can be performed using Hilbert series method. In Table \ref{tab::dim6} and Table \ref{tab::dim8} we show the content of the various operators and their multiplicities as shown in reference \cite{Henning:2015alf} and the corresponding independent minimal amplitudes, respectively for the dimension-six and the dimension-eight effective interactions.

The running of the Wilson coefficients
\begin{equation}
    \dot{c}_i = 16\pi^2 \mu \frac{\partial}{\partial\mu} c_i\ ,
\end{equation}
of the thirteen dimension-six operators is
\begin{table}[!t]
    \centering
    \begin{tabular}{c | c | c}
    \# & Hilbert series & Minimal amplitude \\ \hline
    1 & $\bar{H}^3 H^3$ & $\mathcal{Y}_{\begin{ytableau} {\scriptstyle 1} &  {\scriptstyle 2} & {\scriptstyle 3} \end{ytableau}} \circ \delta_{j_1}^{i_4} \delta_{j_2}^{i_5} \delta_{j_3}^{i_6}$ \\
    2 & \multirow{2}{*}{$2 D^2 \bar{H}^2 H^2$} & $\mathcal{Y}_{\begin{ytableau} {\scriptstyle 1} &  {\scriptstyle 2} \end{ytableau}} \circ \mathcal{Y}_{\begin{ytableau} {\scriptstyle 3} &  {\scriptstyle 4} \end{ytableau}} \circ \agl{1}{3} \sqr{1}{3} \delta_{j_1}^{i_3} \delta_{j_2}^{i_4}$ \\
    3 &  & $\mathcal{Y}_{\begin{ytableau} {\scriptstyle 1} &  {\scriptstyle 2} \end{ytableau}} \circ \agl{1}{2} \sqr{1}{2} \delta_{j_1}^{i_3} \delta_{j_2}^{i_4}$ \\
    4 & \multirow{2}{*}{$2 D \bar{Q} Q \bar{H} H$} & $\agl{1}{3} \sqr{2}{3}  \delta _{j_1}^{i_2} \delta _{j_3}^{i_4} \delta _{b_1}^{a_2}$ \\
    5 &  & $\agl{1}{3} \sqr{2}{3} \delta_{j_3}^{i_2} \delta_{j_1}^{i_4} \delta_{b_1}^{a_2}$ \\
    6 & $B_{-} B_{-} \bar{H} H$ & $\agl{1}{2}^2 \delta_{j_3}^{i_4}$ \\
    7 & $B_{+} B_{+} \bar{H} H$ & $\sqr{1}{2}^2 \delta_{j_3}^{i_4}$ \\
    8 & $W_{-} W_{-} \bar{H} H$ & $\agl{1}{2}^2 \delta^{I_1 I_2} \delta _{j_3}^{i_4}$ \\
    9 & $W_{+} W_{+} \bar{H} H$ & $\sqr{1}{2}^2 \delta^{I_1 I_2} \delta _{j_3}^{i_4}$ \\
    10 & $G_{-} G_{-} \bar{H} H$ & $\agl{1}{2}^2 \delta^{A_1 A_2} \delta _{j_3}^{i_4}$ \\
    11 & $G_{+} G_{+} \bar{H} H$ & $\sqr{1}{2}^2 \delta^{A_1 A_2} \delta _{j_3}^{i_4}$ \\
    12 & $B_{-} W_{-} \bar{H} H$ & $\agl{1}{2}^2 \sigma\indices{^{I_2 i_4}_{j_3}}$ \\
    13 & $B_{+} W_{+} \bar{H} H$ & $\sqr{1}{2}^2 \sigma\indices{^{I_2 i_4}_{j_3}}$ 
    \end{tabular}
    \caption{The table shows the thirteen dimension-6 operators and their multiplicity as a result of the Hilbert series method. To each independent operator we associate and enumerate a set of independent minimal amplitudes.}
    \label{tab::dim6}
\end{table}

\begin{table}[!t]
    \centering
    \resizebox{\textwidth}{!}{
    \begin{tabular}{c | c | c || c | c | c}
        \# & Hilbert series & Minimal amplitude & \# & Hilbert series & Minimal amplitude \\ \hline
    1 & $\bar{H}^4 H^4$ & $\mathcal{Y}_{\begin{ytableau} {\scriptstyle 1} &  {\scriptstyle 2} & {\scriptstyle 3}  & {\scriptstyle 4} \end{ytableau}} \circ \delta_{j_1}^{i_4} \delta_{j_2}^{i_5} \delta_{j_3}^{i_6} \delta_{j_4}^{i_8}$ & 34 & \multirow{2}{*}{$2 D^2 B_{-} W_{-} \bar{H} H$} & $\agl{1}{2}^3 \sqr{1}{2} \sigma\indices{^{I_2 i_4}_{j_3}}$ \\[.1em]
    2 & $B_{-}^2 \bar{H}^2 H^2$ & $\mathcal{Y}_{\begin{ytableau} {\scriptstyle 3}  & {\scriptstyle 4} \end{ytableau}} \circ \agl{1}{2}^2 \delta_{j_4}^{i_5} \delta_{j_3}^{i_6}$ & 35 &  & $\agl{1}{2}^2 \agl{2}{3} \sqr{2}{3} \sigma\indices{^{I_2 i_4}_{j_3}}$ \\[.1em]
    3 & $B_{+}^2 \bar{H}^2 H^2$ & $\mathcal{Y}_{\begin{ytableau} {\scriptstyle 3}  & {\scriptstyle 4} \end{ytableau}} \circ \sqr{1}{2}^2 \delta_{j_4}^{i_5} \delta_{j_3}^{i_6}$ & 36 & \multirow{2}{*}{$2 D^2 B_{+} W_{+} \bar{H} H$} & $\sqr{1}{2}^3 \agl{1}{2} \sigma\indices{^{I_2 i_4}_{j_3}}$ \\[.1em]
    4 & $B_{-} W_{-} \bar{H}^2 H^2$ & $ \mathcal{Y}_{\begin{ytableau} {\scriptstyle 3}  & {\scriptstyle 4} \end{ytableau}} \circ \mathcal{Y}_{\begin{ytableau} {\scriptstyle 5}  & {\scriptstyle 6} \end{ytableau}} \circ \agl{1}{2}^2 \delta_{j_4}^{i_6} \sigma\indices{^{I_2 i_5}_{j_3}}$ & 37 &  & $\sqr{1}{2}^2 \agl{2}{3} \sqr{2}{3} \sigma\indices{^{I_2 i_4}_{j_3}}$ \\[.1em]
    5 & $B_{+} W_{+} \bar{H}^2 H^2$ & $ \mathcal{Y}_{\begin{ytableau} {\scriptstyle 3}  & {\scriptstyle 4} \end{ytableau}} \circ \mathcal{Y}_{\begin{ytableau} {\scriptstyle 5}  & {\scriptstyle 6} \end{ytableau}} \circ \sqr{1}{2}^2 \delta_{j_4}^{i_6} \sigma\indices{^{I_2 i_5}_{j_3}}$ & 38 & $D^2 B_{-} W_{+} \bar{H} H$ & $\agl{1}{3}^2 \sqr{2}{3}^2 \sigma\indices{^{I_2 i_4}_{j_3}}$  \\[.1em]
    6 & \multirow{2}{*}{$2\, W_{-}^2 \bar{H}^2 H^2$} & $\mathcal{Y}_{\begin{ytableau} {\scriptstyle 3}  & {\scriptstyle 4} \end{ytableau}} \circ \agl{1}{2}^2 \delta^{I_1 I_2} \delta_{j_4}^{i_5} \delta_{j_3}^{i_6}$ & 39 & $D^2 W_{-} B_{+} \bar{H} H$ & $\agl{1}{3}^2 \sqr{2}{3}^2 \sigma\indices{^{I_1 i_4}_{j_3}}$ \\[.1em]
    7 &  & $\mathcal{Y}_{\begin{ytableau} {\scriptstyle 1}  & {\scriptstyle 2} \end{ytableau}} \circ \agl{1}{2}^2 \sigma^{I_2 i_5 i_6} \sigma\indices{^{I_1}_{j_3 j_4}}$ & 40 & \multirow{2}{*}{$2 D^2 W_{-} \bar{H}^2 H^2$} & $\mathcal{Y}_{\begin{ytableau} {\scriptstyle 2}  & {\scriptstyle 3} \end{ytableau}} \circ \mathcal{Y}_{\begin{ytableau} {\scriptstyle 4}  & {\scriptstyle 5} \end{ytableau}} \circ \agl{1}{2} \agl{1}{4} \sqr{2}{4} \epsilon ^{i_4i_5} \sigma\indices{^{I_1}_{j_2j_3}}$ \\[.1em]
    8 & \multirow{2}{*}{$2\, W_{+}^2 \bar{H}^2 H^2$} & $\mathcal{Y}_{\begin{ytableau} {\scriptstyle 3}  & {\scriptstyle 4} \end{ytableau}} \circ \sqr{1}{2}^2 \delta^{I_1 I_2} \delta_{j_4}^{i_5} \delta_{j_3}^{i_6}$ & 41 &  & $\mathcal{Y}_{\begin{ytableau} {\scriptstyle 2}  & {\scriptstyle 3} \end{ytableau}} \circ \mathcal{Y}_{\begin{ytableau} {\scriptstyle 4}  & {\scriptstyle 5} \end{ytableau}} \circ \agl{1}{2} \agl{1}{3} \sqr{2}{3} \delta _{j_3}^{i_5} \sigma\indices{^{I_1 i_4}_{j_2}} $ \\[.1em]
    9 &  & $\mathcal{Y}_{\begin{ytableau} {\scriptstyle 1}  & {\scriptstyle 2} \end{ytableau}} \circ \sqr{1}{2}^2 \sigma^{I_2 i_5 i_6} \sigma\indices{^{I_1}_{j_3 j_4}}$ & 42 & \multirow{2}{*}{$2 D^2 W_{+} \bar{H}^2 H^2$} & $\mathcal{Y}_{\begin{ytableau} {\scriptstyle 2}  & {\scriptstyle 3} \end{ytableau}} \circ \mathcal{Y}_{\begin{ytableau} {\scriptstyle 4}  & {\scriptstyle 5} \end{ytableau}} \circ \sqr{1}{2} \sqr{1}{4} \agl{2}{4} \epsilon ^{i_4i_5} \sigma\indices{^{I_1}_{j_2j_3}}$ \\[.1em]
    10 & $G_{-}^2 \bar{H}^2 H^2$ & $\mathcal{Y}_{\begin{ytableau} {\scriptstyle 3}  & {\scriptstyle 4} \end{ytableau}} \circ \agl{1}{2}^2 \delta^{A_1 A_2} \delta_{j_4}^{i_5} \delta_{j_3}^{i_6}$ & 43 &  & $\mathcal{Y}_{\begin{ytableau} {\scriptstyle 2}  & {\scriptstyle 3} \end{ytableau}} \circ \mathcal{Y}_{\begin{ytableau} {\scriptstyle 4}  & {\scriptstyle 5} \end{ytableau}} \circ \sqr{1}{2} \sqr{1}{3} \agl{2}{3} \delta _{j_3}^{i_5} \sigma\indices{^{I_1 i_4}_{j_2}} $ \\[.1em]
    11 & $G_{+}^2 \bar{H}^2 H^2$ & $\mathcal{Y}_{\begin{ytableau} {\scriptstyle 3}  & {\scriptstyle 4} \end{ytableau}} \circ \sqr{1}{2}^2 \delta^{A_1 A_2} \delta_{j_4}^{i_5} \delta_{j_3}^{i_6}$ & 44 & \multirow{3}{*}{$3 D^4 \bar{H}^2 H^2$} & $\mathcal{Y}_{\begin{ytableau} {\scriptstyle 1}  & {\scriptstyle 2} \end{ytableau}} \circ \agl{1}{2}^2 \sqr{1}{2}^2\delta _{j_1}^{i_3} \delta _{j_2}^{i_4}$ \\[.1em]
    12 & $B_{-} W_{-}^2 \bar{H} H$ & $\agl{1}{2} \agl{2}{3} \agl{1}{3} \epsilon^{I_2 I_3 X_{6}} \sigma\indices{^{X_{6} i_5}_{j_4}}$ & 45 &  & $\mathcal{Y}_{\begin{ytableau} {\scriptstyle 1}  & {\scriptstyle 2} \end{ytableau}} \circ \mathcal{Y}_{\begin{ytableau} {\scriptstyle 3}  & {\scriptstyle 4} \end{ytableau}} \circ \agl{1}{3}^2 \sqr{1}{3}^2\delta _{j_1}^{i_3} \delta _{j_2}^{i_4}$ \\[.1em]
    13 & $B_{+} W_{+}^2 \bar{H} H$ & $\sqr{1}{2} \sqr{2}{3} \sqr{1}{3} \epsilon^{I_2 I_3 X_{6}} \sigma\indices{^{X_{6} i_5}_{j_4}}$ & 46 &  & $\mathcal{Y}_{\begin{ytableau} {\scriptstyle 1}  & {\scriptstyle 2} \end{ytableau}} \circ \mathcal{Y}_{\begin{ytableau} {\scriptstyle 3}  & {\scriptstyle 4} \end{ytableau}} \circ \agl{1}{2}\agl{1}{3} \sqr{1}{2} \sqr{1}{3} \delta _{j_1}^{i_3} \delta _{j_2}^{i_4}$ \\[.1em]
    14 & $W_{-}^3 \bar{H} H$ & $\agl{1}{2} \agl{2}{3} \agl{1}{3} \epsilon^{I_1I_2I_3} \delta_{j_4}^{i_5}$ & 47 & \multirow{4}{*}{$4 D \bar{Q} Q \bar{H}^2 H^2$} & $\mathcal{Y}_{\begin{ytableau} {\scriptstyle 3}  & {\scriptstyle 4} \end{ytableau}} \circ \mathcal{Y}_{\begin{ytableau} {\scriptstyle 5}  & {\scriptstyle 6} \end{ytableau}} \circ \agl{1}{3} \sqr{2}{3} \epsilon _{j_3j_4} \delta _{j_1}^{i_5} \delta _{b_1}^{a_2}$ \\[.1em]
    15 & $W_{+}^3 \bar{H} H$ & $\sqr{1}{2} \sqr{2}{3} \sqr{1}{3} \epsilon^{I_1I_2I_3} \delta_{j_4}^{i_5}$ & 48 &  & $\mathcal{Y}_{\begin{ytableau} {\scriptstyle 3}  & {\scriptstyle 4} \end{ytableau}} \circ \mathcal{Y}_{\begin{ytableau} {\scriptstyle 5}  & {\scriptstyle 6} \end{ytableau}} \circ \agl{1}{3} \sqr{2}{3} \delta _{j_3}^{i_2} \delta _{j_1}^{i_5} \delta _{j_4}^{i_6} \delta _{b_1}^{a_2}$ \\[.1em]
    16 & $G_{-}^3 \bar{H} H$ & $\agl{1}{2} \agl{2}{3} \agl{1}{3} f^{A_1A_2A_3} \delta_{j_4}^{i_5}$ & 49 &  & $\mathcal{Y}_{\begin{ytableau} {\scriptstyle 3}  & {\scriptstyle 4} \end{ytableau}} \circ \mathcal{Y}_{\begin{ytableau} {\scriptstyle 5}  & {\scriptstyle 6} \end{ytableau}} \circ \agl{1}{5} \sqr{2}{5} \delta _{j_3}^{i_2} \delta _{j_4}^{i_5} \delta _{j_1}^{i_6} \delta _{b_1}^{a_2}$ \\[.1em]
    17 & $G_{+}^3 \bar{H} H$ & $\sqr{1}{2} \sqr{2}{3} \sqr{1}{3} f^{A_1A_2A_3} \delta_{j_4}^{i_5}$ & 50 &  & $\mathcal{Y}_{\begin{ytableau} {\scriptstyle 3}  & {\scriptstyle 4} \end{ytableau}} \circ \mathcal{Y}_{\begin{ytableau} {\scriptstyle 5}  & {\scriptstyle 6} \end{ytableau}} \circ \agl{1}{3} \sqr{2}{3} \delta _{j_1}^{i_2} \delta _{j_4}^{i_5} \delta _{j_3}^{i_6} \delta _{b_1}^{a_2}$ \\[.1em]
    18 & \multirow{2}{*}{$2 D^2 \bar{H}^3 H^3$} & $\mathcal{Y}_{\begin{ytableau} {\scriptstyle 1} & {\scriptstyle 2} & {\scriptstyle 3} \end{ytableau}} \circ \mathcal{Y}_{\begin{ytableau} {\scriptstyle 4} & {\scriptstyle 5} & {\scriptstyle 6} \end{ytableau}} \circ \agl{1}{2} \sqr{1}{2} \delta_{j_1}^{i_4} \delta_{j_2}^{i_5} \delta_{j_3}^{i_6}$ & 51 & \multirow{6}{*}{$6 D W_{-} \bar{Q} Q \bar{H} H$} & $\agl{1}{2}^2 \sqr{2}{3} \delta _{j_4}^{i_5} \delta _{b_2}^{a_3} \sigma\indices{^{I_1 i_3}_{j_2}}$ \\[.1em]
    19 &  & $\mathcal{Y}_{\begin{ytableau} {\scriptstyle 1} & {\scriptstyle 2} & {\scriptstyle 3} \end{ytableau}} \circ \mathcal{Y}_{\begin{ytableau} {\scriptstyle 4} & {\scriptstyle 5} & {\scriptstyle 6} \end{ytableau}} \circ \agl{1}{4} \sqr{1}{4} \delta_{j_1}^{i_4} \delta_{j_2}^{i_5} \delta_{j_3}^{i_6}$ & 52 &  & $\agl{1}{2}\agl{1}{4}\sqr{3}{4} \delta _{j_4}^{i_5} \delta _{b_2}^{a_3} \sigma\indices{^{I_1 i_3}_{j_2}}$ \\[.1em]
    20 & $D^2 B_{-} B_{+} \bar{H} H$ & $\agl{1}{3}^2 \sqr{2}{3}^2 \delta_{j_3}^{i_4}$ & 53 &  & $\agl{1}{2}^2 \sqr{2}{3} \delta _{j_4}^{i_3} \delta _{b_2}^{a_3} \sigma\indices{^{I_1 i_5}_{j_2}}$ \\[.1em]
    21 & \multirow{2}{*}{$2 D^2 W_{-} W_{+} \bar{H} H$} & $\agl{1}{3}^2 \sqr{2}{3}^2 \delta^{I_1 I_2} \delta_{j_3}^{i_4}$ & 54 &  & $\agl{1}{2}\agl{1}{4}\sqr{3}{4} \delta _{j_4}^{i_3} \delta _{b_2}^{a_3} \sigma\indices{^{I_1 i_5}_{j_2}}$ \\[.1em]
    22 & & $\agl{1}{3}^2 \sqr{2}{3}^2 \epsilon^{I_1 I_2 X_6} \sigma\indices{^{X_6 i_4}_{j_3}}$ & 55 &  & $\agl{1}{2}^2 \sqr{2}{3} \delta _{j_2}^{i_3} \delta _{b_2}^{a_3} \sigma\indices{^{I_1 i_5}_{j_4}}$ \\[.1em]
    23 & $D^2 G_{-} G_{+} \bar{H} H$ & $\agl{1}{3}^2 \sqr{2}{3}^2 \delta^{A_1 A_2} \delta_{j_3}^{i_4}$ & 56 &  & $\agl{1}{2}\agl{1}{4}\sqr{3}{4} \delta _{j_2}^{i_3} \delta _{b_2}^{a_3} \sigma\indices{^{I_1 i_5}_{j_4}}$ \\[.1em]
    24 & $D^2 B_{-} \bar{H}^2 H^2$ & $\mathcal{Y}_{\begin{ytableau} {\scriptstyle 2}  & {\scriptstyle 3} \end{ytableau}} \circ \mathcal{Y}_{\begin{ytableau} {\scriptstyle 4}  & {\scriptstyle 5} \end{ytableau}} \circ \agl{1}{2} \agl{1}{4} \sqr{2}{4} \delta _{j_3}^{i_4} \delta _{j_2}^{i_5}$ & 57 & \multirow{6}{*}{$6 D W_{+} \bar{Q} Q \bar{H} H$} & $\sqr{1}{3}^2 \agl{2}{3} \delta _{j_4}^{i_5} \delta _{b_2}^{a_3} \sigma\indices{^{I_1 i_3}_{j_2}}$ \\[.1em]
    25 & $D^2 B_{+} \bar{H}^2 H^2$ & $\mathcal{Y}_{\begin{ytableau} {\scriptstyle 2}  & {\scriptstyle 3} \end{ytableau}} \circ \mathcal{Y}_{\begin{ytableau} {\scriptstyle 4}  & {\scriptstyle 5} \end{ytableau}} \circ \sqr{1}{2} \sqr{1}{4} \agl{2}{4} \delta _{j_3}^{i_4} \delta _{j_2}^{i_5}$ & 58 &  & $\sqr{1}{3}\sqr{1}{4}\agl{2}{4} \delta _{j_4}^{i_5} \delta _{b_2}^{a_3} \sigma\indices{^{I_1 i_3}_{j_2}}$ \\[.1em]
    26 & $D^2 B_{-}^2 \bar{H} H$ & $\agl{1}{2}^3 \sqr{1}{2} \delta _{j_3}^{i_4}$ & 59 &  & $\sqr{1}{3}^2 \agl{2}{3} \delta _{j_4}^{i_3} \delta _{b_2}^{a_3} \sigma\indices{^{I_1 i_5}_{j_2}}$ \\[.1em]
    27 & $D^2 B_{+}^2 \bar{H} H$ & $\sqr{1}{2}^3 \agl{1}{2} \delta _{j_3}^{i_4}$ & 60 &  & $\sqr{1}{3}\sqr{1}{4}\agl{2}{4} \delta _{j_4}^{i_3} \delta _{b_2}^{a_3} \sigma\indices{^{I_1 i_5}_{j_2}}$ \\[.1em]
    28 & \multirow{2}{*}{$2 D^2 W_{-}^2 \bar{H} H$} & $\agl{1}{2}^3 \sqr{1}{2} \delta^{I_1 I_2} \delta _{j_3}^{i_4}$ & 61 &  & $\sqr{1}{3}^2 \agl{2}{3} \delta _{j_2}^{i_3} \delta _{b_2}^{a_3} \sigma\indices{^{I_1 i_5}_{j_4}}$ \\[.1em]
    29 & & $\mathcal{Y}_{\begin{ytableau} {\scriptstyle 1}  & {\scriptstyle 2} \end{ytableau}} \circ \agl{1}{2}^2 \agl{2}{3} \sqr{2}{3} \epsilon^{I_1 I_2 X_6} \sigma\indices{^{X_6 i_4}_{j_3}}$ & 62 &  & $\sqr{1}{3}\sqr{1}{4}\agl{2}{4} \delta _{j_2}^{i_3} \delta _{b_2}^{a_3} \sigma\indices{^{I_1 i_5}_{j_4}}$ \\[.1em]
    30 & \multirow{2}{*}{$2 D^2 W_{+}^2 \bar{H} H$} & $\sqr{1}{2}^3 \agl{1}{2} \delta^{I_1 I_2} \delta _{j_3}^{i_4}$ & 63 & \multirow{4}{*}{$4 D^3 \bar{Q} Q \bar{H} H$} & $\agl{1}{3} \agl{2}{3} \sqr{2}{3}^2 \delta _{j_3}^{i_2} \delta _{j_1}^{i_4} \delta _{b_1}^{a_2}$ \\[.1em]
    31 &  & $\mathcal{Y}_{\begin{ytableau} {\scriptstyle 1}  & {\scriptstyle 2} \end{ytableau}} \circ \sqr{1}{2}^2 \agl{2}{3} \sqr{2}{3} \epsilon^{I_1 I_2 X_6} \sigma\indices{^{X_6 i_4}_{j_3}}$ & 64 & & $\agl{1}{2} \agl{1}{3} \sqr{1}{2} \sqr{2}{3} \delta _{j_3}^{i_2} \delta _{j_1}^{i_4} \delta _{b_1}^{a_2}$\\ 
    32 & $D^2 G_{-}^2 \bar{H} H$ & $\agl{1}{2}^3 \sqr{1}{2} \delta^{A_1 A_2} \delta _{j_3}^{i_4}$ & 65 & & $\agl{1}{3} \agl{2}{3} \sqr{2}{3}^2 \delta _{j_1}^{i_2} \delta _{j_3}^{i_4} \delta _{b_1}^{a_2}$\\
    33 & $D^2 G_{+}^2 \bar{H} H$ & $\sqr{1}{2}^3 \agl{1}{2} \delta^{A_1 A_2} \delta _{j_3}^{i_4}$ & 66 & & $\agl{1}{2} \agl{1}{3} \sqr{1}{2} \sqr{2}{3}\delta _{j_1}^{i_2} \delta _{j_3}^{i_4} \delta _{b_1}^{a_2}$\\
    \end{tabular}
    }
    \caption{The table shows all the dimension-8 operators, their multiplicity and a set of independent minimal amplitudes.}
    \label{tab::dim8}
\end{table}

\begin{align*}
    \dot{c}^{(6)}_1 &= c^{(6)}_1 \left(6 g_1^2 Y_H^2+\frac{9 g_2^2}{2}+108 \lambda \right) + 6 c^{(6)}_1 \gamma_{\rm coll}^{H}\ ,\\
    \dot{c}^{(6)}_2 &= c^{(6)}_5 \left(8 g_1^2 Y_H Y_Q-6 g_2^2+48 \mathcal{Y}_1 \bar{\mathcal{Y}}_1+24 \mathcal{Y}_2 \bar{\mathcal{Y}}_2\right)+c^{(6)}_2 \left(-\frac{8 g_1^2 Y_H^2}{3}+8 g_2^2+24 \lambda \right)\\
    &+c^{(6)}_3 \left(2 g_1^2 Y_H^2+\frac{17 g_2^2}{2}-12 \lambda  \right) + c^{(6)}_4 \left(16 g_1^2 Y_H Y_Q+24 \mathcal{Y}_1 \bar{\mathcal{Y}}_1-24 \mathcal{Y}_2
    \bar{\mathcal{Y}}_2\right)+ 4 c^{(6)}_2 \gamma_{\rm coll}^{H} +\dots\ ,\\
\end{align*}
\begin{align*}
    \dot{c}^{(6)}_3 &= c^{(6)}_3 \left(26 g_1^2 Y_H^2+\frac{33 g_2^2}{2}+12 \lambda \right)+ c^{(6)}_4 \left(32 g_1^2 Y_H Y_Q+48 \mathcal{Y}_1 \bar{\mathcal{Y}}_1-48 \mathcal{Y}_2 \bar{\mathcal{Y}}_2\right)\\
    &+c^{(6)}_5 \left(16 g_1^2 Y_H Y_Q+24 \mathcal{Y}_1 \bar{\mathcal{Y}}_1-24 \mathcal{Y}_2 \bar{\mathcal{Y}}_2\right)-\frac{40}{3} c^{(6)}_2 g_1^2 Y_H^2 + 4 c^{(6)}_3 \gamma_{\rm coll}^{H}+\dots\ ,\\
    \dot{c}^{(6)}_4 & = c^{(6)}_4 \left(\frac{28 g_1^2 Y_H^2}{3}+14 g_1^2 Y_Q^2+\frac{21 g_2^2}{2}+8 g_3^2+12 \mathcal{Y}_1 \bar{\mathcal{Y}}_1\right)+ c^{(6)}_4 \left(2 \gamma_{\rm coll}^{H}  + 2 \gamma_{\rm coll}^{Q}\right)\\
    &+c^{(6)}_5 \left(\frac{2 g_1^2 Y_H^2}{3}+4 g_1^2 Y_Q^2+\frac{11 g_2^2}{6}-4 \mathcal{Y}_1 \bar{\mathcal{Y}}_1+8 \mathcal{Y}_2 \bar{\mathcal{Y}}_2\right)\\
    &+c^{(6)}_3 \left(g_1^2  Y_H Y_Q-\frac{g_2^2}{12}+2 \mathcal{Y}_1 \bar{\mathcal{Y}}_1-\mathcal{Y}_2 \bar{\mathcal{Y}}_2\right)+c^{(6)}_2 \left(-\frac{1}{3} g_1^2 Y_H Y_Q+\frac{g_2^2}{12}-\mathcal{Y}_1 \bar{\mathcal{Y}}_1\right)+\dots\ ,\\
    \dot{c}^{(6)}_5 & = c^{(6)}_5 \left(8 g_1^2 Y_H^2+6 g_1^2 Y_Q^2+\frac{41 g_2^2}{6}+8 g_3^2-4 \mathcal{Y}_1 \bar{\mathcal{Y}}_1+8 \mathcal{Y}_2 \bar{\mathcal{Y}}_2\right)+ c^{(6)}_5 \left(2 \gamma_{\rm coll}^{H}  + 2 \gamma_{\rm coll}^{Q}\right)\\
    &+c^{(6)}_2 \left(-\frac{g_2^2}{6}+\mathcal{Y}_1 \bar{\mathcal{Y}}_1+\mathcal{Y}_2 \bar{\mathcal{Y}}_2\right)+c^{(6)}_3 \left(\frac{g_2^2}{6}-\mathcal{Y}_1 \bar{\mathcal{Y}}_1-\mathcal{Y}_2 \bar{\mathcal{Y}}_2\right)\\
    & +c^{(6)}_4 (12 \mathcal{Y}_2 \bar{\mathcal{Y}}_2-12 \mathcal{Y}_1 \bar{\mathcal{Y}}_1) +\dots\ ,\\
    \dot{c}^{(6)}_6 &= c^{(6)}_6 \left(10 g_1^2 Y_H^2+\frac{3 g_2^2}{2}+12 \lambda \right)+6 c^{(6)}_{12} g_1 g_2 Y_H + c^{(6)}_6 \left(2 \gamma_{\rm coll}^{H}  + 2 \gamma_{\rm coll}^{B}\right)+\dots\ ,\\
    \dot{c}^{(6)}_7 &= c^{(6)}_7 \left(10 g_1^2 Y_H^2+\frac{3 g_2^2}{2}+12 \lambda \right)+6 c^{(6)}_{13} g_1 g_2 Y_H +c^{(6)}_7 \left(2 \gamma_{\rm coll}^{H}  + 2 \gamma_{\rm coll}^{B}\right)+\dots\ ,\\
    \dot{c}^{(6)}_8 &= c^{(6)}_8 \left(2 g_1^2 Y_H^2+\frac{7 g_2^2}{2}+12 \lambda \right)+2 c^{(6)}_{12} g_1 g_2 Y_H + c^{(6)}_8 \left(2 \gamma_{\rm coll}^{H}  + 2 \gamma_{\rm coll}^{W}\right)+\dots\ ,\\
    \dot{c}^{(6)}_9 & =c^{(6)}_9 \left(2 g_1^2 Y_H^2+\frac{7 g_2^2}{2}+12 \lambda \right)+2 c^{(6)}_{13} g_1 g_2 Y_H + c^{(6)}_9 \left(2 \gamma_{\rm coll}^{H}  + 2 \gamma_{\rm coll}^{W}\right)+\dots\ ,\\
    \dot{c}^{(6)}_{10} &=c^{(6)}_{10} \left(2 g_1^2 Y_H^2+\frac{3 g_2^2}{2}+12 \lambda \right) + c^{(6)}_{10} \left(2 \gamma_{\rm coll}^{H}  + 2 \gamma_{\rm coll}^{G}\right)+\dots\ ,\\
\end{align*}
\begin{align*}
    \dot{c}^{(6)}_{11} &=c^{(6)}_{11} \left(2 g_1^2 Y_H^2+\frac{3 g_2^2}{2}+12 \lambda \right) + c^{(6)}_{11} \left(2 \gamma_{\rm coll}^{H}  + 2 \gamma_{\rm coll}^{G}\right)+\dots\ ,\\
    \dot{c}^{(6)}_{12} &=c^{(6)}_{12} \left(6 g_1^2 Y_H^2+\frac{g_2^2}{2}+4 \lambda \right)+4 c^{(6)}_6 g_1 g_2 Y_H+4 c^{(6)}_8 g_1 g_2 Y_H\\
    & + c^{(6)}_{12} \left(2 \gamma_{\rm coll}^{H}  + \gamma_{\rm coll}^{W} + \gamma_{\rm coll}^{B}\right)+\dots\ ,\\
    \dot{c}^{(6)}_{13} &=c^{(6)}_{13} \left(6 g_1^2 Y_H^2+\frac{g_2^2}{2}+4 \lambda \right)+4 c^{(6)}_7 g_1 g_2 Y_H+4 c^{(6)}_9 g_1 g_2 Y_H \\
    &+ c^{(6)}_{13} \left(2 \gamma_{\rm coll}^{H}  + \gamma_{\rm coll}^{W} + \gamma_{\rm coll}^{B}\right)+\dots \ ,
\end{align*}
where the dots indicate that the operator mixes with other operators which we are not considering, {\it i.e.} already at leading order in the couplings the sector we are looking at is not closed. The last term in the RG evolution of each coefficient is needed to isolate the UV contributions from the diagonal IR anomalous dimension. These results fully match with previous calculations in the literature, after a proper change of basis, and we take this as a cross-check for the on-shell methods techniques in this paper.

Then we present the result for the running of the Wilson coefficients of the dimension-eight operators associated to the minimal amplitudes. Since most of the operators mix with operators outside the sector we are investigating, we are going to omit the dots, as well as the IR subtraction, {\it i.e.} we show $\dot{c}^{\prime (8)}_i = \dot{c}^{(8),{\rm UV}}_i - \dot{c}^{(8),{\rm IR}}_i$.

\begin{align*}
    \dot{c}^{\prime (8)}_{1} & = \left(6g_2^2 + 8 g_1^2 Y_H^2 + 192 \lambda\right) c^{(8)}_{1}\ ,\\ 
    \dot{c}^{\prime (8)}_{2} & =\left(3 g_2^2+20 g_1^2 Y_H^2+48 \lambda \right) c_2^{(8)}+8 g_1 g_2 Y_H c_4^{(8)}\ ,\\
    \dot{c}^{\prime (8)}_{3} & =\left(3
    g_2^2+20 g_1^2 Y_H^2+48 \lambda \right) c_3^{(8)}+8 g_1 g_2 Y_H c_5^{(8)}\ ,\\
    \dot{c}^{\prime (8)}_{4} & =8 g_1 g_2 Y_H
   c_2^{(8)}+\left(13 g_2^2+12 g_1^2 Y_H^2+40 \lambda \right) c_4^{(8)}+8 g_1 g_2 Y_H c_6^{(8)}+2
   g_1 g_2 Y_H c_7^{(8)}\ ,\\
   \dot{c}^{\prime (8)}_{5} & =8 g_1 g_2 Y_H c_3^{(8)}+\left(13 g_2^2+12 g_1^2 Y_H^2+40 \lambda
   \right) c_5^{(8)}+8 g_1 g_2 Y_H c_8^{(8)}+2 g_1 g_2 Y_H c_9^{(8)}\ ,\\ 
   \dot{c}^{\prime (8)}_{6} & =4 g_1 g_2 Y_H
   c_4^{(8)}+\left(7 g_2^2+4 g_1^2 Y_H^2+48 \lambda \right) c_6^{(8)}+\left(4 g_2^2-4 \lambda \right)
   c_7^{(8)}\ ,\\
   \dot{c}^{\prime (8)}_{7} & =\left(31 g_2^2+4 g_1^2 Y_H^2+24 \lambda \right) c_7^{(8)}+8 g_1 g_2 Y_H c_4^{(8)}\ ,\\ 
   \dot{c}^{\prime (8)}_{8} & =4
   g_1 g_2 Y_H c_5^{(8)}+\left(7 g_2^2+4 g_1^2 Y_H^2+48 \lambda \right) c_8^{(8)}+\left(4 g_2^2-4 \lambda \right) c_9^{(8)}\ ,\\ 
   \dot{c}^{\prime (8)}_{9} & =\left(31 g_2^2+4 g_1^2 Y_H^2+24 \lambda \right) c_9^{(8)}+8 g_1 g_2 Y_H
   c_5^{(8)}\ ,\\ 
   \dot{c}^{\prime (8)}_{10} & =\left(3 g_2^2+4 g_1^2 Y_H^2+48 \lambda \right) c_{10}^{(8)}\ ,\\ 
   \dot{c}^{\prime (8)}_{11} & =\left(3 g_2^2+4 g_1^2 Y_H^2+48 \lambda  \right) c_{11}^{(8)}\ ,\\ 
   \dot{c}^{\prime (8)}_{12} & =\left(\frac{39 g_2^2}{2}+6 g_1^2 Y_H^2+4 \lambda \right) c_{12}^{(8)}+4 g_1 g_2
   Y_H c_{14}^{(8)}\ ,\\ 
   \dot{c}^{\prime (8)}_{13} & =\left(\frac{39 g_2^2}{2}+6 g_1^2 Y_H^2+4 \lambda \right) c_{13}^{(8)}+4 g_1 g_2 Y_H
   c_{15}^{(8)}\ ,\\ 
   \dot{c}^{\prime (8)}_{14} & =\left(\frac{57 g_2^2}{2}+2 g_1^2 Y_H^2+12 \lambda \right) c_{14}^{(8)}+3 g_1 g_2 Y_H
   c_{12}^{(8)}\ ,\\ 
   \dot{c}^{\prime (8)}_{15} & =\left(\frac{57 g_2^2}{2}+2 g_1^2 Y_H^2+12 \lambda \right) c_{15}^{(8)}+3 g_1 g_2 Y_H
   c_{13}^{(8)}\ ,\\ 
   \dot{c}^{\prime (8)}_{16} & =\left(\frac{3 g_2^2}{2}+36 g_3^2+2 g_1^2 Y_H^2+12 \lambda \right) c_{16}^{(8)}\ ,\\ 
   \dot{c}^{\prime (8)}_{17} & =\left(\frac{3
   g_2^2}{2}+36 g_3^2+2 g_1^2 Y_H^2+12 \lambda \right) c_{17}^{(8)}\ ,\\
   \dot{c}^{\prime (8)}_{18} & =\left(10 g_2^2+\frac{116 g_1^2 Y_H^2}{3}+72
   \lambda \right) c_{18}^{(8)}+\left(\frac{17 g_2^2}{6}-26 g_1^2 Y_H^2-4 \lambda \right) c_{19}^{(8)}\\
   &+\left(-18
   g_2^2+108 \mathcal{Y}_1 \bar{\mathcal{Y}}_1+108 \mathcal{Y}_2 \bar{\mathcal{Y}}_2\right) c_{47}^{(8)}+\left(18 Y_H Y_Q g_1^2+\frac{45 g_2^2}{2}-108
   \mathcal{Y}_1 \bar{\mathcal{Y}}_1-162 \mathcal{Y}_2 \bar{\mathcal{Y}}_2\right) c_{48}^{(8)}\\
   &+\left(-18 Y_H Y_Q g_1^2-\frac{9 g_2^2}{2}+54 \mathcal{Y}_2
   \bar{\mathcal{Y}}_2\right) c_{49}^{(8)}+\left(36 Y_H Y_Q g_1^2+54 \mathcal{Y}_1 \bar{\mathcal{Y}}_1-54 \mathcal{Y}_2 \bar{\mathcal{Y}}_2\right)
   c_{50}^{(8)}\ ,\\
\end{align*}

\begin{align*}
    \dot{c}^{\prime (8)}_{20} & =g_1^2 c_{44}^{(8)} Y_H^2+\frac{1}{3} g_1^2
   c_{45}^{(8)} Y_H^2-\frac{1}{3} g_1^2 c_{46}^{(8)} Y_H^2+3 g_1 g_2 c_{38}^{(8)} Y_H+3 g_1 g_2
   c_{39}^{(8)} Y_H\\[.2em]
    & +\left(9 g_2^2+20 g_1^2 Y_H^2\right) c_{20}^{(8)}-4 g_1^2 Y_Q^2 c_{63}^{(8)}-8 g_1^2
    Y_Q^2 c_{65}^{(8)}\ ,\\
    \dot{c}^{\prime (8)}_{19} & =\left(-\frac{34 g_2^2}{3}-\frac{8 g_1^2 Y_H^2}{3}+16 \lambda \right) c_{18}^{(8)}+\left(\frac{145
    g_2^2}{6}+2 g_1^2 Y_H^2+52 \lambda \right) c_{19}^{(8)}\\
    &+\left(-27 g_2^2+162 \mathcal{Y}_1 \bar{\mathcal{Y}}_1+162 \mathcal{Y}_2
    \bar{\mathcal{Y}}_2\right) c_{47}^{(8)}+\left(-18 Y_H Y_Q g_1^2+\frac{45 g_2^2}{2}-162 \mathcal{Y}_1 \bar{\mathcal{Y}}_1-108 \mathcal{Y}_2 \bar{\mathcal{Y}}_2\right)
    c_{48}^{(8)}\\
    &+\left(18 Y_H Y_Q g_1^2+\frac{9 g_2^2}{2}-54 \mathcal{Y}_2 \bar{\mathcal{Y}}_2\right) c_{49}^{(8)}+\left(-36 Y_H Y_Q
    g_1^2-54 \mathcal{Y}_1 \bar{\mathcal{Y}}_1+54 \mathcal{Y}_2 \bar{\mathcal{Y}}_2\right) c_{50}^{(8)}\ ,\\
    \dot{c}^{\prime (8)}_{21} & =\frac{1}{4} c_{44}^{(8)} g_2^2+\frac{1}{12} c_{45}^{(8)} g_2^2-\frac{1}{12} c_{46}^{(8)}
    g_2^2-c_{63}^{(8)} g_2^2-2 c_{65}^{(8)} g_2^2+g_1 Y_H c_{38}^{(8)} g_2+g_1 Y_H
    c_{39}^{(8)} g_2\\
    &+\left(\frac{77 g_2^2}{3}+12 g_1^2 Y_H^2\right) c_{21}^{(8)}\ ,\\ 
    \dot{c}^{\prime (8)}_{22} & =\left(\frac{25 g_2^2}{3}+12 g_1^2
    Y_H^2\right) c_{22}^{(8)}-i g_2^2 c_{63}^{(8)}\ ,\\ 
    \dot{c}^{\prime (8)}_{23} & =-\frac{2}{3} c_{63}^{(8)} g_3^2-\frac{4}{3} c_{65}^{(8)}
    g_3^2+\left(9 g_2^2+22 g_3^2+12 g_1^2 Y_H^2\right) c_{23}^{(8)}\ ,\\ 
    \dot{c}^{\prime (8)}_{24} & =\left(\frac{25 g_2^2}{2}+\frac{38 g_1^2
    Y_H^2}{3}+12 \lambda \right) c_{24}^{(8)}\ ,\\ 
    \dot{c}^{\prime (8)}_{25} & =\left(\frac{25 g_2^2}{2}+\frac{38 g_1^2 Y_H^2}{3}+12 \lambda \right)
    c_{25}^{(8)}\ ,\\ 
    \dot{c}^{\prime (8)}_{26} & =\left(\frac{3 g_2^2}{2}+\frac{10 g_1^2 Y_H^2}{3}+12 \lambda \right) c_{26}^{(8)}+g_1 g_2 Y_H
    c_{34}^{(8)}-\frac{1}{2} g_1 g_2 Y_H c_{35}^{(8)}\ ,\\
   \dot{c}^{\prime (8)}_{27} & =\left(\frac{3 g_2^2}{2}+\frac{10 g_1^2 Y_H^2}{3}+12 \lambda  \right) c_{27}^{(8)}+g_1 g_2 Y_H c_{36}^{(8)}-\frac{1}{2} g_1 g_2 Y_H c_{37}^{(8)}\ ,\\ 
   \dot{c}^{\prime (8)}_{28} & =-\frac{25}{6} i
   c_{29}^{(8)} g_2^2+\frac{1}{3} g_1 Y_H c_{34}^{(8)} g_2-\frac{1}{6} g_1 Y_H c_{35}^{(8)}
   g_2+\left(\frac{11 g_2^2}{6}+2 g_1^2 Y_H^2+12 \lambda \right) c_{28}^{(8)}\ ,\\ 
   \dot{c}^{\prime (8)}_{29} & =20 i c_{28}^{(8)} g_2^2-4 i g_1
   Y_H c_{34}^{(8)} g_2+2 i g_1 Y_H c_{35}^{(8)} g_2+\left(\frac{22 g_2^2}{3}+8 g_1^2 Y_H^2\right)
   c_{29}^{(8)}\ ,\\ 
   \dot{c}^{\prime (8)}_{30} & =-\frac{25}{6} i c_{31}^{(8)} g_2^2-\frac{1}{3} g_1 Y_H c_{36}^{(8)} g_2+\frac{1}{6} g_1 Y_H
   c_{37}^{(8)} g_2+\left(\frac{11 g_2^2}{6}+2 g_1^2 Y_H^2+12 \lambda \right) c_{30}^{(8)}\ ,\\ 
   \dot{c}^{\prime (8)}_{31} & =20 i c_{30}^{(8)}
   g_2^2+4 i g_1 Y_H c_{36}^{(8)} g_2-2 i g_1 Y_H c_{37}^{(8)} g_2+\left(\frac{22 g_2^2}{3}+8 g_1^2
   Y_H^2\right) c_{31}^{(8)}\ ,\\ 
   \dot{c}^{\prime (8)}_{32} & =\left(\frac{3 g_2^2}{2}+2 g_1^2 Y_H^2+12 \lambda \right) c_{32}^{(8)}\ ,\\ 
   \dot{c}^{\prime (8)}_{33} & =\left(\frac{3
   g_2^2}{2}+2 g_1^2 Y_H^2+12 \lambda \right) c_{33}^{(8)}\ ,\\ 
   \dot{c}^{\prime (8)}_{34} & =\frac{2}{3} g_1 g_2 Y_H c_{26}^{(8)}+\frac{2}{3}
   g_1 g_2 Y_H c_{28}^{(8)}+\frac{5}{3} i g_1 g_2 Y_H c_{29}^{(8)}+\left(-\frac{g_2^2}{3}+\frac{8 g_1^2
   Y_H^2}{3}+4 \lambda \right) c_{34}^{(8)}\\
   &+\left(-\frac{5 g_2^2}{12}+\frac{11 g_1^2 Y_H^2}{3}-2 \lambda \right)
   c_{35}^{(8)}\ ,\\ 
   \dot{c}^{\prime (8)}_{35} & =\left(10 g_1^2 Y_H^2-\frac{7 g_2^2}{6}\right) c_{35}^{(8)}\ ,\\
\end{align*}

\begin{align*} 
    \dot{c}^{\prime (8)}_{36} & =\frac{2}{3} g_1 g_2 Y_H
    c_{27}^{(8)}+\frac{2}{3} g_1 g_2 Y_H c_{30}^{(8)}+\frac{5}{3} i g_1 g_2 Y_H
    c_{31}^{(8)}+\left(-\frac{g_2^2}{3}+\frac{8 g_1^2 Y_H^2}{3}+4 \lambda \right) c_{36}^{(8)}\\
    &+\left(-\frac{5
    g_2^2}{12}+\frac{11 g_1^2 Y_H^2}{3}-2 \lambda \right) c_{37}^{(8)}\ ,\\
    \dot{c}^{\prime (8)}_{37} & =\left(10 g_1^2 Y_H^2-\frac{7 g_2^2}{6}\right)
    c_{37}^{(8)}\ ,\\
   \dot{c}^{\prime (8)}_{38} & =4 g_1 g_2 Y_H c_{20}^{(8)}+4 g_1 g_2 Y_H c_{21}^{(8)}+\left(16 g_1^2 Y_H^2-2
   g_2^2\right) c_{38}^{(8)}+\frac{1}{3} g_1 g_2 Y_H c_{44}^{(8)}+\frac{1}{3} g_1 g_2 Y_H
   c_{45}^{(8)}\\
   &-\frac{1}{3} g_1 g_2 Y_H c_{46}^{(8)}-4 g_1 g_2 Y_Q c_{63}^{(8)}\ ,\\ 
   \dot{c}^{\prime (8)}_{39} & =4 g_1 g_2 Y_H
   c_{20}^{(8)}+4 g_1 g_2 Y_H c_{21}^{(8)}+\left(16 g_1^2 Y_H^2-2 g_2^2\right) c_{39}^{(8)}+\frac{1}{3}
   g_1 g_2 Y_H c_{44}^{(8)}+\frac{1}{3} g_1 g_2 Y_H c_{45}^{(8)}\\
   &-\frac{1}{3} g_1 g_2 Y_H
   c_{46}^{(8)}-4 g_1 g_2 Y_Q c_{63}^{(8)}\ ,\\ 
   \dot{c}^{\prime (8)}_{40} & =\left(\frac{73 g_2^2}{4}+17 g_1^2 Y_H^2+10 \lambda \right)
   c_{40}^{(8)}+\left(\frac{17 g_2^2}{12}+\frac{g_1^2 Y_H^2}{3}-2 \lambda \right) c_{41}^{(8)}\\
   &+\left(-4 g_2^2+24 \mathcal{Y}_1
   \bar{\mathcal{Y}}_1+24 \mathcal{Y}_2 \bar{\mathcal{Y}}_2\right) c_{51}^{(8)}+\left(2 g_2^2-12 \mathcal{Y}_1 \bar{\mathcal{Y}}_1-12 \mathcal{Y}_2 \bar{\mathcal{Y}}_2\right) c_{52}^{(8)}\\[.2em]
   &+\left(8
   Y_H Y_Q g_1^2+2 g_2^2-24 \mathcal{Y}_2 \bar{\mathcal{Y}}_2\right) c_{53}^{(8)}+\left(-4 Y_H Y_Q g_1^2-g_2^2+12 \mathcal{Y}_2
   \bar{\mathcal{Y}}_2\right) c_{54}^{(8)}\\[.2em]
   &+\left(16 Y_H Y_Q g_1^2+24 \mathcal{Y}_1 \bar{\mathcal{Y}}_1-24 \mathcal{Y}_2 \bar{\mathcal{Y}}_2\right) c_{55}^{(8)}+\left(-8
   Y_H Y_Q g_1^2-12 \mathcal{Y}_1 \bar{\mathcal{Y}}_1+12 \mathcal{Y}_2 \bar{\mathcal{Y}}_2\right) c_{56}^{(8)}\ ,\\ 
   \dot{c}^{\prime (8)}_{41} & =\left(\frac{17 g_2^2}{12}+\frac{g_1^2
   Y_H^2}{3}-2 \lambda \right) c_{40}^{(8)}+\left(\frac{73 g_2^2}{4}+17 g_1^2 Y_H^2+10 \lambda \right)
   c_{41}^{(8)}\\
   &+\left(4 g_2^2-24 \mathcal{Y}_1 \bar{\mathcal{Y}}_1-24 \mathcal{Y}_2 \bar{\mathcal{Y}}_2\right) c_{51}^{(8)}+\left(6 g_2^2-36 \mathcal{Y}_1 \bar{\mathcal{Y}}_1-36
   \mathcal{Y}_2 \bar{\mathcal{Y}}_2\right) c_{52}^{(8)}\\[.2em]
   &+\left(-8 Y_H Y_Q g_1^2+6 g_2^2-48 \mathcal{Y}_1 \bar{\mathcal{Y}}_1-24 \mathcal{Y}_2 \bar{\mathcal{Y}}_2\right)
   c_{53}^{(8)}+\left(-12 Y_H Y_Q g_1^2+9 g_2^2-72 \mathcal{Y}_1 \bar{\mathcal{Y}}_1-36 \mathcal{Y}_2 \bar{\mathcal{Y}}_2\right) c_{54}^{(8)}\\[.2em]
   &+\left(-16
   Y_H Y_Q g_1^2-24 \mathcal{Y}_1 \bar{\mathcal{Y}}_1+24 \mathcal{Y}_2 \bar{\mathcal{Y}}_2\right) c_{55}^{(8)}+\left(-24 Y_H Y_Q g_1^2-36 \mathcal{Y}_1
   \bar{\mathcal{Y}}_1+36 \mathcal{Y}_2 \bar{\mathcal{Y}}_2\right) c_{56}^{(8)}\ ,\\ 
   \dot{c}^{\prime (8)}_{42} & =\left(\frac{73 g_2^2}{4}+17 g_1^2 Y_H^2+10 \lambda \right)
   c_{42}^{(8)}+\left(\frac{17 g_2^2}{12}+\frac{g_1^2 Y_H^2}{3}-2 \lambda \right) c_{43}^{(8)}\\
   & +\left(4 g_2^2-24 \mathcal{Y}_1
   \bar{\mathcal{Y}}_1-24 \mathcal{Y}_2 \bar{\mathcal{Y}}_2\right) c_{57}^{(8)}+\left(2 g_2^2-12 \mathcal{Y}_1 \bar{\mathcal{Y}}_1-12 \mathcal{Y}_2 \bar{\mathcal{Y}}_2\right) c_{58}^{(8)}\\[.2em]
   &+\left(-8
   Y_H Y_Q g_1^2-2 g_2^2+24 \mathcal{Y}_2 \bar{\mathcal{Y}}_2\right) c_{59}^{(8)}+\left(-4 Y_H Y_Q g_1^2-g_2^2+12 \mathcal{Y}_2
   \bar{\mathcal{Y}}_2\right) c_{60}^{(8)}\\[.2em]
   &+\left(-16 Y_H Y_Q g_1^2-24 \mathcal{Y}_1 \bar{\mathcal{Y}}_1+24 \mathcal{Y}_2 \bar{\mathcal{Y}}_2\right) c_{61}^{(8)}+\left(-8
   Y_H Y_Q g_1^2-12 \mathcal{Y}_1 \bar{\mathcal{Y}}_1+12 \mathcal{Y}_2 \bar{\mathcal{Y}}_2\right) c_{62}^{(8)}\ ,\\ 
   \dot{c}^{\prime (8)}_{43} & =\left(\frac{17 g_2^2}{12}+\frac{g_1^2
   Y_H^2}{3}-2 \lambda \right) c_{42}^{(8)}+\left(\frac{73 g_2^2}{4}+17 g_1^2 Y_H^2+10 \lambda \right)
   c_{43}^{(8)}\\
   &+\left(-4 g_2^2+24 \mathcal{Y}_1 \bar{\mathcal{Y}}_1+24 \mathcal{Y}_2 \bar{\mathcal{Y}}_2\right) c_{57}^{(8)}+\left(6 g_2^2-36 \mathcal{Y}_1 \bar{\mathcal{Y}}_1-36
   \mathcal{Y}_2 \bar{\mathcal{Y}}_2\right) c_{58}^{(8)}\\[.2em]
   &+\left(8 Y_H Y_Q g_1^2-6 g_2^2+48 \mathcal{Y}_1 \bar{\mathcal{Y}}_1+24 \mathcal{Y}_2 \bar{\mathcal{Y}}_2\right)
   c_{59}^{(8)}+\left(-12 Y_H Y_Q g_1^2+9 g_2^2-72 \mathcal{Y}_1 \bar{\mathcal{Y}}_1-36 \mathcal{Y}_2 \bar{\mathcal{Y}}_2\right) c_{60}^{(8)}\\[.2em]
   &+\left(16
   Y_H Y_Q g_1^2+24 \mathcal{Y}_1 \bar{\mathcal{Y}}_1-24 \mathcal{Y}_2 \bar{\mathcal{Y}}_2\right) c_{61}^{(8)}+\left(-24 Y_H Y_Q g_1^2-36 \mathcal{Y}_1
   \bar{\mathcal{Y}}_1+36 \mathcal{Y}_2 \bar{\mathcal{Y}}_2\right) c_{62}^{(8)}\ ,\\ 
   \dot{c}^{\prime (8)}_{44} & =12 c_{21}^{(8)} g_2^2-\frac{8}{3} g_1 Y_H c_{38}^{(8)}
   g_2-\frac{8}{3} g_1 Y_H c_{39}^{(8)} g_2+16 g_1^2 Y_H^2 c_{20}^{(8)}+\left(26 g_2^2+\frac{100 g_1^2
   Y_H^2}{3}+\frac{32 \lambda }{3}\right) c_{44}^{(8)}\\
   &+\left(5 g_2^2+\frac{16 \lambda }{3}\right) c_{45}^{(8)}+\left(-\frac{14
   g_2^2}{3}-\frac{28 g_1^2 Y_H^2}{3}-\frac{8 \lambda }{3}\right) c_{46}^{(8)}+\left(2 g_2^2+8 \mathcal{Y}_1 \bar{\mathcal{Y}}_1-8 \mathcal{Y}_2
   \bar{\mathcal{Y}}_2\right) c_{63}^{(8)}\\
   &+\left(-4 g_2^2+24 \mathcal{Y}_1 \bar{\mathcal{Y}}_1+24 \mathcal{Y}_2 \bar{\mathcal{Y}}_2\right) c_{64}^{(8)}+(24 \mathcal{Y}_1 \bar{\mathcal{Y}}_1+24
   \mathcal{Y}_2 \bar{\mathcal{Y}}_2) c_{65}^{(8)}\ ,\\
\end{align*}

\begin{align*} 
    \dot{c}^{\prime (8)}_{45} & =2 c_{21}^{(8)} g_2^2+\frac{2}{3} g_1 Y_H c_{38}^{(8)} g_2+\frac{2}{3} g_1
    Y_H c_{39}^{(8)} g_2+\frac{8}{3} g_1^2 Y_H^2 c_{20}^{(8)}+\left(-\frac{17 g_2^2}{6}-\frac{22 g_1^2
    Y_H^2}{3}+\frac{40 \lambda }{3}\right) c_{44}^{(8)}\\
    &+\left(\frac{g_2^2}{2}-6 g_1^2 Y_H^2+\frac{88 \lambda }{3}\right)
    c_{45}^{(8)}+\left(\frac{11 g_2^2}{3}+4 g_1^2 Y_H^2-\frac{28 \lambda }{3}\right) c_{46}^{(8)}\\
    &+\left(-4 Y_H Y_Q
    g_1^2-g_2^2+16 \mathcal{Y}_2 \bar{\mathcal{Y}}_2\right) c_{63}^{(8)}+\left(8 Y_H Y_Q g_1^2+2 g_2^2-24 \mathcal{Y}_2 \bar{\mathcal{Y}}_2\right)
    c_{64}^{(8)}\\[.2em]
    &+\left(-8 Y_H Y_Q g_1^2-8 \mathcal{Y}_1 \bar{\mathcal{Y}}_1+16 \mathcal{Y}_2 \bar{\mathcal{Y}}_2\right) c_{65}^{(8)}+\left(16 Y_H Y_Q g_1^2+24 \mathcal{Y}_1 \bar{\mathcal{Y}}_1-24 \mathcal{Y}_2 \bar{\mathcal{Y}}_2\right) c_{66}^{(8)}\ ,\\[.2em]
    \dot{c}^{\prime (8)}_{46} & =12 c_{21}^{(8)} g_2^2-\frac{28}{3} g_1 Y_H
    c_{38}^{(8)} g_2-\frac{28}{3} g_1 Y_H c_{39}^{(8)} g_2+16 g_1^2 Y_H^2 c_{20}^{(8)}+\left(\frac{41
    g_2^2}{3}-\frac{28 g_1^2 Y_H^2}{3}+\frac{16 \lambda }{3}\right) c_{44}^{(8)}\\
    &+\left(5 g_2^2-20 g_1^2 Y_H^2+\frac{16
    \lambda }{3}\right) c_{45}^{(8)}+\left(\frac{23 g_2^2}{3}+20 g_1^2 Y_H^2+\frac{8 \lambda }{3}\right)
    c_{46}^{(8)}\\
    &+\left(-8 Y_H Y_Q g_1^2+2 g_2^2+16 \mathcal{Y}_1 \bar{\mathcal{Y}}_1-16 \mathcal{Y}_2 \bar{\mathcal{Y}}_2\right) c_{63}^{(8)}+\left(16
    Y_H Y_Q g_1^2-4 g_2^2+48 \mathcal{Y}_1 \bar{\mathcal{Y}}_1\right) c_{64}^{(8)}\\[.2em]
    &+\left(-16 Y_H Y_Q g_1^2+48 \mathcal{Y}_2
    \bar{\mathcal{Y}}_2\right) c_{65}^{(8)}+\left(32 Y_H Y_Q g_1^2+48 \mathcal{Y}_1 \bar{\mathcal{Y}}_1-48 \mathcal{Y}_2 \bar{\mathcal{Y}}_2\right)
    c_{66}^{(8)}\ ,\\ 
    \dot{c}^{\prime (8)}_{47} & =\left(\frac{2 g_2^2}{27}-\frac{4 \mathcal{Y}_1 \bar{\mathcal{Y}}_1}{9}-\frac{4 \mathcal{Y}_2 \bar{\mathcal{Y}}_2}{9}\right)
    c_{18}^{(8)}+\left(-\frac{g_2^2}{27}+\frac{2 \mathcal{Y}_1 \bar{\mathcal{Y}}_1}{9}+\frac{2 \mathcal{Y}_2 \bar{\mathcal{Y}}_2}{9}\right) c_{19}^{(8)}\\
    &+\left(\frac{47
    g_2^2}{3}+8 g_3^2+\frac{14 g_1^2 Y_H^2}{3}+6 g_1^2 Y_Q^2+8 \mathcal{Y}_1 \bar{\mathcal{Y}}_1+8 \mathcal{Y}_2 \bar{\mathcal{Y}}_2+20 \lambda \right)
    c_{47}^{(8)}\\
    &+\left(-\frac{17 g_2^2}{12}+\frac{19 g_1^2 Y_H^2}{3}-6 \mathcal{Y}_1 \bar{\mathcal{Y}}_1+6 \mathcal{Y}_2 \bar{\mathcal{Y}}_2+2 \lambda \right)
    c_{48}^{(8)}\\
    &+\left(-\frac{17 g_2^2}{12}-7 g_1^2 Y_H^2+6 \mathcal{Y}_1 \bar{\mathcal{Y}}_1-6 \mathcal{Y}_2 \bar{\mathcal{Y}}_2+2 \lambda \right)
    c_{49}^{(8)}+(12 \mathcal{Y}_2 \bar{\mathcal{Y}}_2-12 \mathcal{Y}_1 \bar{\mathcal{Y}}_1) c_{50}^{(8)}\ ,\\
    \dot{c}^{\prime (8)}_{48} & =\left(\frac{4 g_2^2}{27}-\frac{8 \mathcal{Y}_1 \bar{\mathcal{Y}}_1}{9}-\frac{8
    \mathcal{Y}_2 \bar{\mathcal{Y}}_2}{9}\right) c_{18}^{(8)}+\left(-\frac{g_2^2}{27}+\frac{2 \mathcal{Y}_1 \bar{\mathcal{Y}}_1}{9}+\frac{2 \mathcal{Y}_2 \bar{\mathcal{Y}}_2}{9}\right)
    c_{19}^{(8)}\\
    &+\left(\frac{103
    g_2^2}{12}+8 g_3^2+\frac{49 g_1^2 Y_H^2}{3}+6 g_1^2 Y_Q^2-4 \mathcal{Y}_1 \bar{\mathcal{Y}}_1+20 \mathcal{Y}_2 \bar{\mathcal{Y}}_2+30 \lambda
    \right) c_{48}^{(8)}\\
    &+\left(\frac{17 g_2^2}{6}+\frac{2 g_1^2 Y_H^2}{3}-4 \lambda \right) c_{47}^{(8)}+\left(\frac{17 g_2^2}{12}-13 g_1^2 Y_H^2+12 \mathcal{Y}_1 \bar{\mathcal{Y}}_1-12 \mathcal{Y}_2 \bar{\mathcal{Y}}_2-2 \lambda \right)
    c_{49}^{(8)}\\
    &+(24 \mathcal{Y}_2 \bar{\mathcal{Y}}_2-24 \mathcal{Y}_1 \bar{\mathcal{Y}}_1) c_{50}^{(8)}\ ,\\ 
    \dot{c}^{\prime (8)}_{49} & =\left(\frac{g_2^2}{27}-\frac{2 \mathcal{Y}_1 \bar{\mathcal{Y}}_1}{9}-\frac{2
    \mathcal{Y}_2 \bar{\mathcal{Y}}_2}{9}\right) c_{19}^{(8)}+\left(-\frac{17 g_2^2}{6}-\frac{2 g_1^2 Y_H^2}{3}+4 \lambda \right)
    c_{47}^{(8)}\\
    &+\left(\frac{205
    g_2^2}{12}+8 g_3^2+5 g_1^2 Y_H^2+6 g_1^2 Y_Q^2+8 \mathcal{Y}_1 \bar{\mathcal{Y}}_1+8 \mathcal{Y}_2 \bar{\mathcal{Y}}_2+18 \lambda \right)
    c_{49}^{(8)}\\
    &+\left(-\frac{17 g_2^2}{12}-\frac{g_1^2 Y_H^2}{3}+2 \lambda \right) c_{48}^{(8)}\ ,
\end{align*}

\begin{align*}
    \dot{c}^{\prime (8)}_{50} & =\left(\frac{16}{27} Y_H Y_Q g_1^2-\frac{2 g_2^2}{27}+\frac{4 \mathcal{Y}_1 \bar{\mathcal{Y}}_1}{3}-\frac{4 \mathcal{Y}_2
    \bar{\mathcal{Y}}_2}{9}\right) c_{18}^{(8)}\\
    &+\left(-\frac{4}{9} Y_H Y_Q g_1^2+\frac{g_2^2}{27}-\frac{8 \mathcal{Y}_1 \bar{\mathcal{Y}}_1}{9}+\frac{4 \mathcal{Y}_2
    \bar{\mathcal{Y}}_2}{9}\right) c_{19}^{(8)}+\left(-\frac{17 g_2^2}{6}-\frac{2 g_1^2 Y_H^2}{3}+4 \lambda \right)
    c_{47}^{(8)}\\
    &+\left(\frac{11 g_2^2}{6}+\frac{2 g_1^2 Y_H^2}{3}+4 g_1^2 Y_Q^2-4 \mathcal{Y}_1 \bar{\mathcal{Y}}_1+8 \mathcal{Y}_2 \bar{\mathcal{Y}}_2\right)
    c_{48}^{(8)}\\
    &+\left(g_2^2-4 g_1^2 Y_Q^2+4 \mathcal{Y}_1 \bar{\mathcal{Y}}_1-8 \mathcal{Y}_2 \bar{\mathcal{Y}}_2-4 \lambda \right)
    c_{49}^{(8)}\\
    &+\left(\frac{41 g_2^2}{3}+8 g_3^2+18 g_1^2 Y_H^2+14 g_1^2 Y_Q^2+24 \mathcal{Y}_1 \bar{\mathcal{Y}}_1+28 \lambda
    \right) c_{50}^{(8)}\ ,\\
   \dot{c}^{\prime (8)}_{51} & =\left(-\frac{g_2^2}{6}+\mathcal{Y}_1 \bar{\mathcal{Y}}_1+\mathcal{Y}_2 \bar{\mathcal{Y}}_2\right) c_{40}^{(8)}\\
   &+\left(2 Y_H^2 g_1^2+6
   Y_Q^2 g_1^2+4 Y_H Y_Q g_1^2+14 g_2^2+8 g_3^2+6 \mathcal{Y}_1 \bar{\mathcal{Y}}_1+2 \mathcal{Y}_2 \bar{\mathcal{Y}}_2+12 \lambda \right)
   c_{51}^{(8)}\\
   &+\left(-\frac{11}{3} Y_H^2 g_1^2-2 Y_H Y_Q g_1^2-\frac{11 g_2^2}{4}-2 \mathcal{Y}_2 \bar{\mathcal{Y}}_2+6 \lambda
   \right) c_{52}^{(8)}\\
   &+\left(-\frac{g_2^2}{6}-2 \mathcal{Y}_1 \bar{\mathcal{Y}}_1+4 \mathcal{Y}_2 \bar{\mathcal{Y}}_2+4 \lambda \right) c_{53}^{(8)}+\left(-\frac{17
   g_2^2}{12}-\frac{g_1^2 Y_H^2}{3}+2 \lambda \right) c_{54}^{(8)}\\
   &+\left(\frac{16 g_2^2}{3}+4 \mathcal{Y}_2 \bar{\mathcal{Y}}_2\right)
   c_{55}^{(8)}\ ,\\ 
   \dot{c}^{\prime (8)}_{52} & =\left(\frac{g_2^2}{6}-\mathcal{Y}_1 \bar{\mathcal{Y}}_1-\mathcal{Y}_2 \bar{\mathcal{Y}}_2\right) c_{40}^{(8)}+\left(-8 Y_H Y_Q g_1^2-2
   g_2^2-\frac{8 \mathcal{Y}_2 \bar{\mathcal{Y}}_2}{3}\right) c_{51}^{(8)}\\
   &+\left(\frac{28 Y_H^2 g_1^2}{3}+6 Y_Q^2 g_1^2-4 Y_H Y_Q
   g_1^2+\frac{33 g_2^2}{2}+8 g_3^2+6 \mathcal{Y}_1 \bar{\mathcal{Y}}_1+2 \mathcal{Y}_2 \bar{\mathcal{Y}}_2\right) c_{52}^{(8)}\\
   &+\left(\frac{8 g_2^2}{3}+\frac{2
   g_1^2 Y_H^2}{3}-2 \mathcal{Y}_1 \bar{\mathcal{Y}}_1+4 \mathcal{Y}_2 \bar{\mathcal{Y}}_2\right) c_{54}^{(8)} +\left(-2 g_2^2-4 \mathcal{Y}_1 \bar{\mathcal{Y}}_1+\frac{4 \mathcal{Y}_2
   \bar{\mathcal{Y}}_2}{3}\right) c_{55}^{(8)}\\
   &+\left(\frac{7 g_2^2}{3}-6 \mathcal{Y}_1 \bar{\mathcal{Y}}_1+6 \mathcal{Y}_2 \bar{\mathcal{Y}}_2\right)
   c_{56}^{(8)}\ ,\\ 
   \dot{c}^{\prime (8)}_{53} & =\left(\frac{g_2^2}{6}-\mathcal{Y}_1 \bar{\mathcal{Y}}_1-\mathcal{Y}_2 \bar{\mathcal{Y}}_2\right) c_{40}^{(8)}+\left(-2 g_2^2-\frac{20 \mathcal{Y}_1
   \bar{\mathcal{Y}}_1}{3}+4 \mathcal{Y}_2 \bar{\mathcal{Y}}_2\right) c_{51}^{(8)}\\
   &+\left(2 Y_H^2 g_1^2+6 Y_Q^2 g_1^2+4 Y_H Y_Q g_1^2+\frac{37 g_2^2}{3}+8 g_3^2-\frac{8
   \mathcal{Y}_1 \bar{\mathcal{Y}}_1}{3}+4 \mathcal{Y}_2 \bar{\mathcal{Y}}_2+4 \lambda \right) c_{53}^{(8)}\\
   &+\left(g_2^2-\frac{2 \mathcal{Y}_1 \bar{\mathcal{Y}}_1}{3}+2 \mathcal{Y}_2 \bar{\mathcal{Y}}_2\right)
   c_{52}^{(8)}+\left(-3 Y_H^2 g_1^2-2 Y_H Y_Q
   g_1^2+\frac{13 g_2^2}{12}-\frac{2 \mathcal{Y}_1 \bar{\mathcal{Y}}_1}{3}+2 \lambda \right) c_{54}^{(8)}\\
   &+\left(-\frac{26 g_2^2}{3}-\frac{16
   \mathcal{Y}_1 \bar{\mathcal{Y}}_1}{3}\right) c_{55}^{(8)}+\left(-g_2^2+\frac{2 \mathcal{Y}_1 \bar{\mathcal{Y}}_1}{3}-2 \mathcal{Y}_2 \bar{\mathcal{Y}}_2\right)
   c_{56}^{(8)}\ ,\\
\end{align*}

\begin{align*}   
   \dot{c}^{\prime (8)}_{54} & =\left(-\frac{g_2^2}{6}+\mathcal{Y}_1 \bar{\mathcal{Y}}_1+\mathcal{Y}_2 \bar{\mathcal{Y}}_2\right) c_{40}^{(8)}+\left(\frac{g_2^2}{6}-\mathcal{Y}_1
   \bar{\mathcal{Y}}_1-\mathcal{Y}_2 \bar{\mathcal{Y}}_2\right) c_{41}^{(8)}\\
   &+\left(4 g_2^2+\frac{8 \mathcal{Y}_1 \bar{\mathcal{Y}}_1}{3}+\frac{8 \mathcal{Y}_2 \bar{\mathcal{Y}}_2}{3}\right)
   c_{51}^{(8)}+\left(2 g_2^2-\frac{4 \mathcal{Y}_1 \bar{\mathcal{Y}}_1}{3}+4 \mathcal{Y}_2 \bar{\mathcal{Y}}_2\right) c_{52}^{(8)}\\
   &+\left(8 Y_H^2 g_1^2+6 Y_Q^2 g_1^2-4 Y_H Y_Q
   g_1^2+\frac{79 g_2^2}{6}+8 g_3^2+\frac{8 \mathcal{Y}_1 \bar{\mathcal{Y}}_1}{3}+4 \mathcal{Y}_2 \bar{\mathcal{Y}}_2\right) c_{54}^{(8)}\\
   &+\left(-8 Y_H Y_Q g_1^2+2
   g_2^2+\frac{8 \mathcal{Y}_1 \bar{\mathcal{Y}}_1}{3}\right) c_{53}^{(8)}+\left(\frac{16 \mathcal{Y}_1
   \bar{\mathcal{Y}}_1}{3}-\frac{16 \mathcal{Y}_2 \bar{\mathcal{Y}}_2}{3}\right) c_{55}^{(8)}\\
   &+\left(-\frac{20 g_2^2}{3}+\frac{4 \mathcal{Y}_1 \bar{\mathcal{Y}}_1}{3}-4 \mathcal{Y}_2
   \bar{\mathcal{Y}}_2\right) c_{56}^{(8)}\ ,\\ 
   \dot{c}^{\prime (8)}_{55} & =\left(\frac{1}{3} Y_H Y_Q g_1^2-\frac{g_2^2}{12}+\mathcal{Y}_1 \bar{\mathcal{Y}}_1\right)
   c_{40}^{(8)}+\left(2 g_2^2+\frac{2 \mathcal{Y}_1 \bar{\mathcal{Y}}_1}{3}+2 \mathcal{Y}_2 \bar{\mathcal{Y}}_2\right) c_{51}^{(8)}\\
   &+\left(-g_2^2+\frac{2 \mathcal{Y}_1
   \bar{\mathcal{Y}}_1}{3}-2 \mathcal{Y}_2 \bar{\mathcal{Y}}_2\right) c_{52}^{(8)}+\left(-\frac{11 g_2^2}{6}+4 g_1^2 Y_Q^2-\frac{4 \mathcal{Y}_1 \bar{\mathcal{Y}}_1}{3}+2
   \mathcal{Y}_2 \bar{\mathcal{Y}}_2\right) c_{53}^{(8)}\\
   &+\left(2
   Y_H^2 g_1^2+14 Y_Q^2 g_1^2+4 Y_H Y_Q g_1^2+\frac{62 g_2^2}{3}+8 g_3^2+\frac{16 \mathcal{Y}_1 \bar{\mathcal{Y}}_1}{3}+4
   \mathcal{Y}_2 \bar{\mathcal{Y}}_2+4 \lambda \right) c_{55}^{(8)}\\
   & +\left(-g_2^2+\frac{2 \mathcal{Y}_1 \bar{\mathcal{Y}}_1}{3}-2 \mathcal{Y}_2 \bar{\mathcal{Y}}_2\right) c_{54}^{(8)}+\left(-3 Y_H^2 g_1^2-2 Y_H Y_Q g_1^2+\frac{13
   g_2^2}{12}-\frac{2 \mathcal{Y}_1 \bar{\mathcal{Y}}_1}{3}+2 \lambda \right) c_{56}^{(8)}\ ,\\ 
   \dot{c}^{\prime (8)}_{56} & =\left(-\frac{1}{3} Y_H Y_Q
   g_1^2+\frac{g_2^2}{12}-\mathcal{Y}_1 \bar{\mathcal{Y}}_1\right) c_{40}^{(8)}+\left(-\frac{1}{3} Y_H Y_Q g_1^2-\frac{g_2^2}{12}+\mathcal{Y}_2
   \bar{\mathcal{Y}}_2\right) c_{41}^{(8)}\\
   &+\left(-4 g_2^2-\frac{8 \mathcal{Y}_1 \bar{\mathcal{Y}}_1}{3}-\frac{8 \mathcal{Y}_2 \bar{\mathcal{Y}}_2}{3}\right) c_{51}^{(8)}+\left(-2
   g_2^2-\frac{14 \mathcal{Y}_1 \bar{\mathcal{Y}}_1}{3}+2 \mathcal{Y}_2 \bar{\mathcal{Y}}_2\right) c_{52}^{(8)}\\
   &+\left(-2 g_2^2+\frac{4 \mathcal{Y}_1 \bar{\mathcal{Y}}_1}{3}-4 \mathcal{Y}_2
   \bar{\mathcal{Y}}_2\right) c_{53}^{(8)}+\left(-\frac{17 g_2^2}{6}+4 g_1^2 Y_Q^2-\frac{2 \mathcal{Y}_1 \bar{\mathcal{Y}}_1}{3}\right) c_{54}^{(8)}\\
   &+\left(8 Y_H^2 g_1^2+14
   Y_Q^2 g_1^2-4 Y_H Y_Q g_1^2+\frac{37 g_2^2}{2}+8 g_3^2+\frac{14 \mathcal{Y}_1 \bar{\mathcal{Y}}_1}{3}+6 \mathcal{Y}_2 \bar{\mathcal{Y}}_2\right)
   c_{56}^{(8)}\\
   &+\left(-8
   Y_H Y_Q g_1^2-\frac{4 \mathcal{Y}_1 \bar{\mathcal{Y}}_1}{3}+\frac{4 \mathcal{Y}_2 \bar{\mathcal{Y}}_2}{3}\right) c_{55}^{(8)}\ ,\\
   \dot{c}^{\prime (8)}_{57} & =\left(2 Y_H^2 g_1^2+6 Y_Q^2
   g_1^2-4 Y_H Y_Q g_1^2+14 g_2^2+8 g_3^2+2 \mathcal{Y}_1 \bar{\mathcal{Y}}_1+6 \mathcal{Y}_2 \bar{\mathcal{Y}}_2+12 \lambda \right)
   c_{57}^{(8)}\\
   &+\left(\frac{g_2^2}{6}-\mathcal{Y}_1 \bar{\mathcal{Y}}_1-\mathcal{Y}_2 \bar{\mathcal{Y}}_2\right) c_{42}^{(8)}+\left(\frac{11 Y_H^2 g_1^2}{3}-2 Y_H Y_Q g_1^2+\frac{11 g_2^2}{4}+2 \mathcal{Y}_1 \bar{\mathcal{Y}}_1-6 \lambda
   \right) c_{58}^{(8)}\\
   &+\left(-\frac{11 g_2^2}{2}-2 \mathcal{Y}_2 \bar{\mathcal{Y}}_2+4 \lambda \right) c_{59}^{(8)}+\left(\frac{17
   g_2^2}{12}+\frac{g_1^2 Y_H^2}{3}-2 \lambda \right) c_{60}^{(8)}\\
   &+\left(-\frac{16 g_2^2}{3}-4 \mathcal{Y}_1 \bar{\mathcal{Y}}_1\right)
   c_{61}^{(8)}\ ,\\ 
   \dot{c}^{\prime (8)}_{58} & =\left(\frac{g_2^2}{6}-\mathcal{Y}_1 \bar{\mathcal{Y}}_1-\mathcal{Y}_2 \bar{\mathcal{Y}}_2\right) c_{42}^{(8)}+\left(-8 Y_H Y_Q g_1^2+2
   g_2^2+\frac{8 \mathcal{Y}_1 \bar{\mathcal{Y}}_1}{3}\right) c_{57}^{(8)}\\
   &+\left(\frac{28 Y_H^2 g_1^2}{3}+6 Y_Q^2 g_1^2+4 Y_H Y_Q
   g_1^2+\frac{33 g_2^2}{2}+8 g_3^2+2 \mathcal{Y}_1 \bar{\mathcal{Y}}_1+6 \mathcal{Y}_2 \bar{\mathcal{Y}}_2\right) c_{58}^{(8)}\\
   &+\left(-2 g_2^2+\frac{4 \mathcal{Y}_1
   \bar{\mathcal{Y}}_1}{3}-4 \mathcal{Y}_2 \bar{\mathcal{Y}}_2\right) c_{59}^{(8)}+\left(\frac{g_2^2}{3}+\frac{2 g_1^2 Y_H^2}{3}-2 \mathcal{Y}_1 \bar{\mathcal{Y}}_1+4 \mathcal{Y}_2
   \bar{\mathcal{Y}}_2\right) c_{60}^{(8)}\\
   &+\left(-2 g_2^2+\frac{4 \mathcal{Y}_1 \bar{\mathcal{Y}}_1}{3}-4 \mathcal{Y}_2 \bar{\mathcal{Y}}_2\right) c_{61}^{(8)}+\left(-\frac{7
   g_2^2}{3}-6 \mathcal{Y}_1 \bar{\mathcal{Y}}_1+6 \mathcal{Y}_2 \bar{\mathcal{Y}}_2\right) c_{62}^{(8)}\ ,\\
\end{align*}

\begin{align*} 
    \dot{c}^{\prime (8)}_{59} & =\left(-\frac{g_2^2}{6}+\mathcal{Y}_1 \bar{\mathcal{Y}}_1+\mathcal{Y}_2 \bar{\mathcal{Y}}_2\right)
    c_{42}^{(8)}+\left(-2 g_2^2+4 \mathcal{Y}_1 \bar{\mathcal{Y}}_1-\frac{20 \mathcal{Y}_2 \bar{\mathcal{Y}}_2}{3}\right) c_{57}^{(8)}\\
    &+\left(2 Y_H^2 g_1^2+6 Y_Q^2 g_1^2-4 Y_H Y_Q g_1^2+21
    g_2^2+8 g_3^2+4 \mathcal{Y}_1 \bar{\mathcal{Y}}_1+\frac{8 \mathcal{Y}_2 \bar{\mathcal{Y}}_2}{3}+4 \lambda \right) c_{59}^{(8)}\\
    &+\left(3 Y_H^2 g_1^2-2
    Y_H Y_Q g_1^2-\frac{25 g_2^2}{12}-2 \mathcal{Y}_1 \bar{\mathcal{Y}}_1+\frac{4 \mathcal{Y}_2 \bar{\mathcal{Y}}_2}{3}-2 \lambda \right)
    c_{60}^{(8)}\\
    &+\left(-g_2^2-2 \mathcal{Y}_1
    \bar{\mathcal{Y}}_1+\frac{2 \mathcal{Y}_2 \bar{\mathcal{Y}}_2}{3}\right) c_{58}^{(8)}+\left(\frac{26 g_2^2}{3}+\frac{16 \mathcal{Y}_2 \bar{\mathcal{Y}}_2}{3}\right) c_{61}^{(8)}\\
    &+\left(-g_2^2-2 \mathcal{Y}_1 \bar{\mathcal{Y}}_1+\frac{2
    \mathcal{Y}_2 \bar{\mathcal{Y}}_2}{3}\right) c_{62}^{(8)}\ ,\\ 
    \dot{c}^{\prime (8)}_{60} & =\left(-\frac{g_2^2}{6}+\mathcal{Y}_1 \bar{\mathcal{Y}}_1+\mathcal{Y}_2 \bar{\mathcal{Y}}_2\right)
    c_{42}^{(8)}+\left(\frac{g_2^2}{6}-\mathcal{Y}_1 \bar{\mathcal{Y}}_1-\mathcal{Y}_2 \bar{\mathcal{Y}}_2\right) c_{43}^{(8)}\\
    &+\left(-4 g_2^2-\frac{8 \mathcal{Y}_1
    \bar{\mathcal{Y}}_1}{3}-\frac{8 \mathcal{Y}_2 \bar{\mathcal{Y}}_2}{3}\right) c_{57}^{(8)}+\left(2 g_2^2+4 \mathcal{Y}_1 \bar{\mathcal{Y}}_1-\frac{4 \mathcal{Y}_2 \bar{\mathcal{Y}}_2}{3}\right)
    c_{58}^{(8)}\\
    &+\left(-8 Y_H Y_Q g_1^2-2 g_2^2-\frac{16 \mathcal{Y}_1 \bar{\mathcal{Y}}_1}{3}+\frac{8 \mathcal{Y}_2 \bar{\mathcal{Y}}_2}{3}\right)
    c_{59}^{(8)}\\
    &+\left(8 Y_H^2 g_1^2+6 Y_Q^2 g_1^2+4 Y_H Y_Q g_1^2+\frac{119 g_2^2}{6}+8 g_3^2+8 \mathcal{Y}_1
    \bar{\mathcal{Y}}_1+\frac{4 \mathcal{Y}_2 \bar{\mathcal{Y}}_2}{3}\right) c_{60}^{(8)}\\
    &+\left(\frac{16 \mathcal{Y}_2 \bar{\mathcal{Y}}_2}{3}-\frac{16 \mathcal{Y}_1 \bar{\mathcal{Y}}_1}{3}\right)
    c_{61}^{(8)}+\left(\frac{20 g_2^2}{3}+4 \mathcal{Y}_1 \bar{\mathcal{Y}}_1-\frac{4 \mathcal{Y}_2 \bar{\mathcal{Y}}_2}{3}\right) c_{62}^{(8)}\ ,\\ 
    \dot{c}^{\prime (8)}_{61} & =\left(-\frac{1}{3} Y_H
    Y_Q g_1^2+\frac{g_2^2}{12}-\mathcal{Y}_1 \bar{\mathcal{Y}}_1\right) c_{42}^{(8)}+(6 \mathcal{Y}_2 \bar{\mathcal{Y}}_2-6 \mathcal{Y}_1 \bar{\mathcal{Y}}_1)
    c_{57}^{(8)}\\
    &+\left(\frac{3 g_2^2}{2}+4 g_1^2 Y_Q^2-2 \mathcal{Y}_1 \bar{\mathcal{Y}}_1+4 \mathcal{Y}_2 \bar{\mathcal{Y}}_2\right) c_{59}^{(8)}+\left(3 Y_H^2 g_1^2-2 Y_H Y_Q g_1^2-\frac{g_2^2}{12}+2 \mathcal{Y}_1 \bar{\mathcal{Y}}_1-2 \lambda \right)
    c_{62}^{(8)}\\
    &+\left(2
    Y_H^2 g_1^2+14 Y_Q^2 g_1^2-4 Y_H Y_Q g_1^2+12 g_2^2+8 g_3^2+4 \mathcal{Y}_1 \bar{\mathcal{Y}}_1+4 \lambda \right)
    c_{61}^{(8)}\ ,\\ 
    \dot{c}^{\prime (8)}_{62} & =\left(-\frac{1}{3} Y_H Y_Q g_1^2+\frac{g_2^2}{12}-\mathcal{Y}_1 \bar{\mathcal{Y}}_1\right) c_{42}^{(8)}+\left(-\frac{1}{3}
    Y_H Y_Q g_1^2-\frac{g_2^2}{12}+\mathcal{Y}_2 \bar{\mathcal{Y}}_2\right) c_{43}^{(8)}\\
    &+(6 \mathcal{Y}_2 \bar{\mathcal{Y}}_2-6 \mathcal{Y}_1 \bar{\mathcal{Y}}_1)
    c_{58}^{(8)}+\left(\frac{3 g_2^2}{2}+4 g_1^2 Y_Q^2-2 \mathcal{Y}_1 \bar{\mathcal{Y}}_1+4 \mathcal{Y}_2 \bar{\mathcal{Y}}_2\right) c_{60}^{(8)}\\
    &+\left(8 Y_H^2 g_1^2+14 Y_Q^2 g_1^2+4 Y_H
    Y_Q g_1^2+\frac{71 g_2^2}{6}+8 g_3^2+2 \mathcal{Y}_1 \bar{\mathcal{Y}}_1+6 \mathcal{Y}_2 \bar{\mathcal{Y}}_2\right) c_{62}^{(8)}\\
    &+\left(-8 Y_H
    Y_Q g_1^2+4 \mathcal{Y}_1 \bar{\mathcal{Y}}_1-4 \mathcal{Y}_2 \bar{\mathcal{Y}}_2\right) c_{61}^{(8)}\ ,\\ 
    \dot{c}^{\prime (8)}_{63} & =\frac{4}{3} i
    c_{22}^{(8)} g_2^2-\frac{8}{3} g_1 Y_Q c_{38}^{(8)} g_2-\frac{8}{3} g_1 Y_Q c_{39}^{(8)}
    g_2+\left(\frac{2 \mathcal{Y}_2 \bar{\mathcal{Y}}_2}{3}-\frac{2 \mathcal{Y}_1 \bar{\mathcal{Y}}_1}{3}\right) c_{44}^{(8)}\\
    &+\left(\frac{2 \mathcal{Y}_2 \bar{\mathcal{Y}}_2}{3}-\frac{2 \mathcal{Y}_1
    \bar{\mathcal{Y}}_1}{3}\right) c_{45}^{(8)}+\left(\frac{2 \mathcal{Y}_1 \bar{\mathcal{Y}}_1}{3}-\frac{2 \mathcal{Y}_2 \bar{\mathcal{Y}}_2}{3}\right) c_{46}^{(8)}\\
    &+\left(12 Y_H^2
    g_1^2+\frac{34 Y_Q^2 g_1^2}{3}-4 Y_H Y_Q g_1^2+\frac{19 g_2^2}{6}+\frac{136 g_3^2}{9}-\frac{16 \mathcal{Y}_1
    \bar{\mathcal{Y}}_1}{3}+8 \mathcal{Y}_2 \bar{\mathcal{Y}}_2\right) c_{63}^{(8)}\\
    &+\left(8 Y_H Y_Q g_1^2+2 g_2^2+\frac{16 \mathcal{Y}_1 \bar{\mathcal{Y}}_1}{3}-\frac{8
    \mathcal{Y}_2 \bar{\mathcal{Y}}_2}{3}\right) c_{64}^{(8)}+\left(-2 g_2^2-\frac{32 \mathcal{Y}_1 \bar{\mathcal{Y}}_1}{3}+8 \mathcal{Y}_2 \bar{\mathcal{Y}}_2\right) c_{65}^{(8)}\\
    &+\left(4
    g_2^2+\frac{8 \mathcal{Y}_1 \bar{\mathcal{Y}}_1}{3}+\frac{8 \mathcal{Y}_2 \bar{\mathcal{Y}}_2}{3}\right) c_{66}^{(8)}\ ,
\end{align*}

\begin{align*} 
    \dot{c}^{\prime (8)}_{64} & =\frac{2}{3} i c_{22}^{(8)} g_2^2-\frac{4}{3}
    g_1 Y_Q c_{38}^{(8)} g_2-\frac{4}{3} g_1 Y_Q c_{39}^{(8)} g_2+\left(-\frac{g_2^2}{6}+\frac{2 \mathcal{Y}_1
    \bar{\mathcal{Y}}_1}{3}+\frac{4 \mathcal{Y}_2 \bar{\mathcal{Y}}_2}{3}\right) c_{44}^{(8)}\\
    &+\left(\frac{g_2^2}{6}-\frac{4 \mathcal{Y}_1 \bar{\mathcal{Y}}_1}{3}-\frac{2 \mathcal{Y}_2
    \bar{\mathcal{Y}}_2}{3}\right) c_{45}^{(8)}+\left(\frac{\mathcal{Y}_1 \bar{\mathcal{Y}}_1}{3}-\frac{\mathcal{Y}_2 \bar{\mathcal{Y}}_2}{3}\right) c_{46}^{(8)}\\
    &+\left(2 Y_H^2
    g_1^2+\frac{8 Y_Q^2 g_1^2}{3}+8 Y_H Y_Q g_1^2-\frac{13 g_2^2}{3}+\frac{32 g_3^2}{9}+\frac{8 \mathcal{Y}_1
    \bar{\mathcal{Y}}_1}{3}\right) c_{63}^{(8)}\\
    &+\left(8 Y_H^2 g_1^2+6 Y_Q^2 g_1^2+4 Y_H Y_Q g_1^2+\frac{107 g_2^2}{6}+8
    g_3^2+\frac{16 \mathcal{Y}_1 \bar{\mathcal{Y}}_1}{3}\right) c_{64}^{(8)}\\
    &+\left(4 g_2^2-\frac{8 \mathcal{Y}_1 \bar{\mathcal{Y}}_1}{3}+8 \mathcal{Y}_2 \bar{\mathcal{Y}}_2\right)
    c_{65}^{(8)}+\left(2 g_2^2+\frac{8 \mathcal{Y}_1 \bar{\mathcal{Y}}_1}{3}\right) c_{66}^{(8)}\ ,\\ 
    \dot{c}^{\prime (8)}_{65} & =-4 c_{21}^{(8)} g_2^2-\frac{2}{3} i
    c_{22}^{(8)} g_2^2+\frac{4}{3} g_1 Y_Q c_{38}^{(8)} g_2+\frac{4}{3} g_1 Y_Q c_{39}^{(8)}
    g_2-\frac{16}{3} g_1^2 Y_Q^2 c_{20}^{(8)}\\
    &-\frac{64}{9} g_3^2 c_{23}^{(8)}+\left(\frac{4 \mathcal{Y}_1 \bar{\mathcal{Y}}_1}{3}+\frac{2
    \mathcal{Y}_2 \bar{\mathcal{Y}}_2}{3}\right) c_{44}^{(8)}+\frac{2}{3} \mathcal{Y}_1 \bar{\mathcal{Y}}_1 c_{45}^{(8)}-\frac{2}{3} \mathcal{Y}_1 \bar{\mathcal{Y}}_1
    c_{46}^{(8)}\\
    &+\left(\frac{17 g_2^2}{3}-\frac{8 \mathcal{Y}_1 \bar{\mathcal{Y}}_1}{3}+4 \mathcal{Y}_2 \bar{\mathcal{Y}}_2\right) c_{63}^{(8)}+\left(2 g_2^2-\frac{4
    \mathcal{Y}_1 \bar{\mathcal{Y}}_1}{3}+4 \mathcal{Y}_2 \bar{\mathcal{Y}}_2\right) c_{64}^{(8)}\\
    &+\left(12 Y_H^2 g_1^2+\frac{34 Y_Q^2 g_1^2}{3}-4 Y_H Y_Q
    g_1^2+\frac{37 g_2^2}{2}+\frac{136 g_3^2}{9}+\frac{32 \mathcal{Y}_1 \bar{\mathcal{Y}}_1}{3}\right) c_{65}^{(8)}\\
    &+\left(8 Y_H Y_Q g_1^2-2
    g_2^2-\frac{8 \mathcal{Y}_1 \bar{\mathcal{Y}}_1}{3}\right) c_{66}^{(8)}\ ,\\ 
    \dot{c}^{\prime (8)}_{66} & =-2 c_{21}^{(8)} g_2^2-\frac{1}{3} i c_{22}^{(8)} g_2^2+\frac{2}{3}
    g_1 Y_Q c_{38}^{(8)} g_2+\frac{2}{3} g_1 Y_Q c_{39}^{(8)} g_2-\frac{8}{3} g_1^2 Y_Q^2
    c_{20}^{(8)}\\
    &-\frac{32}{9} g_3^2 c_{23}^{(8)}+\left(-Y_H Y_Q g_1^2+\frac{g_2^2}{12}-\frac{4 \mathcal{Y}_1 \bar{\mathcal{Y}}_1}{3}+\frac{4
    \mathcal{Y}_2 \bar{\mathcal{Y}}_2}{3}\right) c_{44}^{(8)}\\
    &+\left(\frac{1}{3} Y_H Y_Q g_1^2-\frac{g_2^2}{12}+\frac{4 \mathcal{Y}_1 \bar{\mathcal{Y}}_1}{3}\right)
    c_{45}^{(8)}+\left(\frac{2}{3} Y_H Y_Q g_1^2+\frac{2 \mathcal{Y}_1 \bar{\mathcal{Y}}_1}{3}-\mathcal{Y}_2 \bar{\mathcal{Y}}_2\right) c_{46}^{(8)}\\
    &+\left(\frac{83
    g_2^2}{12}-\frac{g_1^2 Y_H^2}{3}-2 g_1^2 Y_Q^2-\frac{8 \mathcal{Y}_1 \bar{\mathcal{Y}}_1}{3}+8 \mathcal{Y}_2 \bar{\mathcal{Y}}_2\right)
    c_{63}^{(8)}\\
    &+\left(-\frac{13 g_2^2}{6}+\frac{2 g_1^2 Y_H^2}{3}+4 g_1^2 Y_Q^2-\frac{4 \mathcal{Y}_1 \bar{\mathcal{Y}}_1}{3}\right)
    c_{64}^{(8)}\\
    &+\left(\frac{4 Y_H^2 g_1^2}{3}-\frac{4 Y_Q^2 g_1^2}{3}+8 Y_H Y_Q g_1^2+\frac{3 g_2^2}{2}+\frac{32
    g_3^2}{9}+\frac{8 \mathcal{Y}_1 \bar{\mathcal{Y}}_1}{3}\right) c_{65}^{(8)}\\
    &+\left(\frac{28 Y_H^2 g_1^2}{3}+14 Y_Q^2 g_1^2+4 Y_H
    Y_Q g_1^2+\frac{19 g_2^2}{2}+8 g_3^2-\frac{8 \mathcal{Y}_1 \bar{\mathcal{Y}}_1}{3}\right) c_{66}^{(8)}\ .
\end{align*}
\newpage

\section{Bootstrapping the tree-level amplitudes}
\label{sec::bootstraphigherpoints}

In order to push the computation of the mixing matrix beyond leading order, higher-point tree-level amplitudes are needed. In the amplitudes literature, the computation of higher-point tree-level amplitudes from on-shell data is usually performed through BFCW recursion relations \cite{Britto:2004nc,Britto:2005fq,Arkani-Hamed:2008bsc}, or its generalisations \cite{Risager:2005vk,Cohen:2010mi,Kampf:2012fn,Cheung:2014dqa,Cheung:2015ota}. The strategy of BCFW-like recursion relations is the following:
\begin{itemize}
    \item $l$ momenta are shifted introducing a complex parameter $z$ ($l$-line shift) in a way which preserves momentum conservation and on-shell conditions.
    \item The original amplitude is computed as a contour integral in the complex $z$-plane: $\Aa_n (0) = \frac{1}{2\pi i}\oint_{z=0} \frac{A_n(z)}{z}$, using Cauchy theorem knowing that, under the assumption of a good behaviour in the $z\to \infty$ limit ({\it i.e.} $\Aa_{n} (z)\to z^\gamma$ with $\gamma \leq -1$), the other poles of the amplitude correspond to factorisation channels and can be computed from \eqref{eq:factorisation}.
\end{itemize}

These recursion relations are particularly well-suited for the computation of amplitudes involving vectors and gravitons, for which the BCFW (2-line) shift gives rather compact results summing over a small subset of the actual factorisation channels. The most general criteria for the shifted amplitude to be well-behaved in the $z\to \infty$ limit are given in \cite{Cheung:2015cba}: all renormalisable theories are shown to be $5$-line constructible and, in particular, theories involving fermions and scalars charged under a $U(1)$ are $3$-line constructible, as in the case of the Standard Model. Moreover, non-renormalisable amplitudes with no-derivative operator insertions are on-shell constructible, but it is not generally true for operators with derivatives. Finally, $n$-line shifts with $n\geq 3$ give rather cumbersome results and in no case locality is manifest in the final amplitude. 

Since in our approach we should consider all kinds of operators, we have to find an alternative approach to recursion relations, which is anyway completely on-shell. The general strategy has been outlined in the Section \ref{sec::4points}, and in the following section we are going to argue that in our framework any effective field theory is fully on-shell constructible from unitarity and locality. In particular, the singularity structure will be manifest in the final result.

\subsection{Higher-point amplitudes in the SM without recursion relations}

The procedure can be roughly divided into two parts: the construction of an ansatz and a matching procedure on the single-particle cuts to fix the free-parameters, which we perform numerically over finite fields to speed up the computation.

\subsubsection{Constructing an ansatz}\label{sec:ansatz}

A generic tree-level amplitude can be schematically written as
\begin{equation}\label{eq:ampansatz}
    \Aa_n(p_1^{a_1, h_1}, \ldots , p_n^{a_n, h_n}) = \sum_{i,j,k} \frac{C_{i,j}^{\, a_1  \cdots a_n}}{\mathcal{D}_{i}} \, c_{i,j,k}\,\NN_{i,j,k} \> + \mathcal{P}^{a_1  \cdots a_n},
\end{equation}
where $p_i^{a_i, h_i}$ represents a generic state with helicity $h_i$ and gauge-group index $a_i$. The tensors $C_{i,j}^{a_1 \cdots a_n}$ are the gauge-group invariant structure of the amplitude, whereas $\mathcal{D}_{j}$ and $\NN_{i,j,k}$ are kinematic denominators and numerators respectively, where the latter carry the dependence on the helicity structure. The $c_{i,j,k}$ are rational coefficients associated to the different helicity structures $\NN_{i,j,k}$. Finally, the $\mathcal{P}^{a_1  \cdots a_n}$ are terms with polynomial dependence in the kinematic variables, in other words contact terms, which vanish whenever we probe any factorisation channel.
We will show that in our framework the contact terms are irrelevant and the tree-level amplitudes are fully determined by lower-point amplitudes from factorisation.

First we motivate this assumption for renormalisable theories through a simple dimensional analysis consideration: due to \eqref{eq:massdimamp}, for $n > 4$ we have $\left[ \Aa_{\, n} \right] < 0$. Moreover, all the couplings in the SM are dimensionless, we are considering only massless states (there are no dimension-full parameters in the amplitude), and by construction $\left[\mathcal{P}^{a_1  \cdots a_n}\right] \geq 0$. These considerations imply necessarily that for renormalisable massless theories for $n>4$ $\mathcal{P}^{a_1  \cdots a_n}=0$ and every term in the amplitude must posses some kinematic denominators $\mathcal{D}_i$. This means that the amplitudes can be fully determined from factorisation, through a recursive procedure described below in this section.
%Thus, any kinematic structure in the numerator of an amplitude must, and can only, be compensated by an appropriate number of kinematic structures in the denominator, which will result in poles.

This argument is somehow subtle for $n = 4$, because it is possible to build terms of mass dimension zero which are ratios of spinor variables but vanish on any cut. An example of such a structure for the all-plus four-gluon amplitudes is
\begin{align}\label{eq:nonsing}
    \frac{\sqr{1}{2}^2\sqr{3}{4}^2}{s_{12}^2} = \frac{\sqr{1}{3}^2\sqr{2}{4}^2}{s_{13}^2} = \frac{\sqr{1}{4}^2\sqr{2}{3}^2}{s_{14}^2}  \> ,
\end{align}
whose residue is zero on any of the three invariants $s_{12}$, $s_{13}$ and $s_{14}$. These structures do not introduce any correction to the factorisation channels of four-point amplitudes (\textit{i.e.} they are polynomial in the kinematic variables). We will systematically ignore such contact terms at four points, except for the four-scalar contact term (corresponding in the Lagrangian formalism to the $\lambda\phi^4$ interaction). Indeed, such terms are usually computed through $d$-dimensional generalised unitarity techniques as one-loop finite rational terms \cite{Bern:1995db,Bern:1996ja,Bern:2005hs,Brandhuber:2005jw,Ellis:2008ir,Badger:2008cm,Nandan:2018ody,AccettulliHuber:2019abj}, hence they must be vanishing at tree-level\footnote{For example, we know that such terms can never be generated by any local Lagrangian interaction at tree-level.}. In particular, in the case of the four-scalar amplitude we will add to the factorisable part a contact term whose kinematic dependence is trivial:

\begin{equation}
    \Aa_4(\bar{H}^{i_1} \bar{H}^{i_2} H^{i_3} H^{i_4}) = - \left(g_1^2\, Y_H^2\,\delta^{i_3}_{i_1} \delta^{i_4}_{i_2} + g_2^2\, \sigma\indices{^{I\, i_3}_{i_1}}\sigma\indices{^{I\, i_4}_{i_2}} \right)\frac{s_{12}-s_{14}}{s_{1 3}} - \lambda\, \delta^{i_3}_{i_1} \delta^{i_4}_{i_2}+ \left(3\leftrightarrow 4\right) \ .
\end{equation}

We stress that for $n>4$ non-singular terms such as \eqref{eq:nonsing} cannot appear: this can be easily seen focusing on real kinematics and by dimensional analysis considerations, which tell us that there must be a singularity for renormalisable amplitudes with more than four external particles.

%This argument cannot be generalised to the case of scattering amplitudes with insertions of effective interactions. For example, if we consider the six-scalar amplitude with an insertion of a $\partial^2 \phi^4$ interaction, there is no equivalent argument to discard a $\phi^6$-like contact term contribution. On the other hand, if we are already considering an effective field theory with both $\partial^2 \phi^4$ and $\phi^6$ interactions, when we consider the $n$-scalar amplitudes with $n\geq 6$, there cannot be any other contact terms and it is not possible to fully disentangle the contribution to the six-point contact term from $\partial^2 \phi^4$ or $\phi^6$. In other words, we can always shift the Wilson coefficient of the $\phi^6$ in such a way that the contribution of $\partial^2 \phi^4$ disappear from the six-scalar contact term.
This argument cannot be generalised to the case of scattering amplitudes with insertions of effective interactions. For example, consider the six-scalar amplitude with an insertion of a $\partial^2 \phi^4$ interaction, call it $\mathcal{F}_{6,6,\partial^2 \phi^4}$ (using the notation introduced in Section \ref{sec::mixing}). There is no equivalent argument to discard a $\phi^6$-like contact term contribution arising in the calculation of this amplitude. On the other hand, any physical process which gets a contribution from $\mathcal{F}_{6,6,\partial^2 \phi^4}$ will also get one from $\mathcal{F}_{6,6,\phi^6}$ which is the contact interaction due to the operator $\phi^6$ itself. Physically, the two contact term contributions cannot be disentangled, because they provide the same description for the interaction between scalars. As a consequence, if we are already considering an effective field theory with both $\partial^2 \phi^4$ and $\phi^6$ interactions in our operator basis, neglecting the $\phi^6$-like contact term in $\mathcal{F}_{6,6,\partial^2 \phi^4}$ can be compensated by appropriately shifting the Wilson coefficient of the $\phi^6$ operator.

This argument can be generalised to more generic theories, like the SMEFT in our case. What we wanted to convey is that, as long as we consider a complete basis of operators up to a given dimension, contact terms can only contribute shifting the Wilson coefficients of a different operator. Then we {\it choose} our basis of EFT interactions such that it does not generate polynomial terms when computing higher-multiplicity amplitudes and thus we can effectively neglect them in the computations, so $\mathcal{P}^{a_1  \cdots a_n} = 0$.
%Combining this physical fact with the observation that higher mass dimension operators cannot ``renormalise'' the Wilson coefficients of lower mass dimension operators in the above described way, which can be seen again by dimensional analysis, we conclude that as long as we consider a complete basis of operators up to a given dimension, contact terms can be included by definition in the Wilson coefficients and thus neglected in the computations.

We present now the algorithm to compute higher-point tree-level amplitudes from factorisation.
\begin{enumerate}
    \item We begin by enumerating all the possible singularity structures of the amplitude consistent with locality, which are provided by all the possible ways the amplitude can consistently factorise into %where by maximal we mean that each lower-point amplitude appearing in the factorisation is itself factored into even lower-point ones, all the way down to the
    trivalent graphs\footnote{When talking about trivalent graphs or three-point amplitudes in this section we always mean the building blocks of our theory, which strictly speaking includes not only the three-point amplitudes but also the four-point scalar interaction $-\lambda (\bar{H} H)^2/4$ (with the corresponding quadrivalent vertices in the graphs) and, if we are considering amplitudes with effective operator insertions, also any of the relevant effective interaction classified in Section \ref{sec:classificationSMEFTops}.}. We enumerate all the possible tree graphs with trivalent and quadrivalent internal vertices, and then a selection criterion is applied to discard channels which are not compatible with Standard Model interactions.
    \item To each trivalent graph a unique kinematic denominator $\mathcal{D}_i$ is associated, this is the product of the propagators corresponding to internal edges in the graphs, {\it i.e.} it is a product of the Mandelstam invariants characterising the channels.
    \item Unitarity also fixes the colour structures associated to each graph $\{C_{i,j}^{a_1 \cdots a_n}\}_{j=1,\dots ,s}$. In particular, different colour structures correspond to different particles propagating in the internal lines. Once the internal particles are determined, the colour structures are obtained from the product of the colour structures in the three-point amplitudes.
    %\item Finally the kinematic numerators are generated with the algorithm presented in Section \ref{sec:kinematics}. The $\{ \mathcal{N}_{i,j,k}\}_{k=1,\dots ,m}$ are the $m$ independent spinor structures in our basis, each of which is multiplied by arbitrary (rational) coefficients $\{c_{i,j,k} \}_{k=1,\dots , m}$, corresponding to the colour structure $C_{i,j}^{a_1 \cdots a_n}$ and the denominator $\mathcal{D}_i$. The basis depends only on the mass dimension of $\mathcal{D}_i$ and the helicity of the external particles, then we have $\mathcal{N}_{i,j_1,k} = \mathcal{N}_{i,j_2,k}$. This coefficients will be fixed by the matching procedure over the different factorisation channels described in details in Section \ref{sec::ansatzsol}
    \item Finally the kinematic numerators are generated with the algorithm presented in Section \ref{sec:kinematics}\footnote{The full algorithm presented in this section can be applied to the case of form factors as well. If this was the case we were interested in, we should consider at this point a simplified version of the algorithm presented in Section \ref{sec:kinematics}, in which we ignore momentum conservation.}. The $\{ \mathcal{N}_{i,j,k}\}_{k=1,\dots ,h}$ are $h$ independent spinor structures in our basis, and a set of these numerators is associated to each of the colour structure $C_{i,j}^{a_1 \cdots a_n}$ corresponding to the denominator $\mathcal{D}_i$. The latter fixes the mass dimension of the numerators through $\left[\NN_{i,j,k} \right] = \left[ \Aa_n  \right] + \left[ \mathcal{D}_i \right]$ whereas the helicity weights are given by the external particles.
    Each of the $ \mathcal{N}_{i,j,k}$ is multiplied by arbitrary (rational) coefficients $c_{i,j,k}$ which will be fixed by the matching procedure over the different factorisation channels described in detail in Section \ref{sec::ansatzsol}. Notice that the basis of numerators does obviously not depend on the colour structures, but only on the mass dimension of the denominator structure: {\it i.e.} $\mathcal{N}_{i_1,j_1,k} = \mathcal{N}_{i_2,j_2,k}$ if $\left[\mathcal{D}_{i_1}\right] = \left[\mathcal{D}_{i_2}\right]$ for any colour structure labelled by $j_1$ and $j_2$. This fact has been exploited heavily to speed up the numerical evaluation of the ansatz when solving for the coefficients $\{c_{i,j,k}$.
    \item Some of the coefficients can be fixed before the matching procedure by demanding that the ansatz is not redundant. In particular, the simplifying observation is that the various coefficients cannot combine in such a way that the sum over the related structures is proportional to any of the Mandelstam invariants appearing in the denominators.
    %\item Before proceeding to the solution, the numerator ansatz is further simplified by imposing that combinations of the numerators which would be proportional to one of the Mandelstam invariants in the denominator do not appear. This leads to linear relations among some of the $\{c_{i,j,k} \}_{k=1,\dots , m}$ which are thus a priori fixed.
    \item Finally we solve for the $\{c_{i,j,k} \}$ by matching over the different factorisation channels as described in \ref{sec::ansatzsol}.
\end{enumerate}

\begin{figure}
    \centering
    \begin{tikzpicture}[scale=18]
    
    \node at (-2pt,2pt) (A) {$Q^{a_1,i_1}$};
    \node at (2pt,2pt) (B) {$H^{i_5}$};
    \node at (2pt,-2pt) (C) {$\overline{H}^{i_3}$};
    \node at (-2pt,-2pt) (D) {$u^{a_2}$};
    \node at (2.5pt,0) (E) {$H^{i_4}$};
    \node at (0,0) (oo) {};
    
    \draw[fermion,>=Stealth] (A) -- (oo);
    \draw[dashed] (B) -- (oo);
    \draw[dashed] (C) -- (oo);
    \draw[dashed] (E) -- (oo);
    \draw[fermionbar,>=Stealth] (oo) -- (D);
    \node at (0,0) [draw, fill = gray, circle, inner sep=2.5mm] (oo) {};
    
    \def\x{7.5pt}
    \def\y{6pt}
    
    \def\xa{\x}
    \def\ya{\y}
    
    \node at (0+\xa,0+\ya) (oo){};
    \node at (-1.8pt+\xa,\ya) (A) {$u$};
    \node at (1.5pt+\xa,\ya) (B) {};
    \node at (\xa,-1.5pt+\ya) (C) {$Q$};
    %\node at (1.8pt+\xa,1pt+\ya) (up) {};
    %\node at (1.8pt+\xa,-1pt+\ya) (down) {};
    %\node at (1.9pt+\xa,0.95pt+\ya) [rotate around={-90:(1pt,2.5pt)}]{\Cutright};
    \node at (1.65pt+\xa,0.8pt+\ya) {$H/\overline{H}$};
    \node at (1.65pt+\xa,-0.8pt+\ya) {$s_{12}$};
    
    \draw (A) -- (oo.center);
    \draw (B) -- (oo.center);
    \draw (C) -- (oo.center);
    %\draw[dashed] (up) -- (down);
    
    \def\xa{\x+3pt}
    \def\ya{\y}
    
    \node at (0+\xa,0+\ya) (oo){};
    \node at (-1.5pt+\xa,\ya) (A) {};
    \node at (1.5pt+\xa,\ya) (B) {};
    \node at (\xa,-1.5pt+\ya) (C) {$H$};
    %\node at (1.8pt+\xa,1pt+\ya) (up) {};
    %\node at (1.8pt+\xa,-1pt+\ya) (down) {};
    %\node at (1.9pt+\xa,0.95pt+\ya) [rotate around={-90:(1pt,2.5pt)}]{\Cutright};
    \node at (1.65pt+\xa,1.5pt+\ya) {$W^{\pm}/W^{\mp}$};
    \node at (1.65pt+\xa,0.8pt+\ya) {$B^{\pm}/B^{\mp}$};
    \node at (1.65pt+\xa,-0.8pt+\ya) {$s_{35}$};
    
    \draw (A) -- (oo.center);
    \draw (B) -- (oo.center);
    \draw (C) -- (oo.center);
    %\draw[dashed] (up) -- (down);
    
    \def\xa{\x+6pt}
    \def\ya{\y}
    
    \node at (0+\xa,0+\ya) (oo){};
    \node at (-1.5pt+\xa,\ya) (A) {};
    \node at (1.8pt+\xa,\ya) (B) {$H$};
    \node at (\xa,-1.5pt+\ya) (C) {$\overline{H}$};
    
    \draw (A) -- (oo.center);
    \draw (B) -- (oo.center);
    \draw (C) -- (oo.center);
    
    \node at (\x+11.5pt,\y) {$\mapsto \hspace{0.2cm}\dfrac{C_{1,1}}{\mathcal{D}_1}\>, \dfrac{C_{1,2}}{\mathcal{D}_1}$};
    
    \def\x{7.5pt}
    \def\y{2pt}
    
    \def\xa{\x}
    \def\ya{\y}
    
    \node at (0+\xa,0+\ya) (oo){};
    \node at (-1.8pt+\xa,\ya) (A) {$H$};
    \node at (1.5pt+\xa,\ya) (B) {};
    \node at (\xa,-1.5pt+\ya) (C) {$Q$};
    %\node at (1.8pt+\xa,1pt+\ya) (up) {};
    %\node at (1.8pt+\xa,-1pt+\ya) (down) {};
    %\node at (1.9pt+\xa,0.95pt+\ya) [rotate around={-90:(1pt,2.5pt)}]{\Cutright};
    \node at (1.65pt+\xa,0.8pt+\ya) {$u/\overline{u}$};
    \node at (1.65pt+\xa,-0.8pt+\ya) {$s_{14}$};
    
    \draw (A) -- (oo.center);
    \draw (B) -- (oo.center);
    \draw (C) -- (oo.center);
    %\draw[dashed] (up) -- (down);
    
    \def\xa{\x+3pt}
    \def\ya{\y}
    
    \node at (0+\xa,0+\ya) (oo){};
    \node at (-1.5pt+\xa,\ya) (A) {};
    \node at (1.5pt+\xa,\ya) (B) {};
    \node at (\xa,-1.5pt+\ya) (C) {$u$};
    %\node at (1.8pt+\xa,1pt+\ya) (up) {};
    %\node at (1.8pt+\xa,-1pt+\ya) (down) {};
    %\node at (1.9pt+\xa,0.95pt+\ya) [rotate around={-90:(1pt,2.5pt)}]{\Cutright};
    \node at (1.65pt+\xa,0.8pt+\ya) {$B^{\pm}/B^{\mp}$};
    \node at (1.65pt+\xa,-0.8pt+\ya) {$s_{35}$};
    
    \draw (A) -- (oo.center);
    \draw (B) -- (oo.center);
    \draw (C) -- (oo.center);
    %\draw[dashed] (up) -- (down);
    
    \def\xa{\x+6pt}
    \def\ya{\y}
    
    \node at (0+\xa,0+\ya) (oo){};
    \node at (-1.5pt+\xa,\ya) (A) {};
    \node at (1.8pt+\xa,\ya) (B) {$H$};
    \node at (\xa,-1.5pt+\ya) (C) {$\overline{H}$};
    
    \draw (A) -- (oo.center);
    \draw (B) -- (oo.center);
    \draw (C) -- (oo.center);
    
    \draw[loosely dotted,very thick] (\x+3pt,\y-2.5pt) -- (\x+3pt,\y-4pt);
    
    \node at (\x+11.5pt,\y) {$\mapsto \hspace{0.2cm}\dfrac{C_{2,1}}{\mathcal{D}_2}$};

    \def\x{9pt}
    \def\y{-4pt}
    
    \def\xa{\x}
    \def\ya{\y}
    
    \node at (0+\xa,0+\ya) (oo){};
    \node at (-1.8pt+\xa,\ya) (A) {$u$};
    \node at (1.5pt+\xa,\ya) (B) {};
    \node at (\xa,-1.5pt+\ya) (C) {$Q$};
    %\node at (1.8pt+\xa,1pt+\ya) (up) {};
    %\node at (1.8pt+\xa,-1pt+\ya) (down) {};
    %\node at (1.9pt+\xa,0.95pt+\ya) [rotate around={-90:(1pt,2.5pt)}]{\Cutright};
    \node at (1.65pt+\xa,0.8pt+\ya) {$H/\overline{H}$};
    \node at (1.65pt+\xa,-0.8pt+\ya) {$s_{12}$};
    
    \draw (A) -- (oo.center);
    \draw (B) -- (oo.center);
    \draw (C) -- (oo.center);
    %\draw[dashed] (up) -- (down);
    
    \def\xa{\x+3.5pt}
    \def\ya{\y}
    
    \node at (0+\xa,0+\ya) (oo){};
    \node at (-1.5pt+\xa,\ya) (A) {};
    \node at (1.8pt+\xa,\ya) (B) {$H$};
    \node at (\xa,-1.5pt+\ya) (C) {$\overline{H}$};
    \node at (\xa,1.5pt+\ya) (D) {$H$};
    %\node at (1.8pt+\xa,1pt+\ya) (up) {};
    %\node at (1.8pt+\xa,-1pt+\ya) (down) {};
    %\node at (1.9pt+\xa,0.95pt+\ya) [rotate around={-90:(1pt,2.5pt)}]{\Cutright};
    \node at (1.65pt+\xa,0.8pt+\ya) {$s_{45}$};
    
    \draw (A) -- (oo.center);
    \draw (B) -- (oo.center);
    \draw (C) -- (oo.center);
    \draw (D) -- (oo.center);
    %\draw[dashed] (up) -- (down);
    
    \draw[decorate,decoration={brace,amplitude=10pt,mirror}] (4.5pt,7pt) -- (4.5pt,-6pt);
    
    \node at (\x+10pt,\y) {$\mapsto \hspace{0.2cm}\dfrac{C_{21,1}}{\mathcal{D}_{21}}$};
    
    \def\y{-8.5pt}
    
    \node at (0pt,\y) {$\dfrac{C_{1,1}}{\mathcal{D}_1}= \dfrac{\delta^{a_2}_{a_1}\, \epsilon^{i_1 i_4}\,\delta^{i_5}_{i_3}}{s_{12}\,s_{35}} \>, $};
    
     %\node at (-3pt,\y) {$\dfrac{C_{1,1}}{\mathcal{D}_1}= \dfrac{\delta^{a_2}_{a_1}\, \epsilon^{i_1 i_4}\,\delta^{i_5}_{i_3}}{s_{12}\,s_{35}} \>, \hspace{0.3cm}\dfrac{C_{1,2}}{\mathcal{D}_1}= \dfrac{\delta^{a_2}_{a_1}\,\epsilon^{i_1 i_5}\,\delta^{i_4}_{i_3}}{s_{12}\,s_{35}}\>,\hspace{0.3cm}\dfrac{C_{2,1}}{\mathcal{D}_2}= \dfrac{\delta^{a_2}_{a_1}\, \epsilon^{i_1 i_4}\,\delta^{i_5}_{i_3}}{s_{14}\,s_{35}}\>,\hspace{0.3cm}\dfrac{C_{21,1}}{\mathcal{D}_{21}}=\dfrac{\delta^{a_2}_{a_1}\, (\epsilon^{i_1 i_4}\,\delta^{i_5}_{i_3}+\epsilon^{i_1 i_5}\,\delta^{i_4}_{i_3})}{s_{12}}$};
     
     \node at (6.5pt,\y) {$\dfrac{C_{1,2}}{\mathcal{D}_1}= \dfrac{\delta^{a_2}_{a_1}\,\sigma^{I\, i_1 i_4}\,\sigma\indices{^{I\, i_5}_{i_3}}}{s_{12}\,s_{35}} \> ,$};
    
    \node at (13pt,\y) {$\dfrac{C_{2,1}}{\mathcal{D}_2}= \dfrac{\delta^{a_2}_{a_1}\, \epsilon^{i_1 i_4}\,\delta^{i_5}_{i_3}}{s_{14}\,s_{35}} \> ,$};
    
    \draw[loosely dotted,very thick] (2.5pt,\y-2.5pt) -- (7.5pt,\y-2.5pt);
    
    \node at (16pt,\y-2.5pt) {$\dfrac{C_{21,1}}{\mathcal{D}_{21}}=\dfrac{\delta^{a_2}_{a_1}\, (\epsilon^{i_1 i_4}\,\delta^{i_5}_{i_3}+\epsilon^{i_1 i_5}\,\delta^{i_4}_{i_3})}{s_{12}}$};
    
    \end{tikzpicture}
    \caption{The splitting of $\Aa_5(Q^{a_1,i_1},u^{a_2},\overline{H}^{i_3},H^{i_4},H^{i_5})$ into trivalent graphs and the associated colour factors and kinematic denominators. There are a total of 21 possible trivalent graphs associated with this amplitude, we showed explicitly the first, the second and the last, as significant examples. The second is a trivial instance of trivalent graphs and there is a unique choice compatible with the Standard Model interactions of internal particle propagating. The same is not true for the first factorisation channel, for which we can have both $B$s and $W$s propagating, which give us two different colour structures $C_{1,1}$ and $C_{2,1}$, respectively. The last channel is the only one for this amplitude which involves an insertion of the quadrivalent Higgs interaction.}\label{fig:my_label}
\end{figure}
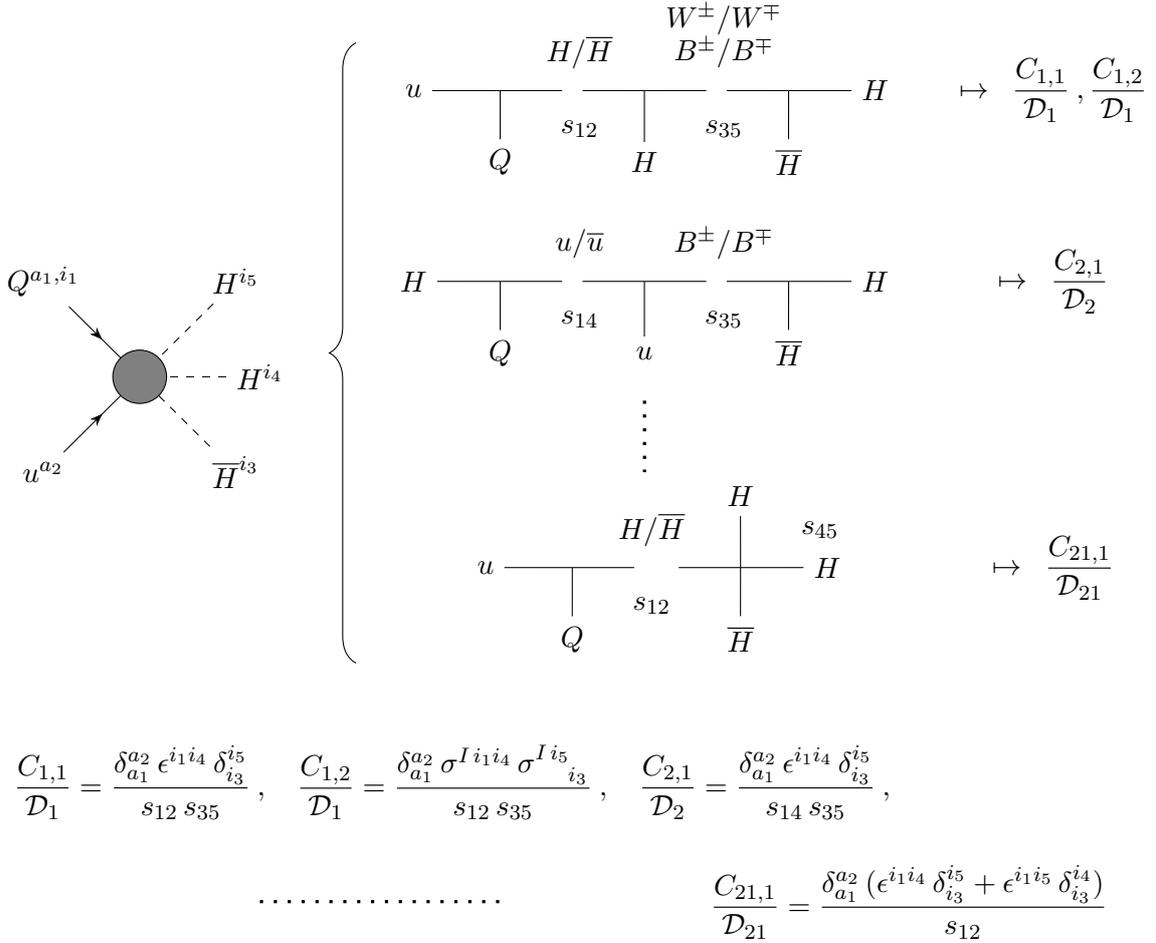

We consider, as an example, the five-point amplitude $\Aa_5(Q^{a_1,i_1},u^{a_2},\bar{H}^{i_3},H^{i_4},H^{i_5})$. There are 21 trivalent graphs compatible with this process, and some of them are shown in Figure \ref{fig:my_label}. Most of the graphs do not involve the scalar quadrivalent interaction, except the last one, we then have $\left[ \mathcal{D}_i\right] = 4$ for $i=1,\dots ,20$ and $\left[\mathcal{D}_{21}\right] = 2$ with:
\begin{equation}
\begin{split}
    \{\mathcal{D}_i\}_{i=1,\dots ,21} = \{&s_{1 2} s_{3 5}, s_{1 4} s_{3 5}, s_{2 4} s_{3 5}, s_{1 2} s_{3 4}, s_{1 5} s_{3 4}, s_{2 5} s_{3 4}, s_{1 3} s_{2 5}, s_{1 4} s_{2 5}, s_{2 5} s_{3 4}, s_{1 3} s_{2 4},\\ &s_{1 5} s_{2 4},
   s_{2 4} s_{3 5}, s_{1 5} s_{2 4}, s_{1 5} s_{3 4}, s_{1 4} s_{2 5}, s_{1 4} s_{3 5}, s_{1 3} s_{2 4}, s_{1 3}
   s_{2 5}, s_{1 2} s_{3 4}, s_{1 2} s_{3 5}, s_{1 2}\}
\end{split}
\end{equation}

Next we build the kinematic numerators whose structure is fixed by the helicity of the external particles along with the mass dimension of the amplitude and of the denominators as
\begin{equation}
    \left[\Aa_n \right] =  \left[ \NN_{i,j,k} \right] - \left[ \mathcal{D}_{j} \right] \hspace{0.3cm} \Rightarrow \hspace{0.3cm} \left[ \NN_{i,j,k} \right]= 4-n+\left[ \mathcal{D}_{j} \right] \>.
\end{equation}
In our example we have then
\begin{align}
    \{\NN_{i,j,k}\}_{k=1,\ldots,6} &=\{ s_{12}\sqr{1}{2},s_{13}\sqr{1}{2},s_{23}\sqr{1}{2},s_{24}\sqr{1}{2},s_{34}\sqr{1}{2}, \agl{3}{4}\sqr{1}{4}\sqr{2}{3}  \} \> , \\
    \{\NN_{21,j,k}\}_{k=1} &= \{ \sqr{1}{2}\} \> ,
\end{align}
for $i=1,\dots , 20$. Computing the amplitude then reduces to fixing the rational coefficients $c_{i,j,k}$. In fact, before proceeding with the system solution we can fine tune the ansatz in order to remove combinations which would lead to cancellations in the denominators. In particular, since there are two Mandelstam invariants for the first twenty denominators, this would fix a priori two coefficients for each denominator and for each colour structures. We consider, for example, the first two trivalent graphs, shown in Figure \ref{fig:my_label}. The general algorithm to fix the coefficient is the following: 
\begin{itemize}
    \item We have a set of independent helicity structures with a specified mass dimension $d$, {\it i.e.} $\{ \mathcal{N}_{i,j,k}\}_{k=1,\dots ,h_1}$, and we assume the existence of a set of structures with the same helicity configuration and mass dimension $d-2$, {\it i.e.} $\{ \mathcal{M}_{i,j,l}\}_{l=1,\dots ,h_2}$. If the latter do not exist, this procedure can be skipped.
    \item For each Mandelstam invariant $s_{i_1\cdots i_n}$ appearing in the denominator $\mathcal{D}_i$ we fix some coefficients $d^{(p)}_{i,j,k}$ through
    \begin{equation}
        \sum_{k=1}^{h_1} d^{(p)}_{i,j,k}\, \mathcal{N}_{i,j,k} = s_{i_1\cdots i_n}\, \mathcal{M}_{i,j,l} \hspace{.5cm} \forall\, l\ .
    \end{equation}
    These conditions provide us with $p=1,\dots ,\frac{\left[\mathcal{D}_i \right]}{2}\cdot h_2$ vectors $d^{(p)}_{i,j,k}$.
    \item Finally, we impose the orthogonality condition for the $c$'s with respect to the $d$'s
    \begin{equation}
        \sum_{k=1}^{h_1} c_{i,j,k}\, d^{(p)}_{i,j,k} = 0 \hspace{.5cm} \forall\, p\ ,
    \end{equation}
    which fixes some of the $c_{i,j,k}$, as anticipated.
\end{itemize}
In our specific example, for $\mathcal{D}_1$ we find $c_{1,j,1}=0$ and $c_{1,j,5} = -c_{1,j,2} -c_{1,j,3}$ with $j=1,2$ and for $\mathcal{D}_2$ we find $c_{2,1,4} = -c_{2,1,1}$ and, again, $c_{2,1,5} = -c_{2,1,2} -c_{2,1,3}$.

\subsubsection{The case of external vector bosons}\label{sec:vectorbosons}

The procedure described so far works very well when we are dealing with amplitudes with scalars and fermions as external particles. But when vector bosons are involved, or more in general massless particles with $| h | \geq 1$, an extension of the method is required. One has to take into account that these particles provide further kinematic denominators which are not due to intermediate particle exchanges. A simple example has already been shown in Section \ref{sec:4gluon}, where we considered the four-gluon amplitude. Indeed, the four-point amplitude has mass-dimension zero, the helicity structure with the smallest mass dimension is $\agl{1}{2}^2 \sqr{3}{4}^2$ which has mass-dimension 4, and consequently a single $\frac{1}{s_{i j}}$ (associated to a trivalent graph) is not enough to get the mass-dimensions right\footnote{When we think of the problem in terms of a Feynman diagrammatic approach for $|h|=1$, this additional kinematic dependence is hidden in the polarisation vectors which in terms of spinor-helicity variables can be written as
\begin{equation}
    \epsilon_{\alpha\dot{\alpha}}^+(p,\xi) = \sqrt{2} \, \frac{\xi_\alpha \tilde{\lambda}_{\dot{\alpha}}}{\agl{\xi}{\lambda}} \> , \hspace{0.5cm}
    \epsilon_{\alpha\dot{\alpha}}^-(p,\xi) = \sqrt{2} \, \frac{\lambda_\alpha \tilde{\xi}_{\dot{\alpha}}}{\sqr{\tilde{\lambda}}{\tilde{\xi}}} \> ,
\end{equation}
where $p_{\alpha \dot{\alpha}}=\lambda_{\alpha}\tilde{\lambda}_{\dot{\alpha}}$ and $\xi$ is an arbitrary reference spinor. In our approach, it is either a simple dimensional analysis as for the four-gluon amplitude which forces us to add more denominators, or for higher-point amplitudes it will be unitarity itself that does so.}. %In fact when  solving the systems obtained when studying the singularity structure of the amplitudes, our factorisation approach might not lead to any solution when external vector bosons are involved (more about this in section \ref{sec::ansatzsol}).
Typically, once a set of denominators has been generated as described in the previous section, we need to add at least one Mandelstam invariant to each denominator or possibly more in case of higher-point amplitudes. %, up to at most the number of external vectors.
This is done in iterated steps: we first add to every denominator a single Mandelstam invariant $s_{ij}$ in all the possible ways compatible with locality\footnote{By this we mean exhausting the combinatorics of possible invariants without however adding those already present in the denominator, which would of course lead to unphysical higher order poles.}, then we build the complete ansatz and try to solve it. If the number of invariants considered for the denominators is insufficient we will find no solution for the $c$'s, so we add all the possible terms with a further invariant in the denominator and try to solve again. At every step clearly the number of possible denominators grows quite drastically, and so does the number of possible numerators since higher and higher mass-dimensions become available. The latter effect is however counteracted by discarding those numerators which cancel any power of Mandelstam invariants from the denominator, which would indeed reproduce a term of the ansatz already present from previous iterations. This part of the method proves to be the bottleneck when it comes to computing higher-multiplicity amplitudes.

This procedure of adding Mandelstam invariants to the kinematic denominators is clearly responsible for the ``mixing'' process between different factorisation channels which brought us to the identities between colour structures at the level of the four-point amplitudes in Section \ref{sec::4points}.

\subsubsection{Solution of the ansatz}
\label{sec::ansatzsol}

So far we have built an ansatz of the form \eqref{eq:ampansatz}, where each of the $\NN_{i,j,k}$ has an associated coefficient $c_{i,j,k}$. In order to fix these coefficients we impose the validity of \eqref{eq:factorisation} in every single kinematic channel, and we do so through repeated numerical evaluations:

\begin{equation}\label{eq:factorisation2}
        -i\, \underset{s_{i_1 \dots i_m}}{\rm Res}\, \underbrace{\mathcal{A}_n (p_1^{h_1} \dots p_n^{h_n})}_{\text{ansatz}} = f \sum_{s_{\rm I},h_{\rm I}} \underbrace{\mathcal{A}_{m+1} (p_{i_1}^{h_{i_1}} \dots p_{i_m}^{h_{i_m}}, p_I^{h_I}) \mathcal{A}_{n-m+1} (p_{I}^{h_I} \to p_{i_{m+1}}^{h_{i_{m+1}}} \dots p_{i_n}^{h_{i_n}})}_{\text{lower point on-shell amplitudes}}\ .
\end{equation}
The lower point amplitudes in the RHS of \eqref{eq:factorisation2} is known, because our algorithm is recursive. On the LHS we take the residue on the ansatz, which selects a subset of the denominator structures.% in the latter and
Next we decompose, through the algorithms described in Section \ref{sec::gaugestructures}, the colour structures on both sides of \eqref{eq:factorisation2} in a suitable basis $\{C_{l}^{\, a_1  \cdots a_n}\}$:
\begin{equation}
    C_{i,j}^{\, a_1  \cdots a_n} = \sum_{l} b_{i,j,l}\, C_{l}^{\, a_1  \cdots a_n}\ .
\end{equation}
Next, we impose the matching of the coefficients of the colour structures in this basis on both sides of the equality \eqref{eq:factorisation2} so we end up with a set of equations of the type
\begin{equation}\label{eq:kinematicfit}
   -i \, \sum_{i^\prime,j,k}\frac{b_{i^\prime,j,l}}{\widetilde{\mathcal{D}}_{i^\prime}}\, c_{i^\prime,j,k}\, \NN_{i^\prime,j,k} = \mathcal{K}_{l} \> .
\end{equation}
Here $i^\prime$ runs over the trivalent graph structures for which the specified Mandelstam invariant $s_{i_1 \dots i_m}$ appears, the $\widetilde{\mathcal{D}}_{i^\prime}$ are the $\mathcal{D}_{i^\prime}$ stripped of a factor $s_{i_1\ldots i_m}$ and the $c_{i,j,k}$ are the rational coefficients to be fixed. The $\mathcal{K}_{l}$ are kinematic coefficients
defined by the product of lower point amplitudes as
\begin{equation}\label{eq:factorizationagain}
    f \sum_{s_{\rm I},h_{\rm I}} \mathcal{A}_{m+1} (p_{i_1}^{h_{i_1}} \dots p_{i_m}^{h_{i_m}}, p_I^{h_I}) \mathcal{A}_{n-m+1} (p_{I}^{h_I} \to p_{i_{m+1}}^{h_{i_{m+1}}} \dots p_{i_n}^{h_{i_n}}) \coloneqq \sum_{l} C_{l}^{\, a_1  \cdots a_n} \mathcal{K}_{l} 
\end{equation}
where the colour structures $C_{l}^{\, a_1  \cdots a_n}$ are elements of the chosen colour basis. The $\mathcal{K}_l$ are known analytic functions of the spinor invariants and Mandelstam invariants, and they also contain the dependence on the couplings $g_k$, $\mathcal{Y}^{(f)}$ and $\lambda$.
Each equation \eqref{eq:kinematicfit} now only contains kinematic invariants, the $c_{i,j,k}$ for which we want to solve and products of couplings. Thus we repeatedly evaluate the kinematics numerically and so obtain a linear system in the $c_{i^\prime,j,k}$ which upon solution yields a subset of the $c_{i^\prime,j,k}$ as functions of the couplings and possibly other $c$'s.
%It is precisely the undetermined solutions which lead to constraints coming from factorisation, since the same unresolved coefficients $c_{i,j,k}$ might appear in the residue of a different kinematic channel, exactly as in \eqref{eq:4gluonconsttraints}.
Since numerical evaluations are performed on very special kinematic points where intermediate states go on-shell, %some of the coefficients appearing in the various systems of the form \eqref{eq:kinematicfit},
some of the coefficients $c_{i^\prime,j,k}$ might in principle drop out of the system. These coefficients are identified by an a priori numerical, which then allows to only solve the system in the actually relevant variables.

Repeating this procedure in every kinematic channel might still not completely fix the ansatz, since some of the $c_{i,j,k}$ might be spurious in the sense that using momentum conservation and Schouten identities appropriately they actually drop out altogether from the final result. In particular this happens when we consider amplitudes with external vectors. % where this redundancy is a consequence of the relations which exist among the colour factors.%\todo{questo accade solo nel caso dei gluoni dove per forza l'ansatz è ridondante a causa delle relazioni tra i fattori di colore}.
At the very end of the calculation, we take advantage of the arbitrary nature of these coefficients to set them, for example, either to a value which makes the final result more compact or to zero.

In order to get exact solutions and avoid possible issues tied to precision loss in floating point arithmetic, we make use of finite fields arithmetic\footnote{The use of finite fields in high-energy physics has been introduced in \cite{vonManteuffel:2014ixa} in the context of IBP reductions, and further pioneered in \cite{Peraro:2016wsq} where a much wider range of applications was explored. A brief overview of the topic can be found in Appendix \ref{sec:finitefields}.} which is made possible by the fact that at tree-level the kinematic dependence of the amplitudes in the spinor variables is rational. More specifically for each subamplitude we generate a set of momentum-twistors \cite{Hodges:2009hk,Badger:2016uuq} with components on $\mathbb{Z}_p$, where twistors associated to different subamplitudes but to the same internal momentum are by construction taken to be on the same plane\footnote{In twistor space, two intersecting lines define a null momentum, and a closed contour with $n$ edges defines $n$ conserved null momenta. When generating kinematics for the two subamplitudes $\Aa_{m+1}$ and $\Aa_{n-m+1}$ in \eqref{eq:factorizationagain}, $p_{I}$ is defined by the same intersecting lines for both of them.}. From these components then we compute the kinematic invariants and from there the products of the tree-amplitudes, all of which naturally live on the field $\mathbb{Z}_p$. This approach in general greatly speeds up the calculations, having as single minor drawback the fact that to obtain the solution to the linear system on $\mathbb{Q}$ once it has been computed on $\mathbb{Z}_p$ would generally require repeated sampling for different values of the prime $p$ (see appendix \ref{sec:finitefields}). However, since the coefficients involved in our calculations are typically very small compared to the prime $p$ we consider, the use of a single field is usually enough, further strengthened by checking the solutions a posteriori on rational kinematic points. The system solution itself is done through row reduction: the matrix $A$ to be reduced is obtained from numerically evaluating \eqref{eq:kinematicfit} $t+1$ times, with $t$ being the number of $c_{i',j,k}$ appearing in the latter linear equation\footnote{Generating and solving a system with an additional redundant equation ensures that when a determined solution is found this is kinematics-independent and thus a true solution. Impossible systems might still admit determined kinematic-dependent solutions which are clearly unacceptable.}, and can be schematically written as
%\begin{equation}
%    \sum_{I}a_I \, c_{i,j,k} \, g_{1/2/3}^m \, \mathcal{Y}^l \, \lambda^t = a_0 \, g_{1/2/3}^{m'} \, \mathcal{Y}^{l'} \, \lambda^{t'}
%    \hspace{0.5cm} \mapsto \hspace{0.5cm}
%\end{equation}

\begin{equation}\label{eq:system}
    \begin{cases}
        \sum_{s=1}^S a_{0,s}m_s=0 \\
        \sum_{s=1}^S a_{1,s}m_s=0 \\
        \hspace{1cm}\vdots \\
        \sum_{s=1}^S a_{t+1,s}m_s=0
    \end{cases}
    \hspace{0.5cm}
    \mapsto
    \hspace{0.5cm}
    \underbrace{
    \begin{pmatrix}
    a_{0,0} & \cdots & a_{0,S} \\
    \vdots & & \vdots \\
    a_{t+1,0} & \cdots & a_{t+1,S}
    \end{pmatrix}
    }_{A}
    \underbrace{
    \begin{pmatrix}
    m_0 \\
    \vdots \\
    m_S
    \end{pmatrix}}_{V}
    =0 \> ,
\end{equation}
where the $a_{i,j}$ are numeric constants (from the numerical evaluations of the kinematic parts) and the $m_s$ are the unknowns $c_{i,j,k}$ or monomials in the couplings $g$, $\mathcal{Y}$ and $\lambda$ and the imaginary unit $i$. The explicit mention of the imaginary unit is due to the fact that these need to be treated with some care when using finite fields. Imaginary units are almost ubiquitous in our construction and we decided to treat them as symbolic objects on the same footing as the coupling constants. Square roots would in principle require a similar treatment, but these are easily removed by choosing appropriate normalisations of the colour factors, and thus are never present in our calculation. Getting back to the system solution, upon row-reducing the numeric matrix $A$ on finite fields one gets to a matrix $B$ in row echelon form, which of course still satisfies $V' \equiv B\, V=0$, with $V$ the vector of constants $c_{i,j,k}$ and couplings. The relation $V'=0$ can then be trivially solved for the couplings $c_{i,j,k}$ corresponding to the leading (the first non-vanishing) entries in each row of $B$. These relations provide the solution to the system.
%As usual when using finite fields, square roots and imaginary units need to be treated with some care. Our approach is entirely free from square roots, the only possible source being the normalisations of the colour structures which have been appropriately chosen to avoid these. Imaginary units on the other hand are almost ubiquitous, and instead of opting for removing them we decided to treat them as symbolic objects on the same footing as the constants $c_i$ and the couplings $g$ and $\mathcal{Y}$.

It is worth stressing that, differently from either a Feynman diagrammatic approach or a BCFW-like calculation where consistency tests need to be performed a posteriori, through factorisation every step of the calculation is in itself a consistency check on the code. The systems of equations we obtain in the end always have a (possibly vanishing) solution, unless there is some physical obstruction. This is indeed the case when vector bosons are present among the external states (or more in general massless particles with helicity $ |h|\geq 1$) and not enough invariants have been considered in the denominator construction, see Section \ref{sec:vectorbosons}. An impossible solution is symptomatic of unitarity breaking telling us that the ansatz was not general enough.

Thanks to many small, but at times significant, expedients\footnote{These include, for example, recycling numeric data whenever possible, storing and reusing directly the exact invariant products making up the numerators instead of the single invariants, and generating a minimal parametrization of the kinematic points first, reducing thus the numerical kinematic generation to evaluations of polynomials in one/two variables.} the construction of the numeric system is rather fast despite our use of \texttt{Mathematica} rather than dedicated low-level language implementations, for example in C, which are usually better suited for the task. As a consequence, the main bottleneck of the system-solving procedure is the system solution itself. As an aside, we note that our ansatz construction is of course independent of the ansatz solution method. More specifically, if the reader was interested in getting analytic expressions for tree-level amplitudes and already had at her/his disposal a routine for numerically evaluating the amplitude itself, say Berends-Giele \cite{BERENDS1988759} recursion for example, then the ansatz solution could be clearly done in one go solving a single large system in all the $c_{i,j,k}$. Despite being viable, we consider our approach far more appealing, not only conceptually because of the use of just on-shell quantities but also practically: solving the ansatz on the different factorisation channels leads to many small systems whose solution is faster than a single large one and furthermore lends itself to effective parallelisation.

\section{Conclusions}

In this paper we have computed for the first time the one-loop UV mixing matrix $\gamma^{\rm UV}_{8\to 8}$ in the SMEFT at leading order in the SM couplings. Such a calculation requires two main ingredients: the tree-level four-point amplitudes in the SM and the identification of a complete (but not redundant) basis of EFT interactions/operators.

%In recent years the power of on-shell methods in accomplishing both of these tasks has become clear, but m
Most of the time this approach builds on top of foundations made of a Lagrangian and the associated quantum fields. We devoted the first part of this paper to reviewing a whole range of results which allow to completely rid ourselves of such foundations, and build the Standard Model S-matrix from a set of simple physical assumptions
and on-shell quantities.
Once we established the SM particle content, little-group scaling and mass-dimension considerations provide a set of fundamental Poincaré invariant ``minimal'' three-point amplitudes. These by themselves do not provide enough information for a consistent theory to be defined: in fact, unitarity and locality enter the game when four-point amplitudes are considered. Upon writing down the possible manifestly local structures, unitarity (in the guise of factorisation) imposes constraints on the three-point amplitudes forcing the appearance of Jacobi identities \cite{Benincasa:2007xk}, Lie algebras \cite{Arkani-Hamed:2017jhn}, relations among the couplings and charge conservation. This interplay between unitarity and locality further manifests itself at one-loop where the cancellation of inconsistent rational terms imposes additional constraints on the hypercharges, which are obtained as anomaly cancellation conditions in a Lagrangian setting (following the method of \cite{Huang:2013vha,Chen:2014eva}).

Then we discussed how little-group scaling and mass-dimension considerations can provide a set of EFT amplitudes, which correspond to the irrelevant operator basis in a Lagrangian construction of SMEFT. This on-shell analysis provides a crucially efficient way of classifying all the possible interactions. In this paper, we proposed an original take on the construction of the kinematic invariants which enter such minimal amplitudes, making use of multigraphs which allow to effectively build a non-redundant basis accounting for Schouten identities and momentum conservation. After briefly describing the main ingredients in the construction of a set of appropriate colour singlets, which along with the kinematic structures make up a basis of invariant structures, we described how Bose-Einstein and Dirac-Fermi statistics are accounted for.

We then reviewed the on-shell methods to compute the one-loop mixing, which led us to the main result of the paper \cite{Caron-Huot:2016cwu}, {\it i.e.} the mixing matrix of dimension 8 operators with themselves at leading order in the SM couplings.

Extending our results to higher orders requires the knowledge of higher-multiplicity amplitudes which enter when operators of different lengths mix.
In the final part of the paper we presented a completely on-shell algorithm for the construction of arbitrary multiplicity amplitudes (and non-minimal form factors). This algorithm has the advantage of being applicable to any generic renormalisable and non-renormalisable theory, differently to standard BCFW-like recursions which require the theory to be suitably well behaved for large values of the shift parameter. Furthermore, again differing from the standard on-shell recursions, our method also produces manifestly local results, which makes the computed amplitude expressions very well suited for generalised unitarity applications. On the other hand, being based on factorisation properties of the amplitudes, it also retains the desirable feature of making use of lower-point amplitudes only. Our method is based on an ansatz construction, where again locality, mass-dimension considerations and little-group scaling are the main guidelines. To every term in the ansatz a rational coefficient is associated, whose value is fixed by analysing all the possible factorisation channels. In other words, the condition that the residue on a given channel has to be given by the product of lower-point amplitudes is exploited in order to build a system, which is then solved by repeated numerical evaluations over finite fields.

\acknowledgments

%We would like to thank our supervisors Andreas Brandhuber and Gabriele Travaglini, who suggested us to work on this project and supported us at any time, Adrian Keyo Shan Padellaro, Rajath Radhakrishna, Lorenzo Ricci  for many helpful discussions, and Congkao Wen and Chris White for useful discussions and comments on the draft version of the paper. This work  was supported by the European Union's Horizon 2020 research and innovation programme under the Marie Sk\l{}odowska-Curie grant agreement No.~764850 {\it ``\href{https://sagex.org}{SAGEX}''}.

We would like to thank our supervisors Andreas Brandhuber and Gabriele Travaglini, for encouraging us to undertake this project and for their continuous support, Adrian Keyo Shan Padellaro, Rajath Radhakrishna, Lorenzo Ricci for many helpful discussions, and Pierpaolo Mastrolia, Congkao Wen and Chris White for useful discussions and comments on the draft version of the paper. This work  was supported by the European Union's Horizon 2020 research and innovation programme under the Marie Sk\l{}odowska-Curie grant agreement No.~764850 {\it ``\href{https://sagex.org}{SAGEX}''}.

\appendix

\section{Conventions and notations}
\label{sec::conventions}

\subsection{Spinor Helicity Formalism}
\label{sec::spinorhelicity}
Spinor helicity variables\footnote{The \texttt{Mathematica} implementation of the Spinor Helicity Formalism used in this paper has been coded by one of the authors and a beta version is available at the link \url{https://github.com/accettullihuber/SpinorHelicity}} are the most suited object to describe scattering amplitudes. In fact, these are Lorentz invariant functions of the momenta $p_i^\mu$, {\it i.e.} they depend on the momenta through their product $s_{i_1 \dots i_n} = \left( p_{i_1}^\mu + \cdots + p_{i_n}^\mu \right)^2$. By multiplying with gamma matrices, momenta can also be written as tensor transforming in the $d$-dimensional representation of ${\rm Spin}(1,d-1)$, $p_{\alpha \dot{\alpha}} = \gamma^\mu_{\alpha \dot{\alpha}} p_{\mu} \equiv \lambda_{\alpha I} \widetilde{\lambda}^{I}_{\dot{\alpha}}$, where $\lambda_{\alpha I}$ and $\widetilde{\lambda}^{I}_{\dot{\alpha}}$ for the moment are generic rectangular matrices. By definition the Little Group transformations are those which leave the momentum invariant, then without any restriction we can consider the indices $I$ as transforming in the fundamental (spinor) representation of ${\rm Spin}(d-2)$. In $d=4$ and for massless momenta $p_{\alpha \dot{\alpha}}$ has rank 1:
\begin{equation}
    p_{\alpha \dot{\alpha}} = \lambda_{\alpha} \widetilde{\lambda}_{\dot{\alpha}}\ ,
\end{equation}
and the spinor helicity variables transform under $\widetilde{{\rm SO}(2)}$ with a complex phase and helicity weight $\pm \frac{1}{2}$:
\begin{align}
    \lambda_{\alpha} & \to e^{-i\phi/2} \lambda_{\alpha}\ ,\\
    \widetilde{\lambda}_{\dot{\alpha}} &\to e^{i\phi/2} \widetilde{\lambda}_{\dot{\alpha}}\ .
\end{align}
The Lorentz invariant structures that we can form with this variables take the form
\begin{align}
    \agl{i}{j} &= - \epsilon^{\alpha \beta} \lambda_{i\, \alpha} \lambda_{j\, \beta} = \epsilon_{\alpha \beta} \lambda_{i}^{\alpha} \lambda_{j}^{\beta}\ ,\\
    \sqr{i}{j} &= \epsilon^{\dot{\alpha} \dot{\beta}} \lambda_{i\, \dot{\alpha}} \lambda_{j\, \dot{\beta}} = -\epsilon_{\dot{\alpha} \dot{\beta}} \lambda_{i}^{\dot{\alpha}} \lambda_{j}^{\dot{\beta}}\ ,\\
    s_{i j} &= \agl{i}{j} \sqr{j}{i}\ .
\end{align}
Notice that in the case of real momenta $\lambda$ and $\tilde{\lambda}$ are related by complex conjugation:
\begin{equation}
    (\lambda_{\alpha})^* = \pm \widetilde{\lambda}_{\dot{\alpha}}\ ,
\end{equation}
where the sign corresponds to positive or negative energy respectively. However the very definition of three-point amplitudes requires us to work with complex momenta (or alternatively move from a $(1,3)$ signature of space-time to $(2,2)$ \cite{Witten:2003nn}), so that $\lambda$ and $\tilde{\lambda}$ become independent. This allows to write non-vanishing structures as in \eqref{eq:threeptansatz} while having $s_{ij}=0$ satisfied as well.
Finally, when flipping the sign of the momentum $p$ we adopt the symmetric convention on the associated spinors
\begin{equation}
    \lambda_{-p \, \alpha}=i \, \lambda_{p\, \alpha} \>, \hspace{0.5cm} \tilde{\lambda}_{-p\,\dot{\alpha}} = i \, \tilde{\lambda}_{p\,\dot{\alpha}} \>,
\end{equation}
this convention enters also when performing the crossing of fermions from in to out state, leading to a factor $\frac{1}{i}$ for every crossed fermion.

\subsection{The Standard Model gauge group}
\label{sec::SMparticles}

In Table \ref{tab::particlecontent} we write explicitly the representations under which each particle in the infrared spectrum of the Standard Model transforms, for the gauge group $U(1)\times SU(2) \times SU(3)$.

\begin{table}[!ht]
\centering
\begin{tabular}{c|c|c|c|}
          & $U(1)$         & $SU(2)$            & $SU(3)$            \\ \hline
$B_{\pm}$ & $0$        & $\mathbf{1}$            & $\mathbf{1}$            \\ \hline
$W_{\pm}$ & $0$        & $\mathbf{3}$       & $\mathbf{1}$            \\ \hline
$G_{\pm}$ & $0$        & $\mathbf{1}$            & $\mathbf{8}$       \\ \hline
$\bar{Q}$ & $-\frac{1}{6}$ & $\bar{\mathbf{2}}$ & $\bar{\mathbf{3}}$ \\ \hline
$\bar{u}$ & $+\frac{2}{3}$  & $\mathbf{1}$            & $\mathbf{3}$       \\ \hline
$\bar{d}$ & $-\frac{1}{3}$ & $\mathbf{1}$            & $\mathbf{3}$       \\ \hline
$\bar{L}$ & $+\frac{1}{2}$  & $\bar{\mathbf{2}}$ & $\mathbf{1}$            \\ \hline
$\bar{e}$ & $-1$           & $\mathbf{1}$            & $\mathbf{1}$            \\ \hline
$Q$       & $+\frac{1}{6}$  & $\mathbf{2}$       & $\mathbf{3}$       \\ \hline
$u$       & $-\frac{2}{3}$ & $\mathbf{1}$            & $\bar{\mathbf{3}}$ \\ \hline
$d$       & $+\frac{1}{3}$  & $\mathbf{1}$            & $\bar{\mathbf{3}}$ \\ \hline
$L$       & $-\frac{1}{2}$ & $\mathbf{2}$       & $\mathbf{1}$            \\ \hline
$e$       & $+1$            & $\mathbf{1}$            & $\mathbf{1}$            \\ \hline
$\bar{H}$ & $-\frac{1}{2}$ & $\bar{\mathbf{2}}$ & $\mathbf{1}$            \\ \hline
$H$       & $+\frac{1}{2}$  & $\mathbf{2}$       & $\mathbf{1}$   \\  \hline       
\end{tabular}
\label{tab::particlecontent}
\caption{The spectrum of the Standard Model and the transformation properties of all the fields.}
\end{table}

%\subsection{Conventions on the SU(2) and SU(3) structures}
Our convention on the colour factor are completely specified by the decomposition of the contraction of two generators for both the $SU(N)$ and $SU(2)$ groups respectively:
\begin{equation}
    \tau\indices{^{A\, a}_{c}} \tau\indices{^{B\, c}_{b}} = \frac{1}{2 N}\, \delta^{A B} \,\delta^{a}_{b} + \frac{i}{2}\, f^{A B C}\, \tau\indices{^{C\, a}_{b}} + \frac{1}{2} \,d^{A B C}\, \tau\indices{^{C\, a}_{b}}\ ,
\end{equation}
where $f^{A B C}$ are the structure constants and $d^{A B C}$ is the traceless completely symmetry $d$-tensor, and
\begin{equation}
    \sigma\indices{^{I\, i}_{k}} \sigma\indices{^{J\, k}_{j}} = \frac{1}{4}\, \delta^{I J} \,\delta^{i}_{j} + \frac{i}{2}\, \epsilon^{I J K}\, \sigma\indices{^{K\, i}_{j}}\ .
\end{equation}
For the $SU(2)$ group we also need to specify how indices in the fundamental are raised and lowered by the $\epsilon$-tensor:
\begin{equation}
    x_{i} = \epsilon_{i j}\, x^{j} = \epsilon_{i j}\, \epsilon^{j k}\, x_{k}\ .
\end{equation}

\section{3-point amplitudes in the Standard Model}
\label{sec::3points}

In this section we present the complete set of non-vanishing three-point amplitudes in the Standard Model. As already mentioned in section \ref{sec:4pt}, consistent factorisation of the four-point amplitudes imposes constraints which not only fix the colour structures but also relate the couplings of the various three-point amplitudes among each other. Once these constraints are taken into account a small set of the numerical coefficients in front of the amplitudes is still arbitrary and up to convention.

\begin{equation*}
    \Aa (W_{-}^{I},W_{-}^{J},W_{+}^{K}) = g_{2}\, \epsilon^{I J K} \frac{\agl{1}{2}^3}{\agl{2}{3}\agl{3}{1}}\ , \hspace{1cm} \Aa (W_{-}^{I},W_{+}^{J},W_{+}^{K}) = - g_{2}\, \epsilon^{I J K} \frac{\sqr{2}{3}^3}{\sqr{1}{2}\sqr{3}{1}}\ ,
\end{equation*}
\begin{equation*}
    \Aa (G_{-}^{A},G_{-}^{B},G_{+}^{C}) =g_{3} \, f^{A B C} \frac{\agl{1}{2}^3}{\agl{2}{3}\agl{3}{1}}\ , \hspace{1cm} \Aa (G_{-}^{A},G_{+}^{B},G_{+}^{C}) =- g_{3} \, f^{A B C} \frac{\sqr{2}{3}^3}{\sqr{1}{2}\sqr{3}{1}}\ ,
\end{equation*}

\begin{equation*}
    \Aa (B_{-},\bar{e}_m,e_n) = i\, g_{1}\,\delta_{n m}\, \frac{\agl{1}{2}^2}{\agl{2}{3}}\ , \hspace{1cm} \Aa (B_{+},\bar{e}_m,e_n) = i\, g_{1}\,\delta_{n m}\, \frac{\sqr{1}{3}^2}{\sqr{2}{3}}\ ,
\end{equation*}
\begin{equation*}
    \Aa (B_{-},\bar{L}^i_m,L^j_n) =-i\, \frac{g_{1}}{2}\,\delta_{m n}\,\delta_{i}^{j} \frac{\agl{1}{2}^2}{\agl{2}{3}}\ , \hspace{1cm} \Aa (B_{+},\bar{L}^i,L^j) = -i\, \frac{g_{1}}{2}\,\delta_{m n}\,\delta_{i}^{j} \frac{\sqr{1}{3}^2}{\sqr{2}{3}}\ ,
\end{equation*}
\begin{equation*}
    \Aa (B_{-},\bar{u}^a_m,u^b_n) = -i\, \frac{2\, g_{1}}{3}\,\delta_{n m}\,\delta^{a}_{b} \frac{\agl{1}{2}^2}{\agl{2}{3}}\ , \hspace{1cm} \Aa (B_{+},\bar{u}^a_m,u^b_n) = - i\, \frac{2\, g_{1}}{3}\,\delta_{n m}\,\delta^{a}_{b} \frac{\sqr{1}{3}^2}{\sqr{2}{3}}\ ,
\end{equation*}
\begin{equation*}
    \Aa (B_{-},\bar{d}^a_m,d^b_n) = i\, \frac{g_{1}}{3}\,\delta_{n m}\,\delta^{a}_{b} \frac{\agl{1}{2}^2}{\agl{2}{3}}\ , \hspace{1cm} \Aa (B_{+},\bar{d}^a_m,d^b_n) = i\, \frac{g_{1}}{3}\,\delta_{n m}\,\delta^{a}_{b} \frac{\sqr{1}{3}^2}{\sqr{2}{3}}\ ,
\end{equation*}
\begin{equation*}
    \Aa (B_{-},\bar{Q}_m^{a,i},Q_n^{b,j}) = i\, \frac{g_{1}}{6}\,\delta_{m n}\, \delta_{i}^{j} \,\delta^{b}_{a} \frac{\agl{1}{2}^2}{\agl{2}{3}}\ , \hspace{1cm} \Aa (B_{+},\bar{Q}_m^{a,i},Q_n^{b,j}) = i\,\frac{g_{1}}{6}\,\delta_{m n}\, \delta_{i}^{j} \,\delta^{b}_{a} \frac{\sqr{1}{3}^2}{\sqr{2}{3}}\ 
\end{equation*}
\begin{equation*}
    \Aa (B_{-},\bar{H}^{i},H^{j}) = i\, \frac{g_{1}}{2}\,\delta_{i}^{j} \frac{\agl{1}{2}\agl{3}{1}}{\agl{2}{3}}\ , \hspace{1cm} \Aa (B_{+},\bar{H}^{i},H^{j}) = -i\, \frac{g_{1}}{2}\,\delta_{i}^{j} \frac{\sqr{1}{2}\sqr{3}{1}}{\sqr{2}{3}}\ ,
\end{equation*}

\begin{equation*}
    \Aa (W^I_{-},\bar{L}^i_m,L^j_n) = i\,g_{2}\,\delta_{m n}\,\sigma\indices{^{I\, j}_{i}} \frac{\agl{1}{2}^2}{\agl{2}{3}}\ , \hspace{1cm} \Aa (W^I_{+},\bar{L}^i_m,L^j_n) = i\,g_{2}\,\delta_{m n}\,\sigma\indices{^{I\, j}_{i}} \frac{\sqr{1}{3}^2}{\sqr{2}{3}}\ ,
\end{equation*}
\begin{equation*}
    \Aa (W^I_{-},\bar{Q}^{a,i}_m,Q^{b,j}_n) = i\,g_{2}\,\delta_{m n}\,\sigma\indices{^{I\, j}_{i}}\, \delta^b_a \frac{\agl{1}{2}^2}{\agl{2}{3}}\ , \hspace{1cm} \Aa (W^I_{+},\bar{Q}^{a,i}_m,Q^{b,j}_n) = i\,g_{2}\,\delta_{m n}\,\sigma\indices{^{I\, j}_{i}}\, \delta^b_a \frac{\sqr{1}{3}^2}{\sqr{2}{3}}\ ,
\end{equation*}
\begin{equation*}
    \Aa (W^I_{-},\bar{H}^{i},H^{j}) = i\, g_{2}\,\sigma\indices{^{I\, j}_{i}} \frac{\agl{1}{2}\agl{3}{1}}{\agl{2}{3}}\ , \hspace{1cm} \Aa (W^I_{+},\bar{H}^{i},H^{j}) = -i\, g_{2}\,\sigma\indices{^{I\, j}_{i}} \frac{\sqr{1}{2}\sqr{3}{1}}{\sqr{2}{3}}\ ,
\end{equation*}

\begin{equation*}
    \Aa (G^A_{-},\bar{u}_m^a,u_n^b) = -i\,g_{3}\,\delta_{n m}\,\tau\indices{^{A\, a}_{b}} \frac{\agl{1}{2}^2}{\agl{2}{3}}\ , \hspace{1cm} \Aa (G^A_{+},\bar{u}_m^a,u_n^b) = -i\,g_{3}\,\delta_{n m}\,\tau\indices{^{A\, a}_{b}} \frac{\sqr{1}{3}^2}{\sqr{2}{3}}\ ,
\end{equation*}
\begin{equation*}
    \Aa (G^A_{-},\bar{d}^a_m,d^b_n) = -i\,g_{3}\,\delta_{n m}\,\tau\indices{^{A\, a}_{b}} \frac{\agl{1}{2}^2}{\agl{2}{3}}\ , \hspace{1cm} \Aa (G^A_{+},\bar{d}_m^a,d_n^b) = -i\,g_{3}\,\delta_{n m}\, \delta_{i}^{j} \,\tau\indices{^{A\, a}_{b}} \frac{\sqr{1}{3}^2}{\sqr{2}{3}}\ ,
\end{equation*}
\begin{equation*}
    \Aa (G^A_{-},\bar{Q}^{a,i}_m,Q^{b,j}_n) = i\, g_{3}\,\delta_{m n}\,\tau\indices{^{A\, b}_{a}}\, \delta_{i}^{j} \frac{\agl{1}{2}^2}{\agl{2}{3}}\ , \hspace{1cm} \Aa (G^A_{+},\bar{Q}^{a,i}_m,Q^{b,j}_n) = i\,g_{3}\,\delta_{m n}\, \delta_{i}^{j} \,\tau\indices{^{A\, b}_{a}}\frac{\sqr{1}{3}^2}{\sqr{2}{3}}\ ,
\end{equation*}

\begin{equation*}
    \Aa (\bar{Q}^{a,i}_m,\bar{u}^{b}_n,\bar{H}^{j}) = i\,\mathcal{Y}^{(1)}_{m n}\, \epsilon_{i j} \delta^{b}_{a}\, \agl{1}{2}\ , \hspace{1cm} \Aa (Q^{a,i}_m,u^{b}_n,H^{j}) = - i\,\bar{\mathcal{Y}}^{(1)}_{n m}\, \epsilon^{i j} \delta^{a}_{b}\, \sqr{1}{2}\ ,
\end{equation*}
\begin{equation*}
    \Aa (Q^{a,i}_m,d^{b}_n,\bar{H}^{j}) = i\,\mathcal{Y}^{(2)}_{n m}\, \delta_{j}^{i}\, \delta^{a}_{b}\, \sqr{1}{2}\ , \hspace{1cm} \Aa (\bar{Q}^{a,i}_m,\bar{d}^{b}_n,H^{j}) = i\,\bar{\mathcal{Y}}^{(2)}_{m n} \, \delta_{i}^{j}\, \delta^{b}_{a}\, \agl{1}{2}\ ,
\end{equation*}
\begin{equation*}
    \Aa (L^{i}_m,e_n,\bar{H}^{j}) = i\,\mathcal{Y}^{(3)}_{n m}\, \delta_{j}^{i}\, \sqr{1}{2}\ , \hspace{1cm}\Aa (\bar{L}^{i}_m,\bar{e}_n,H^{j}) = i\,\bar{\mathcal{Y}}^{(3)}_{m n} \, \delta_{i}^{j}\, \agl{1}{2}\ .
\end{equation*}

\section{One-loop scalar integrals}\label{sec:integrals}

The expression for the following scalar integrals have been taken from \cite{Bern:1994cg}. Defining
%\begin{equation}
%    I_m=(-1)^{m+1} \, i \, (4 \pi )^{2-\epsilon}\, \int \frac{d^{4-2\epsilon}l}{(2\pi)^{4-2\epsilon}} \, \frac{1}{l^2 \, (l-p_1)^2 \,(l-p_1-p_2)^2\cdots (l+p_m)^2 } \>,
%\end{equation}
\begin{equation}
    r_{\Gamma}=\frac{\Gamma(1+\epsilon)\Gamma^2(1-\epsilon)}{\Gamma(1-2\epsilon)}
\end{equation}
one has for the bubble
\begin{equation}
    I_2(s_{12})=\int \frac{d^{4-2\epsilon}l}{(2\pi)^{4-2\epsilon}} \, \frac{1}{l^2\,(l-p_1-p_2)^2}=\frac{i}{(4 \pi )^{2-\epsilon}}\frac{r_{\Gamma}}{\epsilon(1-2\epsilon)}(-s_{12})^{-\epsilon}\> ,
\end{equation}
for the one-mass triangle
\begin{equation}
    I_3(s_{12})= \int \frac{d^{4-2\epsilon}l}{(2\pi)^{4-2\epsilon}} \, \frac{1}{l^2 \, (l-p_1)^2 \,(l-p_1-p_2)^2 }=-\frac{i}{(4 \pi )^{2-\epsilon}}\frac{r_{\Gamma}}{\epsilon^2}(-s_{12})^{-1-\epsilon} \> ,
\end{equation}
and finally the massless box
\begin{equation}
\def\arraystretch{2}
\begin{array}{rl}
I_4(s_{12},s_{14})&=\displaystyle\int \frac{d^{4-2\epsilon}l}{(2\pi)^{4-2\epsilon}} \, \dfrac{1}{l^2 \, (l-p_3)^2 \,(l-p_3-p_4)^2 (l+p_2)^2 } \\ &=\dfrac{i}{(4 \pi )^{2-\epsilon}} \displaystyle\frac{r_{\Gamma}}{s_{12}s_{14}}\left[\frac{2}{\epsilon^2}\left((-s_{12})^{-\epsilon}+(-s_{14})^{-\epsilon} \right) -\log^2\left( \frac{-s_{12}}{-s_{14}}\right) -\pi^2 \right] \> .
\end{array}
\end{equation}

\section{Infrared collinear anomalous dimensions in the Standard Model}
\label{sec::collinear}

In this section we are going to show an example of the computation of the collinear anomalous dimension for the $W$ bosons in the Standard Model and we will give the result for all the particles in the spectrum of the theory.

We start by giving the stress-tensor form factor \cite{Arkani-Hamed:2017jhn} following the normalisation procedure given in \cite{Caron-Huot:2016cwu} for generic complex scalars, fermions and vectors respectively\footnote{The different overall minus sign with respect to \cite{Caron-Huot:2016cwu} comes from our different convention choice for $\lambda^{\alpha}_{-k} = i \lambda^{\alpha}_{k}$ and $\widetilde{\lambda}^{\dot{\alpha}}_{-k} = i \widetilde{\lambda}^{\dot{\alpha}}_{k}$, while the authors in \cite{Caron-Huot:2016cwu} chose $\lambda^{\alpha}_{-k} = \lambda^{\alpha}_{k}$ and $\widetilde{\lambda}^{\dot{\alpha}}_{-k} = - \widetilde{\lambda}^{\dot{\alpha}}_{k}$.}:
\begin{equation}
\begin{split}
    \langle \bar{\phi}^{\mathbb{A}} \phi^{\mathbb{B}} | T^{\alpha \dot{\alpha} \beta \dot{\beta}} | 0 \rangle &=\frac{1}{3} \delta_{\mathbb{A}}^{\mathbb{B}} \Big( \lambda^{\alpha}_{1} \lambda^{\beta}_{1} \widetilde{\lambda}^{\dot{\alpha}}_{1} \widetilde{\lambda}^{\dot{\beta}}_{1} - \lambda^{\alpha}_{1} \lambda^{\beta}_{2} \widetilde{\lambda}^{\dot{\alpha}}_{1} \widetilde{\lambda}^{\dot{\beta}}_{2} - \lambda^{\alpha}_{1} \lambda^{\beta}_{2} \widetilde{\lambda}^{\dot{\alpha}}_{2} \widetilde{\lambda}^{\dot{\beta}}_{1} - \lambda^{\alpha}_{2} \lambda^{\beta}_{1} \widetilde{\lambda}^{\dot{\alpha}}_{1} \widetilde{\lambda}^{\dot{\beta}}_{2}\\
    &\hspace{1.5cm}- \lambda^{\alpha}_{2} \lambda^{\beta}_{1} \widetilde{\lambda}^{\dot{\alpha}}_{2} \widetilde{\lambda}^{\dot{\beta}}_{1} + \lambda^{\alpha}_{2} \lambda^{\beta}_{2} \widetilde{\lambda}^{\dot{\alpha}}_{2} \widetilde{\lambda}^{\dot{\beta}}_{2} \Big)\\
    \langle \bar{\psi}^{\mathbb{A}} \psi^{\mathbb{B}} | T^{\alpha \dot{\alpha} \beta \dot{\beta}} | 0 \rangle &= \frac{1}{2} \delta_{\mathbb{A}}^{\mathbb{B}} \left( \lambda^{\alpha}_{1} \lambda^{\beta}_{1} \widetilde{\lambda}^{\dot{\alpha}}_{2} \widetilde{\lambda}^{\dot{\beta}}_{1} +\lambda^{\alpha}_{1} \lambda^{\beta}_{1} \widetilde{\lambda}^{\dot{\alpha}}_{1} \widetilde{\lambda}^{\dot{\beta}}_{2}  -\lambda^{\alpha}_{1} \lambda^{\beta}_{2} \widetilde{\lambda}^{\dot{\alpha}}_{2} \widetilde{\lambda}^{\dot{\beta}}_{2} - \lambda^{\alpha}_{2} \lambda^{\beta}_{1} \widetilde{\lambda}^{\dot{\alpha}}_{2} \widetilde{\lambda}^{\dot{\beta}}_{2} \right)\\
    \langle v_{-}^{\mathbb{I}} v_{+}^{\mathbb{J}} | T^{\alpha \dot{\alpha} \beta \dot{\beta}} | 0 \rangle &= -2\, \delta^{\mathbb{I}\, \mathbb{J}}\, \lambda^{\alpha}_{1} \lambda^{\beta}_{1} \widetilde{\lambda}^{\dot{\alpha}}_{2} \widetilde{\lambda}^{\dot{\beta}}_{2}\ ,\\
\end{split}
\end{equation}
where $\mathbb{A},\,\mathbb{B},\,\mathbb{I},\,\mathbb{J}$ are generic colour indices. Once we fix the minimal form factor for the stress tensor, we can apply the formula \eqref{eq::collinear}:
\begin{equation}
\label{eq::collinearW}
    \begin{split}
        \langle W_{-}^{I} W_{+}^{J} | T^{\mu \nu} | 0 \rangle \cdot \sum_{l=1}^{2}\frac{\gamma_{\rm coll}^{(l)}}{16\pi^2}  = \frac{1}{\pi} \sum_{\{l_1,l_2\}} \int& \frac{{\rm d} \Omega_2}{32 \pi^2} \Bigg[\Aa_4 (p_{l_1}^{h_{l_1}} p_{l_2}^{h_{l_2}}\to W_{-}^{I} W_{+}^{J})\\
        &-\sum_{k=1}^{3}\frac{ g_k^2\ T_{k,l_1}\cdot T_{k,l_2}}{\cos^2\theta\, \sin^2\theta}\Bigg] \cdot \langle p_{l_1}^{h_{l_1}} p_{l_2}^{h_{l_2}} | T^{\mu \nu} | 0 \rangle\ ,
    \end{split}
\end{equation}
where the sum over $\{ l_1,l_2\}$ runs over the pairs
\begin{equation}
    \left\{ \{ W_{-},W_{+}\},\{ W_{+},W_{-}\},\{ \bar{Q},Q\},\{Q,\bar{Q}\},\{ \bar{L},L\},\{L,\bar{L}\},\{ \bar{H},H\},\{H,\bar{H}\} \right\}\ .
\end{equation}
Considering that $\gamma_{\rm coll}^{W_{-}} =\gamma_{\rm coll}^{W_{+}} \coloneqq \gamma_{\rm coll}^{W}$, we can rewrite \eqref{eq::collinearW} as
\begin{equation}
        \gamma_{\rm coll}^{W} = 8\pi \sum_{\{l_1,l_2\}} \int \frac{{\rm d} \Omega_2}{32 \pi^2} \Bigg[\Aa_4 (p_{l_1}^{h_{l_1}} p_{l_2}^{h_{l_2}}\to W_{-}^{I} W_{+}^{J})-\sum_{k=1}^{3}\frac{ g_k^2\ T_{k,l_1}\cdot T_{k,l_2}}{\cos^2\theta\, \sin^2\theta}\Bigg] \cdot \frac{\langle p_{l_1}^{h_{l_1}} p_{l_2}^{h_{l_2}} | T^{\mu \nu} | 0 \rangle}{\langle W_{-}^{I} W_{+}^{J} | T^{\mu \nu} | 0 \rangle}\ ,
\end{equation}
We will list now the different contributions from the $W$ bosons (which need the infrared divergence subtraction), the quarks, the leptons and the Higgs doublet, respectively:
\begin{equation}
    \gamma_{\rm coll}^{W} = - g_2^2 \left(\frac{11}{3}\times 2- \frac{N_f}{3}\times 3 - \frac{N_f}{3} - \frac{1}{6}\right)\ ,
\end{equation}
where the factor of $\times  2$ in the first term is the Casimir of the adjoint representation of $SU(2)$, while the factor of $\times 3$ in the second term comes from the sum on different colour of the quarks. This is the usual result for the $SU(2)$ beta function with $N_f$ Weyl fermions and $1$ scalar, both transforming in the fundamental of the gauge group.

Finally, we give the explicit results for the other states in the Standard Model. We start from the vector bosons
\begin{align}
    \gamma_{\rm coll}^{B} &= \frac{2}{3}\, g_1^2 \left[ \left(Y_Q^2 \times 2 + Y_u^2 + Y_d^2\right) \times 3+ \left( Y_L^2 \times 2 + Y_e^2 \right) + Y_H^2 \right]\ ,\\
    \gamma_{\rm coll}^{G} &= - g_3^2 \left(\frac{11}{3}\times 3- \frac{N_f}{3}\times 2 \times 2\right)\ ,
\end{align}
where the first $\times 2$ in the second term of $\gamma_{\rm coll}^{G}$ comes from the sum over $SU(2)$ indices (or equivalently over $d$ and $u$) and the second $\times 2$ factor comes from the fact that $SU(3)$ is not a chiral theory and the quarks behave as a doublet of Dirac fermions. Then we have the collinear anomalous dimensions for the fermions
\begin{align}
    \left(\gamma_{\rm coll}^{Q}\right)_{m n} &= - 3 \left( g_1^2\, Y_Q^2 + \frac{3}{4}\, g_2^2 + \frac{8}{6}\, g_3^2 \right) \delta_{m n} + \mathcal{Y}^{(1)}_{m p}\, \bar{\mathcal{Y}}^{(1)}_{p n} + \mathcal{Y}^{(2)}_{m p}\, \bar{\mathcal{Y}}^{(2)}_{p n}\ ,\\
    \left(\gamma_{\rm coll}^{u}\right)_{m n} &= - 3 \left( g_1^2\, Y_u^2 + \frac{8}{6}\, g_3^2 \right) \delta_{m n} + 2\, \bar{\mathcal{Y}}^{(1)}_{n p}\, \mathcal{Y}^{(1)}_{p m}\ ,\\
    \left(\gamma_{\rm coll}^{d}\right)_{m n} &= - 3 \left( g_1^2\, Y_d^2 + \frac{8}{6}\, g_3^2 \right) \delta_{m n} + 2\, \bar{\mathcal{Y}}^{(2)}_{n p}\, \mathcal{Y}^{(2)}_{p m}\ ,\\
    \left(\gamma_{\rm coll}^{L}\right)_{m n} &= - 3 \left( g_1^2\, Y_L^2 + \frac{3}{4}\, g_2^2 \right) \delta_{m n} + \mathcal{Y}^{(3)}_{m p}\, \bar{\mathcal{Y}}^{(3)}_{p n}\ ,\\
    \left(\gamma_{\rm coll}^{e}\right)_{m n} &= - 3\, g_1^2\, Y_e^2\, \delta_{m n} + 2\, \bar{\mathcal{Y}}^{(3)}_{n p}\, \mathcal{Y}^{(3)}_{p m}\ ,
\end{align}
and, finally, the Higgs
\begin{equation}
    \gamma_{\rm coll}^{H} = - 4\, g_1^2\, Y_H^2 -  4\, g_2^2 \times \frac{3}{4} +  2\, \Tr\, \mathcal{Y}^{(1)} \cdot \bar{\mathcal{Y}}^{(1)} \times 3 +2\,\Tr\, \mathcal{Y}^{(2)} \cdot \bar{\mathcal{Y}}^{(2)} \times 3 + 2\,\Tr\, \mathcal{Y}^{(3)} \cdot \bar{\mathcal{Y}}^{(3)}\ ,
\end{equation}
where $\frac{3}{4}$ and $\frac{8}{6}$ are the Casimir of the fundamental representation of $SU(2)$ and $SU(3)$, respectively.

\section{Finite field arithmetic}\label{sec:finitefields}

In this section we briefly describe the main features of finite field kinematics. Our goal is to give just a taste of the method, motivating its usefulness in our particular context, highlighting at the same time the caveats which come along the benefits. For a more in depth mathematical primer we refer the interested reader to \cite{lidl_niederreiter_1994}  and references therein, whereas for a discussion of applications to modern theoretical physics problems to \cite{Peraro:2016wsq,Peraro:2019svx}.

%Consider the integer numbers $\mathbb{Z}$ endowed with the standard addition and multiplication and choose a natural number $p \in \mathbb{N}$. We define a set of $p$ equivalence classes where $a$ is equivalent to $b$, or  if $a-b= n \, p$ for some $n \in \mathbb{Z}$, we will also say that $a$ equals $b$ modulo $p$ or $a=b \, mod \, p$. The natural numbers
Consider the integer numbers $\mathbb{Z}$ endowed with the standard addition and multiplication and choose a natural number $p \in \mathbb{N}$. We define a set of $p$ equivalence classes through the modulo operation $\rm mod$, we say that $a$ is equivalent to $b$ or $a$ equals $b$ modulo $p$ if
\begin{equation}
    a=b \ \, {\rm mod} \ \, p \hspace{0.5cm} \iff \hspace{0.5cm} \exists \, n \in \mathbb{Z} \hspace{0.3cm} s.t. \hspace{0.3cm} a-b= n \cdot p\ .
\end{equation}
The set of natural numbers
\begin{equation}
 \mathbb{Z}_p\equiv\{0,1,\ldots,p-1\}   
\end{equation}
can be chosen as the most intuitive representatives of these equivalence classes, and it is easy to see that this set endowed with the standard addition (${\rm mod} \, p$) is a representation of the cyclic group of order $p$, hence the use of the symbol $\mathbb{Z}_p$.
It can be shown that if $p$ is a prime number then $\mathbb{Z}_p$ endowed also with the usual multiplication (${\rm mod} \, p$) is a field, which is finite by construction and we thus call it \textit{finite field}. The modulo operation provides clearly a simple map from $\mathbb{Z}$ to $\mathbb{Z}_p$, what is less obvious, but still holds true as long as $p$ is a prime, is that under some restriction there is also a unique and well defined map from the rationals $\mathbb{Q}$ to $\mathbb{Z}_p$. This map is based on the possibility of defining a multiplicative inverse $a^{-1}$ for every $a \in \mathbb{Z}_p$ such that $a \, a^{-1}=a^{-1} \, a=1 \, {\rm mod} \, p$.

%Now that we have given an operative definition of finite fields we can focus on why and how we use them. The starting point is that we want to obtain analytic expressions for tree-level scattering amplitudes, and we want to do so by imposing consistency conditions on our ansatz (\eqref{eq:ampansatz} which has locality built-in) through factorisation (which ensures unitarity). This is precisely what \eqref{eq:kinematicfit} does, however solving it analytically 
%Now that we have given an operative definition of finite fields we can focus on why and how we use them. The starting point is that we want to obtain analytic expressions for tree-level scattering amplitudes. We want to obtain them from lower-point on-shell amplitudes thus avoiding unphysical (and potentially very unhandy) intermediate expressions like those involved in a Feynman diagrammatic calculation.
Now that we have given an operative definition of finite fields we can focus on why and how we use them. Performing some sort of analytic computation on a computer (especially on a laptop) can often prove challenging in that the computational time required is too large for a result to be successfully obtained. In similar situations it might be a good idea to change the perspective on the problem and try to reformulate it using a numeric approach. In our case this amounts to switching from trying to directly obtain the amplitude from simplifying \eqref{eq:factorisation} analytically to building numeric systems to be solved as in \ref{sec:ansatz}. The advantage of numerics is that in principle it is clearly much faster, since potentially large intermediate expressions are replaced with numbers. This is certainly true when dealing for example with floating-point arithmetic. On the other hand, the use of numeric expressions requires carefully keeping track of possible precision loss and makes arbitrary precision arithmetic at times more appealing, which is however slower. Here is where finite fields enter the game, since we can map our problem from $\mathbb{Q}$ to $\mathbb{Z}_p$, which avoids the precision loss of floating point numbers, and then perform the numeric computations on $\mathbb{Z}_p$, which is extremely fast because we can choose $p$ to be a machine-size prime and thus the whole computation will only involve machine-size natural numbers. The obvious issue is that once the problem at hand has been solved on $\mathbb{Z}_p$ we need to map the solution back to $\mathbb{Q}$, through a map which cannot by any means be a bijection.

Despite the fact that the map $\mathbb{Q} \to \mathbb{Z}_p$ is not injective, it is possible under certain circumstances to ``invert it'', or rather to make an educated guess of which element $\frac{n}{d} \in \mathbb{Q}$ corresponds to a given $b \in \mathbb{Z}_p$. In particular it can be shown (see for example \cite{10.1145/800206.806398}) that given $b$ there is only one pair of $n$ and $d$ such that $n^2,d^2<\frac{p}{2}$. In other words, if the correct values of $n$ and $d$ which we are looking for are small enough compared to the prime $p$ which we chose as the order of the field $\mathbb{Z}_p$, then we can uniquely obtain their value from $b \in \mathbb{Z}_p$. The size of $p$ however has an upper bound being the machine-size primes, since the whole point of using finite fields is to deal with machine-size integers. Consequently, one cannot simply choose an arbitrarily large prime so to be confident that the inverted map yields the correct result. Instead, one uses the so called Chinese remainder theorem\footnote{See for example \cite{Peraro:2016wsq}.}, which allows to combine the outcome $X$ of the same calculation on multiple fields $\mathbb{Z}_{p_1}, \ldots, \mathbb{Z}_{p_n}$ to obtain the value of $X \, {\rm mod} \, P$ where $P=\prod_i p_i$. In other words, the Chinese remainder theorem defines a ring isomorphism\footnote{Since $P$ is not a prime $\mathbb{Z}_p$ is not a field but a ring.}
\begin{equation}
\def\arraystretch{1.5}
\begin{array}{rl}
      \mathbb{Z}_{p_1} \times \cdots \times \mathbb{Z}_{p_n} &\to \mathbb{Z}_P \\
      \left( X \, {\rm mod} \, p_1, \ldots , X \, {\rm mod} \, p_n \right) & \mapsto X \, {\rm mod} \, P
\end{array}
\end{equation}
which allows us to access the value $X \, {\rm mod} \, P$ on $\mathbb{Z}_P$ where $P$ is large, through computations on fields with small values of $p_i$. Applying then the ``inverse'' mapping to the value found on $\mathbb{Z}_P$ will very likely return the correct result. This procedure is iterated adding more fields $\mathbb{Z}_{p_{n+1}}$ until the inverse of $b \in \mathbb{Z}_P$ converges to a definite $\frac{n}{d} \in \mathbb{Q}$.

In our specific case it is usually enough to perform the calculation on a single field $\mathbb{Z}_p$: since we are using finite fields to do numeric evaluations aimed at solving the system \eqref{eq:system}, once a solution has been found we can simply test it through a single evaluation of \eqref{eq:kinematicfit} on $\mathbb{Q}$.
As a final remark, we discuss one caveat of the method. Since everything relies on the possibility of mapping $\mathbb{Q}$ to $\mathbb{Z}_p$, one can only apply the so far presented techniques in the case the problem at hand is entirely described by rational functions. This is indeed the scenario we are interested in: in fact, the tree-level scattering amplitudes present an entirely rational dependence on the spinor invariants and thus on the spinor components. Furthermore, one has to take special care of elements appearing in the calculations which do belong to more extended number fields than $\mathbb{Q}$, in particular for us this means square roots and imaginary units. How these are dealt with is often a matter of the specific problem, where our choices have been described in Section \ref{sec::ansatzsol}. 

%Take the set of $p$ natural numbers $\{0,1,\ldots,p-1\}$ and consider them as representatives of the equivalence classes defined by the modulo operation with respect to $p$, where we say that $a$ is equivalent to $b$ if $a-b=n \, p$ for some integer $n$.

\bibliographystyle{JHEP}
\bibliography{remainder}

\end{document}